\documentclass[preprint]{elsarticle}
\usepackage{tabularx}
\usepackage{adjustbox}
\usepackage{wrapfig}

\newcommand\clearrow{\global\let\rowmac\relax}
\clearrow

\usepackage[T1]{fontenc}
\usepackage[utf8]{inputenc}
\usepackage{amsmath}
\usepackage{natbib}
\usepackage{graphicx}
\usepackage{gensymb}
\usepackage{url}
\usepackage{mathtools}
\usepackage[varg]{txfonts}
\usepackage{textcomp}
\usepackage[super]{nth}
\usepackage{caption}
\usepackage{subcaption}
\usepackage{rotating}
\usepackage{siunitx}

\usepackage{float}
\usepackage{booktabs}

\usepackage{wasysym}
\usepackage{esvect}

\usepackage{siunitx}
\usepackage{lscape}
\usepackage{longtable}
\usepackage{threeparttable}

\biboptions{authoryear}

\begin{document}

\begin{frontmatter}

\title{Determining the population of Large Meteoroids in Major Meteor Showers}

\author[uwopa]{K. S. Wisniewski}
\ead{kwisnie2@uwo.ca}

\author[uwopa,wiese]{P. G. Brown}
\author[nasa]{D. E. Moser}
\author[sandia]{R. Longenbaugh}

\address[uwopa]{Department of Physics and Astronomy, University of Western Ontario, London, Ontario, N6A 3K7, Canada}
\address[wiese]{Western Institute for Earth and Space Exploration, University of Western Ontario, London, Ontario, N6A 5B7, Canada}
\address[nasa]{NASA Meteoroid Environment Office, Marshall Space Flight Center, Huntsville, AL 35812 USA}
\address[sandia]{Sandia National Laboratories, Albuquerque, NM 87185, USA}

\begin{abstract}
 We have estimated the largest meteoroids present in major meteor showers from observations conducted between 2019-2022 by the Geostationary Lightning Mapper (GLM) instrument on the GOES-R satellites. Our integrated time area products for the Leonids, Perseids and eta Aquariids are of order 5$\times$10$^{10}$ km$^2$ hours.  We compute photometric masses for shower fireballs using the approach of \cite{vojacek} to correct from narrow-band GLM luminosity to bolometric luminosity and apply the luminous efficiency relation of \cite{Ceplecha_McCrosky_1976} at high speeds. Between 2019 and 2022, the showers definitely observed by GLM were the Leonids, Perseids, and eta Aquariids, with probable detections of the Orionids and Taurids. We find the largest meteoroids to be of order 7 kg for the Leonids, 3 kg for the Perseids, and 3 kg for the eta Aquariids, corresponding to meteoroids of $\approx$ 0.2m diameter. The Orionids and Taurids had maximum meteoroid masses of 4 kg and 150 kg respectively. The Leonids and eta Aquariids are well fit by a single power-law with differential mass exponent, s, of 2.08$\pm$0.08 and 2.00$\pm$0.09 over the mass range 10$^{-7}$ < m < 1 kg. All showers had maximum meteoroid masses compatible with Whipple gas-drag ejection, with the exception of the Perseids which have much larger meteoroids than expected a result also consistent with observations from ground based instruments. This may reflect preferential ejection in narrow jets or possibly some form of mantle erosion/release in the past for the parent comet, 109P/Swift-Tuttle.
\end{abstract}

\end{frontmatter}

\section{Introduction}
\label{sec:intro}

There are several mechanisms which may lead to meteoroid stream formation. The most widely accepted and likely most common is gas-drag sublimation \citep{whipple_1951}. In this process, grains/pebbles are lofted from a cometary nucleus when they are released from the ice-dust matrix of a comet under the effects of solar heating and then entrained with the expanding gas to be ejected from the nucleus. Another potential formation mechanism is nuclear disruption \citep{Belton2015}, whereby essentially the entire mass of a comet nucleus is released at once in a catastrophic disintegration \citep{Jenniskens2006}. Less explored, but still possible formation mechanisms which result in debris release from a parent comet or asteroid, include small meteoroid collisions  \citep{Bottke2020bennumet}, larger, near catastrophic collisions \citep{Hunt1986}, tidal disruption \citep{Richardson1998}, YORP spin-up disruption \citep{Walsh2012} or electrostatic lofting \citep{Hartzell2022}.

Distinguishing between these formation mechanisms for a given stream is challenging. Fortunately, the gas-drag mechanism makes specific predictions about the expected maximum size of particles which can be released \citep{whipple_1951}. Measuring the largest mass meteoroid in a stream is one test of whether gas-drag formed the stream.

The maximum meteoroid mass is also essential for estimating the total meteoroid stream mass as well as the annual mass deposition to Earth by these showers \citep{Hughes1989}. Large stream meteoroids also provide insight into the survival timescales of large cometary boulders detected near active cometary nuclei \citep[e.g.][]{Kelley2013}.

Studying the fragmentation mechanics of a parent comet during each perihelion pass also sheds light on the number and behaviour of large meteoroids. Different models propose different expected maximum meteoroid sizes depending on formation mechanism and parent for a stream, but empirical evidence is lacking.

Here we address this problem by directly measuring the largest meteoroids in several major streams using the Geostationary Lightning Mapper (GLM) onboard the Geostationary Operational Environmental Satellites (GOES).

\section{Previous Work and Theoretical Limits on Maximum Meteoroid Size}
\label{sec:lit_review_theo_work}
\subsection{Telescopic Surveys}

Several previous works have attempted to constrain the upper size limit in streams through telescopic surveys aimed at in-situ detection of large meteoroids. 

\cite{Barabanov_1996} conducted a telescopic survey of the Perseid shower in August of 1995 to search for large bodies within the stream. During the observation period, 4 objects were reported to potentially belong to the Perseids. The size distribution using two different surface reflectivity values ranged from 3 m to 47 m for these bodies, yielding masses up to $10^{8}$ kg.  

Similarly, in a later study, \cite{Barabanov_2005} continued the work from \cite{Barabanov_1996} and reported detection of 124 objects potentially belonging to the Perseid stream from 1995 to 2002. Included also in this study were other showers, such as 18 objects that could potentially belong to the Leonids. Based on these measurements, they predicted a number density of bodies over 1 m in size for the Perseids of $1.3 \times 10^{-6}/ 10^{9} km^3$ based on telescopic observations from 1997 to 2002. For the Leonids, based solely on telescopic observations in 1999, this number density was estimated as $3 \times 10^{-6}/ 10^{9} km^3$.

Motivated by the \cite{Barabanov_1996} observations, \cite{beech_2003} conducted a telescopic survey aimed at confirming the presence of large meteoroids within the Perseids. The telescopic survey failed to confirm the \cite{Barabanov_1996} results. It did provide an overview of potential Perseid detections derived from United States Department of Defence and Department of Energy satellite data spanning 23 years. Within this period, only six fireballs were possibly linked to the Perseids based purely on time of occurrence, resulting in an upper limit to the spatial population density of  $7 \times 10^{-8}$ meteoroids per $10^9 km^3$ to a limiting mass of order 100 kg. However, the lack of detailed velocity information constrains the validity of the resulting flux and detection limits to extreme upper limits. While not statistically significant, this study outlines methodologies for estimating population indices of meteoroids of various sizes within meteoroid streams. It also serves as a foundation for predicting the expected number of large stream meteoroids based on telescopic surveys.

In \cite{Micheli_Tholen_2015}, a telescopic survey searched for meter-sized meteoroids in various meteoroid streams using the Canada-France-Hawai’i telescope's MegaCam and the Pan-STARRS Gigapixel camera. Focus was primarily directed towards the Geminid and Taurid streams due to telescope availability and previous observations of large meteoroids associated with these streams. Additionally, the Quadrantids were considered due to their association with the asteroid 2003 EH1, suggesting the potential presence of larger stream members. The April alpha-Comae Berenicids were also considered, based on the observation of two large fireballs with similar orbits to the stream, indicating the potential for large associated meteoroids.

This telescopic campaign spanned ten observing nights with the two telescopes. While some Near-Earth Objects (NEOs) were observed, none could be correlated with the orbits of the selected meteoroid streams. Using the fluxes proposed by \cite{Barabanov_Smirnov_2005} they showed that the absence of detections of meter-sized meteoroids in these streams in their survey is in tension with the conclusion of \cite{Micheli_Tholen_2015}. They also note more generally that the lack of any detection of meter-sized stream meteoroids implies a relative deficit in meter-sized meteoroids compared to the population of small meteoroids in the same streams. 

Other surveys have focused on attempting to detect large meteoroids in the Taurid meteoroid stream, as several large Near Earth Asteroids (NEAs) have been provisionally linked to the shower complex \citep{Spurny2017}. \cite{Clark_2019} provided a summary of the best viewing opportunity during the Taurid Swarm perihelion pass of June 2019 for observing NEAs associated with the swarm. While no new Taurid NEAs were confirmed during the 2019 surveys, the stream remains a promising target for detection of large Taurids \citep{Egal2022}. 

\subsection{Gas Sublimation Ejection limits on Meteoroid Masses}
\label{sec:gas_sub_mass_ej}

Water gas-drag sublimation occurs when cometary nuclei approach to within 2.5 Astronomical Units  (AU) from the Sun and water ice begins to vigorously sublimate \citep{Jones1995a}, a process which can eject particles embedded in the nucleus. We first summarize the Whipple gas-drag theoretical model which serves as a foundational means of estimating the largest meteoroids expected in showers. Next we will summarize the literature containing observations of fireballs/meteoroids on a per shower basis. These provide independent constraints which we can compare to our final results found using GLM. 

The primary theoretical framework for estimating the maximum meteoroid size ejected from a comet is drawn from Whipple's 1951 model "A Comet Model II: Physical Relations for Comets and Meteors" (\cite{whipple_1951}). In this picture,  comets possess a conglomerate structure where meteoritic material is bound together by various ices that vaporize into gas at typical temperatures (\cite{whipple_1951}). As these structures approach the Sun within approximately 2.5 AU, the icy "glue" sublimates, injecting meteoritic material into a meteoroid stream (\cite{whipple_1951}).

The ejection mass model, as presented by \cite{whipple_1951}, considers materials that have moved considerably from the nucleus and emphasizes gases escaping from the Sunlit hemisphere of the comet. This model assumes rough and irregularly shaped meteoritic material having a gas-drag coefficient approximated as 2(1+4/9) or approximately 2.9. 

By solving equations for net outward acceleration and relative velocity, a relationship governing the maximum diameter ($s_{max}$) of a meteoroid ejected into a meteoroid stream due to gas sublimation is derived (\ref{eqn:whip}).

\begin{equation}
\label{eqn:whip}
s_{max} = \frac{19 \mathrm{cm}}{nr^{9/4}R_C}
\end{equation}

This relation depends solely on the solar radiation efficiency ($n$) which is normally taken to be unity, comet radius ($R_C$) in kilometers, and Sun distance ($r$) in AU. Despite its simplicity, this relationship holds significant predictive power, and it continues to be employed.

By utilizing the parent body density and this maximum diameter value, we can estimate the maximum mass ($M_{max}$) in kilograms for several showers for future comparison (see Table~\ref{tab:whipple}). We have assumed a meteoroid density of 1000 kg/$m^3$ and for Phaethon a density of 1600 kg/$m^3$.

Note that we use the simple, first-order equation \ref{eqn:whip} as a guide to calculate the expected largest meteoroid lofted from a nucleus surface. Modifications and updates to Whipple's work have explicitly accounted for variations in the gas thermal expansion velocity, meteoroid shape/drag and gas flux \citep{Harmon2004} as well as ejection from narrow jets \citep{Jones1995a}. All of these modify the upper size limit, though typically by factors less than two compared to Whipple's original formulation \citep{Beech1999}. We recognize these updates can potentially modify the Whipple values for upper max limit but we will use Whipple's original model as a first-order approximation. Parent body information are taken from the Jet Propulsion Laboratory database for small bodies\footnote{Source: https://ssd.jpl.nasa.gov/} with the exception of C/1861 G1 Thatcher cited from \cite{beech_nikolova_1999}.

\begin{table*}[h]
\centering
\begin{adjustbox}{width=1\textwidth}
    \begin{tabular}{|l|l|l|l|l|l|l|}
    \hline
        \textbf{Shower} & \textbf{Parent Body} & \textbf{Radius $\mathbf{R_C}$ (km)} & \textbf{Perihelion $\mathbf{r}$ (AU)} & \textbf{$\mathbf{D_{max}}$ (m)} & \textbf{$\mathbf{M_{max}}$ (kg) } \\ \hline
        \textbf{Leonids} & 55P/Tempel-Tuttle  & 1.8 & 0.98 & 0.22 & 5.59  \\ \hline
        \textbf{Perseids} & 109P/Swift-Tuttle & 13.0 & 0.97 & 0.03 & 0.02  \\ \hline
        \textbf{Geminids} & 3200 Phaethon & 3.1 & 0.14 & 9.99 & 8.71E+05  \\ \hline
        \textbf{Lyrids} & C/1861 G1 (Thatcher) & 5.5 & 0.95 & 0.08 & 0.25  \\ \hline
        \textbf{SDA} & 96P/Machholz & 3.2 & 0.12 & 12.96 & 1.14E+06  \\ \hline
        \textbf{ETA} & 1P/Halley & 5.5 & 0.57 & 0.24 & 7.37  \\ \hline
        \textbf{Orionids} & 1P/Halley & 5.5 & 0.57 & 0.24 & 7.37  \\ \hline
        \textbf{Taurids} & 2P/Encke & 2.4 & 0.34 & 1.84 & 2.62E+03  \\ \hline
        \textbf{Quadrantids} & 2003 EH1  & 2.0 & 1.19 & 0.13 & 1.12 \\ \hline

    \end{tabular}
\end{adjustbox}
\caption{Estimated maximum meteoroid mass and diameter ($D_{max}$) which can be lofted by gas-drag at perihelion according to Whipple's comet model for parent bodies of showers studied in this work. Note that we assume a bulk meteoroid density of 1000 $kg/m^{3}$ throughout and n=1 for all calculations.}
\label{tab:whipple}
\end{table*}

\subsection{Maximum meteoroid mass in streams from prior meteor observations}
\label{sec:max_mass_obs}

Meteoroid mass influx is a measure of the mass received in a given area per unit time dependant on the masses of the particles m to m + dm as described by \cite{Hughes_1976}. The equation for influx, $\psi$ in $g cm^{-2} s^{-1}$ is: 

\begin{equation}
\psi = C m^{1-s} dm
\label{equ:hughes_differential}
\end{equation}

\noindent where $C$ is a constant, and $s$ is the differential mass index. For a meteoroid stream, we can integrate this equation with respect to $m$ by using the two mass limits $m_1$ for the smallest particle size and $m_2$ for the largest particle size to estimate the total stream mass as 

\begin{equation}
\psi = \int_{m_1}^{m_2} Cm^{1-s} dm
\label{equ:hughes}
\end{equation}

This relation is heavily dependant on the integral bounds for the smallest and largest particle size. We can easily determine an estimate for the smallest particle size for a stream from existing ground observational techniques such as radar or using the known lower limit for a bound orbit for small particles subject to radiation pressure \citep{Moorhead2021}. However, the maximum size of a meteoroid contained within a stream is not a quantity that is well known. Optical systems tend to saturate when recording very bright fireballs. They also don't cover enough of the atmosphere to provide the collection area needed for large and therefore rare bright shower meteors. The following subsections will summarize the available literature where reports of kilogram-sized, shower-associated fireballs belonging to the major streams are examined in this paper. Showers with few reported events will be summarized in Table \ref{table:lit_review}. Magnitudes are for the visual spectrum.  

\subsubsection{Taurids}
\label{sec:tau_mass}
The Taurids for the purpose of this study include both the Northern and Southern Taurids together due to their overlap in activity times and similar physical properties. We do not consider the many smaller showers which may be linked to the broader Taurid Complex \citep{Egal2022}.
One of the largest fireballs reported for the Taurid shower was recorded by the European Fireball Network (EN) on 2015-10-31 18:05:20 UTC during the 2015 resonant outburst of the shower and analysed by \cite{borovicka_spurny_2020}. They estimated an initial mass of 650 kg and a diameter of 0.7 m. Similarly, during the 1995 resonant return, several large Taurids were also recorded by the EN, one with a mass approaching a metric tonne \citep{Jenniskens2006book}. The Taurids are unique as the only shower having multiple observations confirming the presence of multi-hundred kilogram meteoroids.  

Other sources report numerous examples of kilogram-sized Taurids. For example, \cite{madiedo_2014} investigated the Taurids and their associated fireballs as potential meteorite sources, utilizing observations from the Spanish Meteor Network (SPMN). Among these fireballs, one Taurid had a  mass of 29 kg with a diameter of 32 cm. Similarly, two Taurid fireballs recorded by the Prairie Network (PN) by \cite{Ceplecha_McCrosky_1976} from the Northern and Southern Taurids respectively were found to have photometric masses of 8 and 2.5 kg respectively. In their study, \cite{brown_2013} examined the Taurid shower to assess the potential for meteorite recoveries from fireballs. The Southern Ontario Meteor Network (SOMN) observed a Taurid fireball on 2010-10-31 04:43:33 UTC and estimated an initial photometric mass of approximately 4 kg. 
Lastly, conducting a spectral and physical properties analysis of Taurid fireballs using data from the All-Sky Meteor Orbit Network (AMOS), \cite{matlovic_2017} identified several fireballs with photometric masses exceeding 1 kg; the largest was 1.5 kg, featuring an associated magnitude of -8.4. 

\subsubsection{Geminids}
\label{sec:gem_mass}

In their study, \cite{halliday_1988} investigated Geminid meteors from the Meteorite Observation and Recovery Project (MORP). The largest Geminid meteor within their analysis exhibited a photometric mass of 2.4 kg and a magnitude of -11.3. 

In a later study, \cite{Halliday_1996} noted in the MORP a Geminid of 2.5 kg mass. 

\cite{beech_2003} examined a "flickering" Geminid fireball observed in 2002 by the Southern Saskatchewan Fireball Array (SSFA). Through light curve analysis, they determined its photometric mass as 0.429 kg, with a maximum magnitude of -9.2. This study compared the derived mass with the \cite{halliday_1988} study based on brightness, noting similar-magnitude events corresponded to similar orders of mass.

\cite{borovicka_2010} investigated multiple Geminid fireballs observed by the EN. The largest fireball in this study had a magnitude of -9 and a mass of 1 kg. 

Lastly, \cite{madiedo_2013} evaluated the potential for Geminid meteor streams to be sources of meteorites. An observed Geminid fireball in 2009, recorded by the SPMN, displayed a magnitude of -13. Its photometric mass was determined to be 0.757 kg, and its diameter measured 7.3 cm, based on a bulk density of 2900 kg/$m^3$.

\subsubsection{Leonids}
\label{sec:leo_mass}
Observing the 1998 Leonid storm over China, \cite{spurny_2000} used photographic observations with all-sky Canon T-70 cameras. The most massive fireball recorded within this study had a mass of 1.1 kg and a peak magnitude of -13.2.  

In a later study, \cite{shrbeny_2009} investigated Leonid fireballs through the EN 1999 and 2006. The most substantial recorded Leonid fireball in their study has a mass of 2.1 kg and peak magnitude of -14.3. This is the most massive Leonid recorded from the ground.

\subsubsection{Perseids}
\label{sec:per_mass}

One of the largest Perseid fireball was detected from ground-based cameras of the EN and was estimated to have a mass of 0.42 kg \citep{Ceplecha_1977}.

Employing spectral analysis and the luminous efficiency relation, \cite{borovicka_1997} derived a mass of 0.08 kg for a large Perseid fireball using the EN in a later study.

In their study, \cite{kokhirova_2020} report Perseid meteoroids with masses up to 0.02 kg and peak magnitudes of -9 observed by the Tajikistan Fireball Network (TF).

\subsubsection{Estimating the Largest Meteoroids in a Stream from Lunar Impacts}
\label{sec:lunar_impact_mass}

Many showers in our survey have also been detected during lunar impact monitoring, where actual meteoroid impacts are detected as brief flashes on the dark portion of the lunar disk. Data on the flux of kilogram-sized meteoroids based on lunar flashes are discussed by \cite{Suggs_2014}. The moon's extensive surface acts as a large detector, offering several orders of magnitude more collecting area compared to a single ground-based all-sky camera. 

One of the most recent lunar impact studies which includes a summary of many earlier works is that of \cite{Avdellidou_2021}. They outline the methodology for preparing a dedicated telescopic survey system for lunar impact monitoring. From the over 300 lunar impacts in their database, we extract meteoroid masses for several of our study showers. Using a luminous efficiency of $5 \times 10^{-5}$ and their dataset  we find the following largest shower impactors: a 7.1 kg Perseid,  2.7 kg Geminid, 1.6 kg Northern Taurid, 1.1 kg Southern Taurid, 0.2 kg Lyrid, 0.07 kg Leonid, 0.07 kg Quadrantid, a 0.03 kg Orionid and a 0.02 kg Eta Aquariid. 

In addition to this work, several previous surveys help set useful limits on large shower meteoroids based on lunar impacts. 

One of the first studies to be conducted was by \cite{Duennebier_Nakamura_Latham_Dorman_1976} using seismic detectors on the moon placed during the Apollo missions. Most notable was detection of a Taurid Storm, associated with the Beta Taurid or Zeta Perseid showers \citep{Egal2022}, producing impacts in June 1975. The estimated total mass due to meteoroid impacts was put at 320 kg, with individual impactors ranging from 0.05 kg to 50 kg.

\cite{Oberst1991} examined shower signatures among lunar seismic impacts as a whole. They confirmed that the daytime Taurid complex were the only showers associated with "large" (multi-kilogram) impactors among the seismic impact database. They found smaller (sub-kilogram) impactors associated with all other showers listed in Table \ref{tab:whipple} with the exception of Quadrantids and Lyrids.

In an early study involving the Sierra Nevada Observatory, \cite{bellotrubio_2000} published optical detections of Leonids impacting the moon on 1999-11-18 at the peak of the 1999 Leonid storm. They reported five Leonid impactors exceeding a kilogram, the largest being 4.9 kg. Using similar equipment on the same night \cite{ortiz_2000} reported the largest optical flash associated with a Leonid lunar impactor to be approximately 2.5 kg.

\cite{yanagisawa_2006} reported the first confirmed Perseid lunar impact recorded at the Ogawa Observatory. The impactors mass was calculated at 0.012 kg. 

One of the earliest comprehensive surveys of lunar impact frequency was conducted by \cite{Suggs_2014} as part of a project at NASA's Marshall Space Flight Center's (MSFC) devoted to routine lunar monitoring. 

Beginning in 2006, \cite{Suggs_2014} utilized two 0.35m Schmidt-Cassegrain telescopes to monitor lunar impacts over a five year period. During this time over 300 lunar impacts were jointly detected at both stations. They identified flashes most likely associated with the peak activity of established showers by examining the frequency of flashes in 2° solar longitude bins. Although impact flashes cannot be directly linked to showers, they calculated the likelihood of an impact stemming from a stream based on factors including Zenithal Hourly Rate (ZHR), time or solar longitude, and the distance between impact and the moon's sub-radiant point. Results for the largest lunar impacts from this study can be seen in Table \ref{table:lit_review}. 

More recently, \cite{madiedo_2015} published observations of lunar impacts during the Perseids using two 0.3 Schmidt-Cassegrain telescopes during the peak of the 2012 and 2013 Perseids. This study complements \cite{Suggs_2014} by filling in observations during peak shower times where none were previously captured. Thirteen Perseid impacts were observed, the largest of mass 0.19 kg are detailed in Table \ref{table:lit_review}.

From lunar impact observations, the most massive shower-associated meteoroids are listed in Table \ref{table:lit_review}. While most are below one kilogram, these measurements are valuable as they offer similar time-area products to GLM (see section \ref{sec:CA_methods})  and have masses estimated from surface impact luminous efficiency models \citep[e.g.][]{Bouley2012}, providing distinct masses from those found using meteor luminous efficiency measurements in the atmosphere.

\subsection{Recent Meteor Shower Fireballs from the European Fireball Network}

 \cite{Shrbeny_2009_thesis} provided a summary and details of bright fireballs associated with several major showers detected in recent years by the European Fireball Network. Their work summarised fireball observations taken using data from the Czech region of the European Fireball Network, the Australian Desert Network, and Leonids captured during the 1999 outburst from Spanish Network observations.  The showers of interest for this study reported by \cite{Shrbeny_2009_thesis} are the Orionids, Geminids, Southern delta Aquariids, Leonids (which was alrealy summarized in \ref{sec:leo_mass} and the Perseids. Table \ref{shrbeny_thesis_sum} summarizes the largest fireballs for each shower from their study. 

\begin{table}[!ht]
    \centering
    \begin{tabular}{|l|l|l|l|}
    \hline
        \textbf{Shower} & \textbf{Year - Fireball} & \textbf{Mass (kg)} & \textbf{Magnitude} \\ \hline
        \textbf{Orionids} & 2007 - ORI13 & 0.058 & -10.7 \\ \hline
        \textbf{Geminids} & 2006 - GEM08 & 11 & -12.6 \\ \hline
        \textbf{Southern delta Aquariids} & 2006 - SDA03 & 5 & -13.3 \\ \hline
        \textbf{Leonids} & 2001 - LEO31 & 2.1 & -14.3 \\ \hline
        \textbf{Perseids} & 2007 - PER01 & 0.037 & -8.9 \\ \hline
    \end{tabular}
    \caption{Summary of the largest shower fireballs from \cite{Shrbeny_2009_thesis}}
    \label{shrbeny_thesis_sum}
\end{table}

\subsection{Summary: Largest Shower Meteoroids}
\label{sec:lit_rev_sum}

In addition to the datasets previously summarized, we examined several historical fireball network catalogues which have records of bright shower fireballs. This included the Meteorite Observation and Recovery Project (MORP) \citep{Halliday1978}, the European Fireball Network (EN) \citep{Ceplecha_1977} and the Prairie Network (PN) \citep{McCrosky1965}. These are summarized in Table \ref{tab:network}. 

\begin{table*}[h]
\centering
    \begin{tabular}{|l|l|l|l|l|}
    \hline
        \textbf{Shower}  & \textbf{Collection} \textbf{Method} & \textbf{Mass (kg)}  & \textbf{Magnitude}   & \textbf{Source for Citation}   \\ \hline
        \textbf{Northern Taurids}  & MORP  & 4.8 & -11.9  & \citep{Halliday_1996}   \\ \hline
        \textbf{Northern Taurids}  & EN  & 0.73 & -9  & \cite{Spurny_1997}   \\ \hline
        \textbf{Northern Taurids}  & PN  & 7.943 & -11.40  & \cite{Ceplecha_McCrosky_1976}   \\ \hline
        \textbf{Southern Taurids}  & MORP  & 1.2 & -11.2 &  \cite{Halliday_1996}   \\ \hline
        \textbf{Southern Taurids}  & EN  & 2.4 & -14.2 &  \cite{Spurny_1997}   \\ \hline
        \textbf{Southern Taurids}  & PN  & 2.512 & -10.60 &  \cite{Ceplecha_McCrosky_1976}   \\ \hline
        \textbf{Perseids}  & MORP  & 0.21 & -11.9 &  \cite{Halliday_1996}   \\ \hline
        \textbf{Perseids}  & EN  & 0.42 & -12.5 &  \cite{Ceplecha_1977}   \\ \hline
        \textbf{Perseids}  & PN  & 0.1 & N/A &  \cite{Ceplecha_McCrosky_1976}   \\ \hline
        \textbf{Geminids}  & MORP  & 2.5 & -11.6 &  \citep{Halliday_1996}   \\ \hline
        \textbf{Geminids}  & EN  & 1.8 & -10 &  \citep{Spurny_1994}   \\ \hline
        \textbf{Leonids}  & MORP  & 0.23 & -12.3 &  \citep{Halliday_1996}  \\ \hline
    \end{tabular}
    \caption{A summary of the largest shower fireballs reported from the MORP, EN and PN. Magnitude here represents panchromatic peak magnitude.}
    \label{tab:network}
\end{table*}

Table \ref{table:lit_review} summarizes all existing literature data concerning the largest shower meteoroids which have been reported from a variety of different methods.  We also searched for the largest shower members detected by the Southern Ontario Meteor Network (SOMN) \citep{Brown2010a} collected between 2004-2022 and add these to our database for showers otherwise poorly monitored. The masses are as reported in each associated reference. Note that the luminous efficiencies at higher speeds reported for some early EN and PN data use the mass scale and luminous efficiency from \citep{Ceplecha_McCrosky_1976}. This is a factor of several smaller than the luminous efficiency scale used by \citet{Borovicka_2020} for more recent measurements. As a result, the earlier fireball surveys have a mass scale which is several times higher on average than the contemporary scale proposed by \citet{Borovicka_2020}. 

We also emphasize that many ground-based fireball surveys are mainly focused on meteorite-dropping fireballs so are likely incomplete with respect to bright shower fireballs. An Eta Aquariid and Orionid have been added to this table using data reductions from the SOMN, as there are few literature reports of bright fireballs from these showers. Shower three letter codes follow the IAU convention \footnote{Source: www.ta3.sk/IAUC22DB/MDC2022}.

\begin{table}[ht!]
    \centering
    \resizebox{\textwidth}{!}{\begin{tabular}{|l|l|l|l|l|}
        \hline
        \textbf{Shower} & \textbf{Technique/Network} & \textbf{Mass (kg)} & \textbf{Magnitude} & \textbf{Source}\\ 
        \hline
        \hline
        \textbf{eta Aquariids} & Lunar Impact & 0.15 & N/A & \cite{Suggs_2014}   \\ 
        \textbf{eta Aquariids} & Southern Ontario Meteor Network & 0.017 & -7.59 & \\ 
        \hline
        \textbf{Geminids} & Meteorite Observation and Recovery Project & 2.50 & -11.5 & \cite{Halliday_1996}  \\ 
        \textbf{Geminids} & Lunar Impact & 2.70 & N/A & \cite{Avdellidou_2021}   \\ 
        \textbf{Geminids} & European Network & 11.00 & -12.6 & \cite{Shrbeny_2009_thesis}  \\ 
        \hline
        \textbf{Leonids} & Lunar Impact & 4.90 & N/A & \cite{bellotrubio_2000}   \\ 
        \textbf{Leonids} & European Network & 2.10 & -14.3 & \cite{shrbeny_2009}   \\ 
        \textbf{Leonids} & Meteorite Observation and Recovery Project & 0.23 & -12.3 & \cite{Halliday_1996}   \\ 
        \hline
        \textbf{Lyrids} & Lunar Impact & 0.20 & N/A & \cite{Suggs_2014}   \\ 
        \hline
        \textbf{Orionids} & Lunar Impact & 0.04 & N/A & \cite{Suggs_2014}   \\ 
        \textbf{Orionids} & European Network & 0.20 & -10.3 & \cite{Borovicka_2022}   \\
        \textbf{Orionids} & Polish Fireball Network & 1.5 & -14.7 & \cite{Olech2013}   \\ 
        \textbf{Orionids} & Southern Ontario Meteor Network & 0.031 & -9.18  &\\
        \hline
        \textbf{Perseids} & Lunar Impact & 7.10 & N/A & \cite{Avdellidou_2021}   \\ 
        \textbf{Perseids} & Prairie Network & 0.10 & N/A & \cite{Ceplecha_McCrosky_1976}  \\ 
        \textbf{Perseids} & European Network & 0.42 & -12.5 & \cite{Ceplecha_1977}   \\ 
        \textbf{Perseids} & Prairie Network & 0.10 & N/A & \cite{Ceplecha_McCrosky_1976}   \\ 
        \hline
        \textbf{Quadrantids} & Lunar Impact & 0.07 & N/A & \cite{Avdellidou_2021}  \\ 
        \textbf{Quadrantids} & European Network & 2.00 & -11 & \cite{borovicka_2010} \\ 
        \hline
        \textbf{Southern delta Aquariids} & Lunar Impact & 0.16 & N/A & \cite{Suggs_2014}   \\ 
        \textbf{Southern delta Aquariids} & European Network & 5.0 & -13.3 & \cite{Shrbeny_2009_thesis}  \\ 
        \hline
        \textbf{Northern Taurids} & Lunar Impact & 1.60 & N/A & \cite{Avdellidou_2021}  \\ 
        \textbf{Northern Taurids} & Meteorite Observation and Recovery Project & 4.80 & -11.9 & \cite{Halliday_1996}   \\ 
        \textbf{Northern Taurids} & European Network & 0.73 & -9 & \cite{Spurny_1997}   \\ 
        \textbf{Northern Taurids} & Prairie Network & 7.94 & N/A & \cite{Ceplecha_McCrosky_1976}  \\ 
        \hline
        \textbf{Southern Taurids} & Lunar Impact & 1.10 & N/A & \cite{Avdellidou_2021}  \\ 
        \textbf{Southern Taurids} & Meteorite Observation and Recovery Project & 1.20 & -11.2 & \cite{Halliday_1996}   \\ 
        \textbf{Southern Taurids} & European Network & 2.40 & -14.2 & \cite{Spurny_1997}  \\ 
        \textbf{Southern Taurids} & Prairie Network & 2.51 & N/A & \cite{Ceplecha_McCrosky_1976}   \\ 
        \hline
        \textbf{Taurids} & European Network & 650 & -18.6 & \cite{borovicka_spurny_2020} \\ 

        \hline
    \end{tabular}}
    \caption{A summary of the literature values for the largest observed meteoroids among our study showers given in Table \ref{tab:whipple}.}
\label{table:lit_review}
\end{table}

\begin{figure*}[ht!]
  \includegraphics[width=\linewidth]{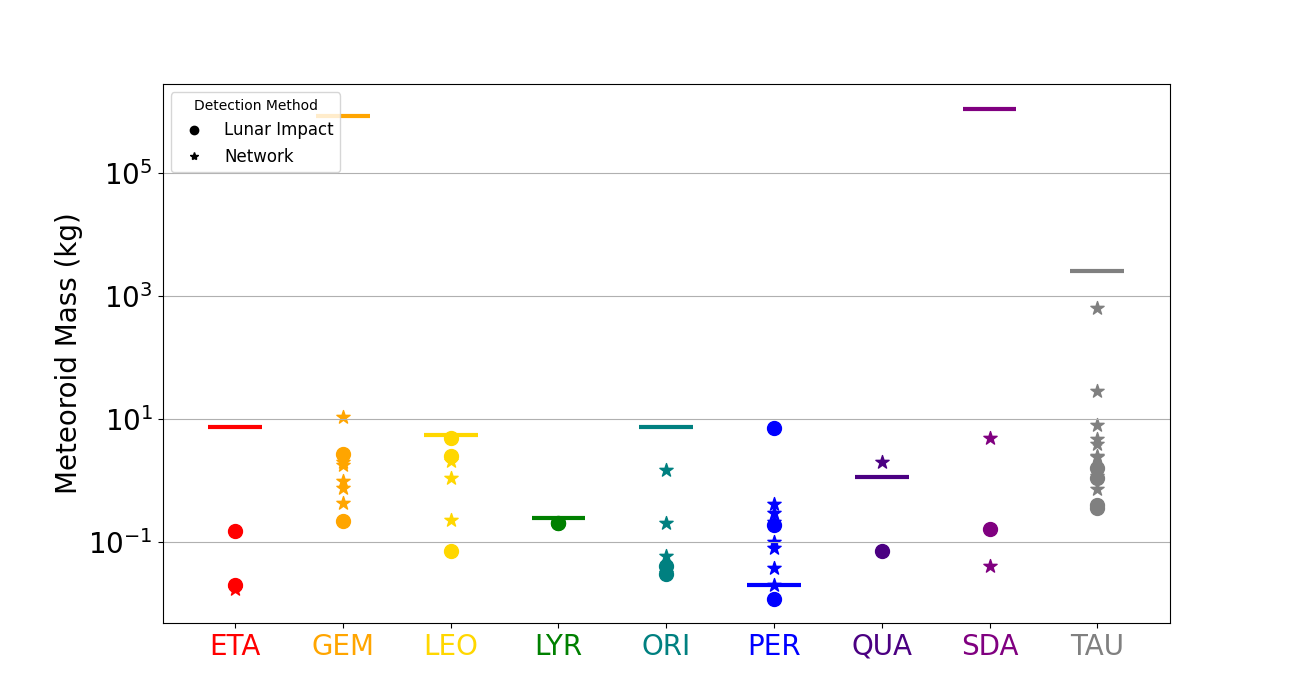}
  \caption{The maximum reported literature mass for meteoroids in major showers from different techniques (shown by symbols). The gas-drag theoretical upper mass limit for each of the major showers is shown by a horizontal line.}
  \label{fig:lit_review}
\end{figure*}

In Figure \ref{fig:lit_review} we compare both the meteoroid mass limit estimated from gas-drag and the observations summarized in Table \ref{table:lit_review}. 
 
\section{Instruments}
\label{sec:instruments}

\subsection{Geostationary Lightning Mapper (GLM)}
\label{sec:glm}

For our goal of detecting massive shower meteoroids, we exploit the large time-area product offered by the Geostationary Lightning Mapper (GLM) instrument onboard the Geostationary Operational Environmental Satellites (GOES) to identify likely shower-related fireballs.

The concept of observing lightning from cloud tops was introduced by \cite{christian_blakeslee_goodman_1989}. Earlier lightning detections from low Earth orbit satellites yielded suboptimal data due to limited spatial resolution and the inability to provide continuous storm monitoring. To enhance data quality, the potential integration of a lightning instrument into National Oceanic and Atmospheric Administration's (NOAA) GOES series was proposed. This system was recognized early in its development for its potential applicability in detecting fireballs during atmospheric entry, owing to the similar luminous characteristics of lightning and fireballs \citep{Tagliaferri94}.

The GOES-R series was initiated by the launch of GOES-16 on 2016-11-19, and GOES-17 on 2019-02-12. The instrument suite onboard each satellite consists of 6 instruments: Advanced Baseline Imager (ABI), Extreme Ultraviolet and X-ray Irradiance Sensors (EXIS), Magnetometer (MAG), Space Environment In-Situ Suite (SEISS), the Solar Ultraviolet Imager (SUVI), and the Geostationary Lightning Mapper (GLM)\footnote{Source: https://www.goes-r.gov/}. The GLM instruments onboard GOES-16 and GOES-17 are referred to as GLM-16 (also known as GLM-East) and GLM-17 (GLM-West), respectively. These sensors are positioned to view in a nadir pointing configuration. 

Encompassing about 40 \% of Earth's surface, the combined GLM sensors cover North and South America, along with substantial portions of the Atlantic and Pacific Oceans, within latitudinal range of 55° N to 55° S and longitudinal range of 16° W to 165° W. These sensors operate throughout the day and night, providing observations even in Sunlit regions.

GLM's significance in fireball detection arises from its measurement of a specific wavelength – the neutral oxygen line at 777 nm. This wavelength is one of the primary emission lines for lightning discharges however it also is produced by meteors, particularly those with high velocity \citep{Ceplecha1998}. The instrument's narrow bandwidth is tailored for lightning detection, but still allows observation of fireballs spanning a range of velocities. The complication in GLM fireball characterization is estimating the true energy of the fireball from GLM's very narrow (1-2 nm) bandpass \citep{jenniskens}.

\cite{jenniskens} was the first to explore GLM's ability to detect fireballs using GLM-16. This initial study found that a minimum detection threshold magnitude was roughly -14 for bright, relatively slow fireballs. More recently, \cite{vojacek} suggests that for higher velocity (>30 km/s) meteors, the threshold might extend up to -8. \cite{jenniskens} also summarized a significant fireball which occurred over the West Atlantic on 2017-03-11 during GLM commissioning which represented the most energetic recorded fireball by GLM up to that time. The United States Government satellite data produced an energy estimate of 2.9 kT which suggested a 2-6 m diameter impactor, consistent with GLM's measurements, though some saturation occurred near the peak of the event. This fireball had a peak magnitude of -22.

Such bright fireballs tend to saturate ground cameras, hindering the extraction of physical properties. Consequently, knowledge gaps persist regarding the upper size limit of meteoroids in showers, both due to the saturation problem for ground-based cameras and the small collection area accessible from ground networks. This is a gap GLM is ideally suited to address.

Moreover, GLM also detects fireballs during both day and night, effectively doubling the observation time compared to ground-based cameras. Additionally, its wide field of view (FOV) surpasses that of ground-based all-sky cameras which have collecting areas of order 500,000 km\textsuperscript{2}. GLM has a collecting area some 400 times this value.

Employing a machine learning algorithm and human oversight, GLM-detected fireball events of the group Level 2 data are made available through NASA's online repository\footnote{Source: https://neo-fireball.ndc.nasa.gov/}. It is from this database that our data is extracted. This site contains a map of detection by both sensors, plots of the light curves and positions of the detections for individual, stereo heights, information added in a sentence or two by the human vetting process, and the downloadable csv files for individual events. Efforts are ongoing to automate the process entirely, distinguishing fireball events from other phenomena as discussed in \cite{rumpf_longenbaugh_henze_chavez_mathias_2019} and \cite{smith_morris_rumpf_longenbaugh_mccurdy_henze_dotson_2021}. For this study we used GLM data from 2019 to 2022, comprising approximately 4500 fireballs. 

\subsection{United States Government Sensors (USG)}
\label{section:usg}

Data from US Government Sensors (USG) are satellite-based measurements capable of detecting fireballs in the silicon bandpass and providing data on their atmospheric entry and ablation \citep{Tagliaferri94}.

NASA's Jet Propulsion Laboratory hosts the dataset through the Center for Near Earth Objects Studies (CNEOS) site\footnote{Source: https://cneos.jpl.nasa.gov/fireballs/}, featuring observations captured by USG. This includes positional information, radiated energy, total impact energy, and occasionally velocity and height at peak brightness measurements. 

This information aids in assessing whether significant fireballs observed by GLM might also be detected by USG. The USG data also includes intensity versus time (light curves) for some USG observations\footnote{Source: https://cneos.jpl.nasa.gov/fireballs/lc/}. These are valuable independent checks for calibrating GLM fireball observations allowing us to compare our approach for computing total energy from GLM measurements to the approach used by USG. USG energies have been validated through comparisons with other techniques such as infrasound \citep[e.g.][]{brown_flux, Ens2012, Gi2017}, meteorite-producing fireballs \citep{Borovicka2015_AST4} and ground-based optical measurements \citep{Devillepoix2019}. 

\section{GLM Shower Identification Methodology}
\label{sec:methods}

We have developed several different processing pipelines for the GLM data to extract fireballs likely related to different meteor showers. From these potential shower fireballs we then use that data to derive properties of the most energetic shower events, both individually and as a population. 

\subsection{The Filtering Processes}
\label{sec:filter}

\begin{figure*}
  \includegraphics[width=\linewidth]{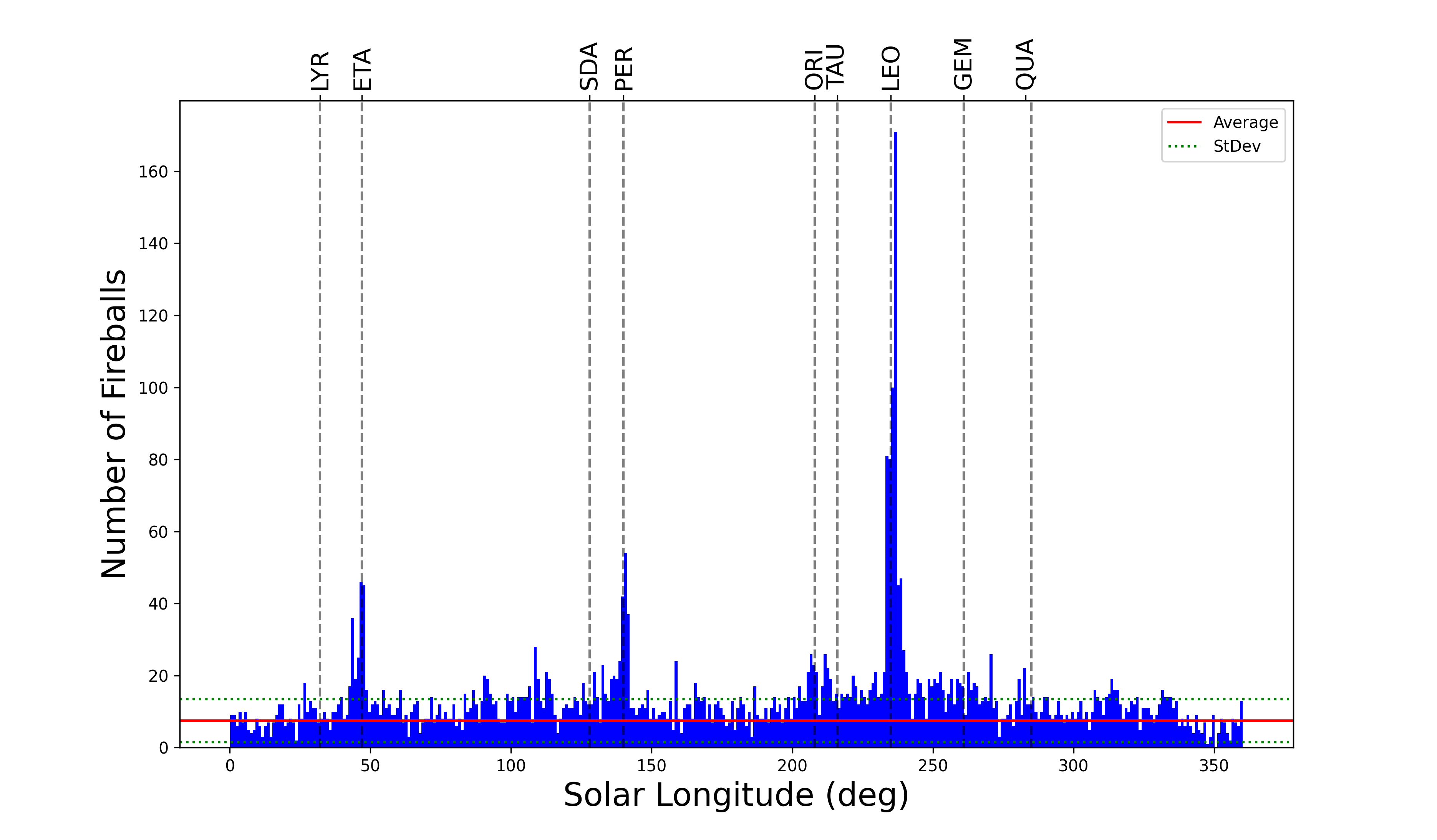}
  \caption{Histogram of all the GLM fireballs detected between 2018 and 2022 in one degree  bins. Also shown are the times of major meteor shower peaks as a function of J2000 solar longitude (vertical dashed lines), the average daily fireball rate (red solid line) and the standard deviation (green dotted line) using days outside major shower activity. Only the LEO, PER and ETA are at least 3$\sigma$ above the average.}
  \label{fig:all_events}
\end{figure*}

The observed frequency of GLM fireball detection with time is shown in Figure \ref{fig:all_events}. From the timing of localized maxima, we infer that the following major showers may be detected by GLM and use these as the starting list for our survey: the Leonids (LEO), Perseids (PER), Taurids (TAU), eta Aquariids (ETA), Geminids (GEM), Southern delta Aquariids (SDA), Lyrids (LYR), and Orionids (ORI). The GLM's 777.4 nm band detection favors high-velocity showers \citep{vojacek}, enhancing the likelihood of detection for less energetic events for faster streams. From Figure \ref{fig:all_events}, it is clear that during the time near their respective maximum, the Leonids and Perseids exhibit substantial detections by GLM, due in part to their high velocities (70.2 and 59.1 km/s, respectively). The eta Aquariids, with a velocity of 65.7 km/s, also display a significant signal. The average daily detections, excluding shower-related events, reveal periods of elevated signal beyond typical background plus one standard deviation (refer to Figure \ref{fig:all_events}). While the Taurids present a weaker signal due to slower velocities, their notably high literature mass values for some fireballs make them an interesting shower for GLM study. 

Two significant filters are applied to the complete dataset as a preliminary step to establish an initial working list of shower-related detections termed "possible fireballs" for a given shower.  The first filter is temporal and the second utilizes the radiant geometry.

With the temporal filter we identify the duration of showers as the time window where a signal (number of events per day) is above the regular GLM fireball rate background by at least 1 $\sigma$. For the Leonids, this one $\sigma$ criteria was exceeded within the solar longitude ($\lambda$\textsubscript{\(\odot\)}) range of 231 to 241$^{\circ}$, for the Perseids $\lambda$\textsubscript{\(\odot\)} = 136 to 142$^{\circ}$, and for the eta Aquariids $\lambda$\textsubscript{\(\odot\)} = 42 to 49$^{\circ}$. The Orionids only just barely met this criteria for 205$^{\circ}$ <$\lambda$\textsubscript{\(\odot\)}< 209$^{\circ}$ and finally the Taurids were just above background from 211$^{\circ}$ <$\lambda$\textsubscript{\(\odot\)}< 218$^{\circ}$. We can see from Figure \ref{fig:all_events}, the Leonids have a peak signal over 15$\sigma$ above background making it the strongest shower signal within the GLM fireball dataset.We have high confidence that most events in this interval are associated with these showers. We also note that the Taurids (TAU) and the Orionids (ORI) show some signal at least one $\sigma$ above the average. 

As we have no way to definitively correlate specific GLM fireballs with showers using velocities at this point in the filtering process, we choose to focus on the dominant three showers for our analysis, namely the LEO, PER and ETA. This selection ensures that our results have minimal contamination from sporadic fireballs.

Analysis of the TAU and ORI are included in the Appendix. The rest of the showers listed in Table \ref{tab:whipple} show such weak temporal signatures that we have chosen to leave them out of the current study. However, as the summary in section \ref{sec:lit_rev_sum} suggests, showers that have a low signal detection from GLM could still contain fireball data. As GLM collects more data in the years to come, these showers should be revisited.

The next filter to establish our list of possible shower related fireballs is a simple geometry filter. It tests those GLM events which pass the temporal filter for association with a potential shower's radiant. By determining the radiant's altitude based on the GLM event's latitude and longitude and time, we check whether the event occurred when the radiant was above the horizon (altitude > 0°).

The resulting list after filtering of "possible shower fireballs" are then used to extract the most energetic GLM shower fireballs for individual analysis to attempt more rigorous further shower association. Where possible, we use individual event data to further refine shower membership by making checks with a velocity estimate or height information, but only for the most energetic possible shower fireballs (typically the 10 most energetic events per shower). 

For these most energetic shower fireballs, data on height can only be computed for "stereo" events observed by both GLM-16 and GLM-17. These account for approximately one-third of all detections. Nonetheless, many of the most energetic possible shower fireballs fall within the stereo region. Here we make use of the fact that higher luminous heights in the atmosphere are associated with faster shower meteoroids \citep{Borovicka2019}. Any GLM fireball observation that extended below 60 km altitude was removed as a possible shower event as this is well below heights for most fireballs from fast (>40 km/s) streams \citep{Halliday_1996}.

\begin{figure}
  \includegraphics[width=\linewidth]{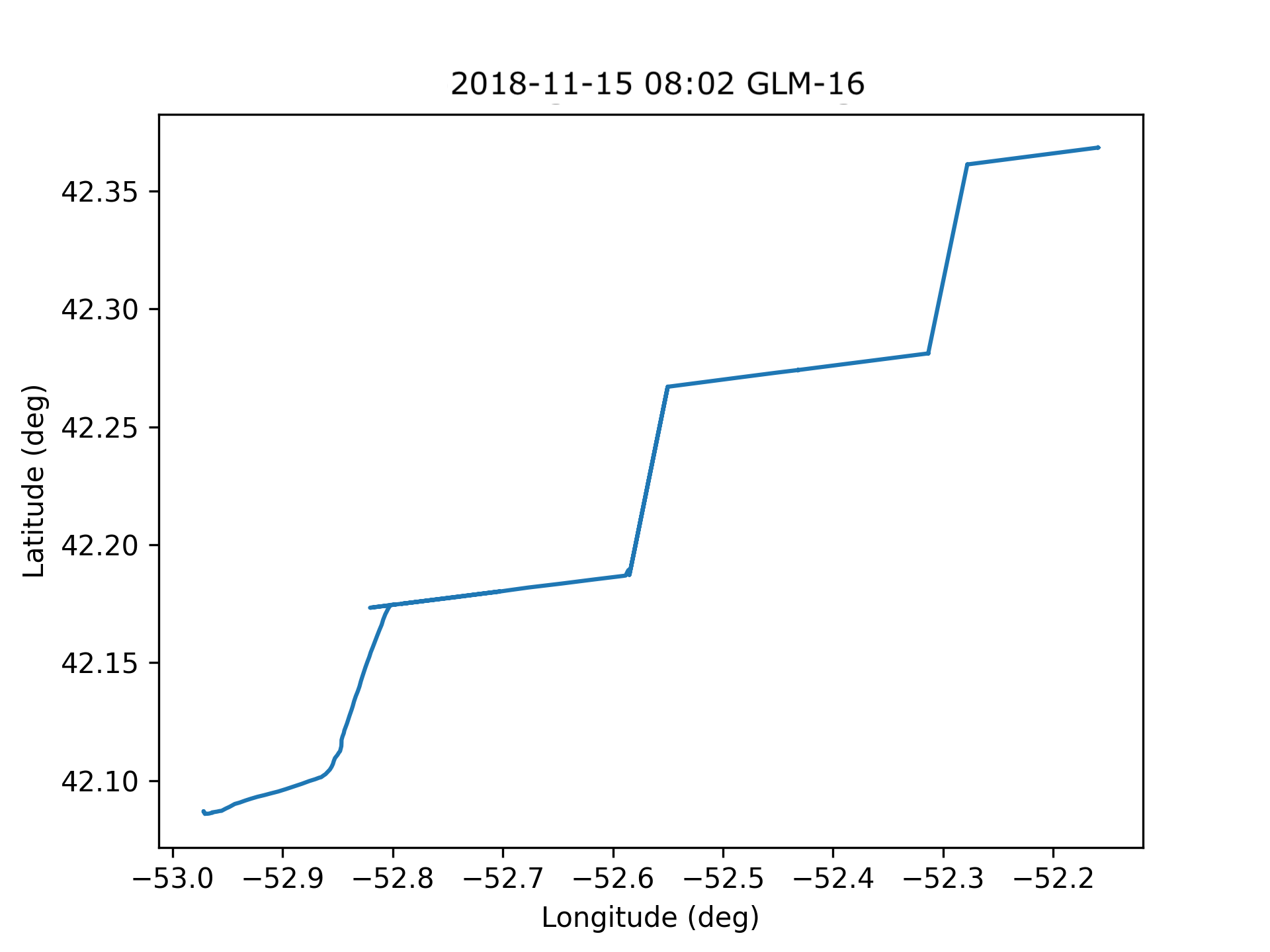}
  \caption{An example of a GLM-16 detection of a possible Leonid with good ground track. Here we see the detection spans multiple pixels, has good separation and overall is linear in nature. The 2D speed is estimated to be 89 km/s, while Leonid meteors observed in this geometry from GLM-16 are expected to have cross-speeds of 65.5 km/s, reasonable agreement at such speeds.}
  \label{fig:ground_track_good}
\end{figure}

Using the GLM event's initial and final latitude/longitude coordinates, the Haversine distance is calculated and divided by the event duration to yield an across-plane speed and travel direction. Events captured across multiple pixels on the CCD detector show a linear on-plane distance progression, allowing a more reliable velocity estimation (see Figure \ref{fig:ground_track_good}). Conversely, events captured in only a single pixel exhibit more erratic spatial displacement (see Figure \ref{fig:ground_track_bad}), therefore producing an unreliable velocity. Some linear detections which encompass point clusters covering limited ground, result in overestimated velocities. The limitations and challenges in using GLM data for estimating velocity are discussed in detail in \cite{smith_morris_rumpf_longenbaugh_mccurdy_henze_dotson_2021} and \cite{jenniskens}. For this study, we selected observations for our biggest fireballs that behaved similarly to the type of ground track seen in Figure~\ref{fig:ground_track_good}. Even in those cases, the velocity is only a crude estimate. Given the known imprecision of GLM velocities \citep{ACM_Thom_2023} we choose a broad acceptance requiring only that the observed shower velocity be within  a factor of ~2 of the established velocity.

\begin{figure}
  \includegraphics[width=\linewidth]{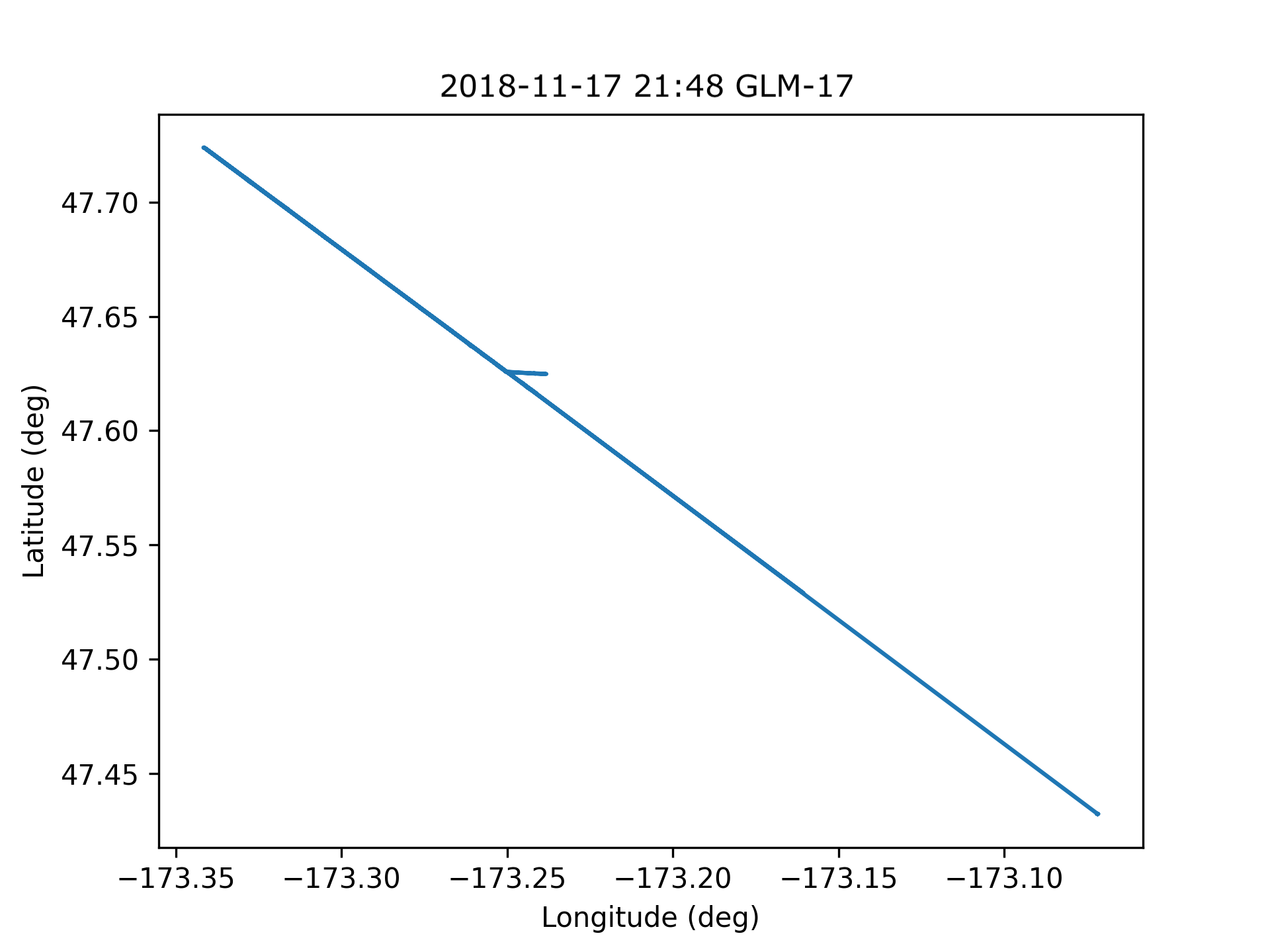}
  \caption{An example of a GLM-17 fireball detection with a linear progression, but which does not span multiple pixels. It does not provide a reliable velocity estimate but may provide a usable bearing estimate.}
  \label{fig:ground_track_linear_bad}
\end{figure}

\begin{figure}
  \includegraphics[width=\linewidth]{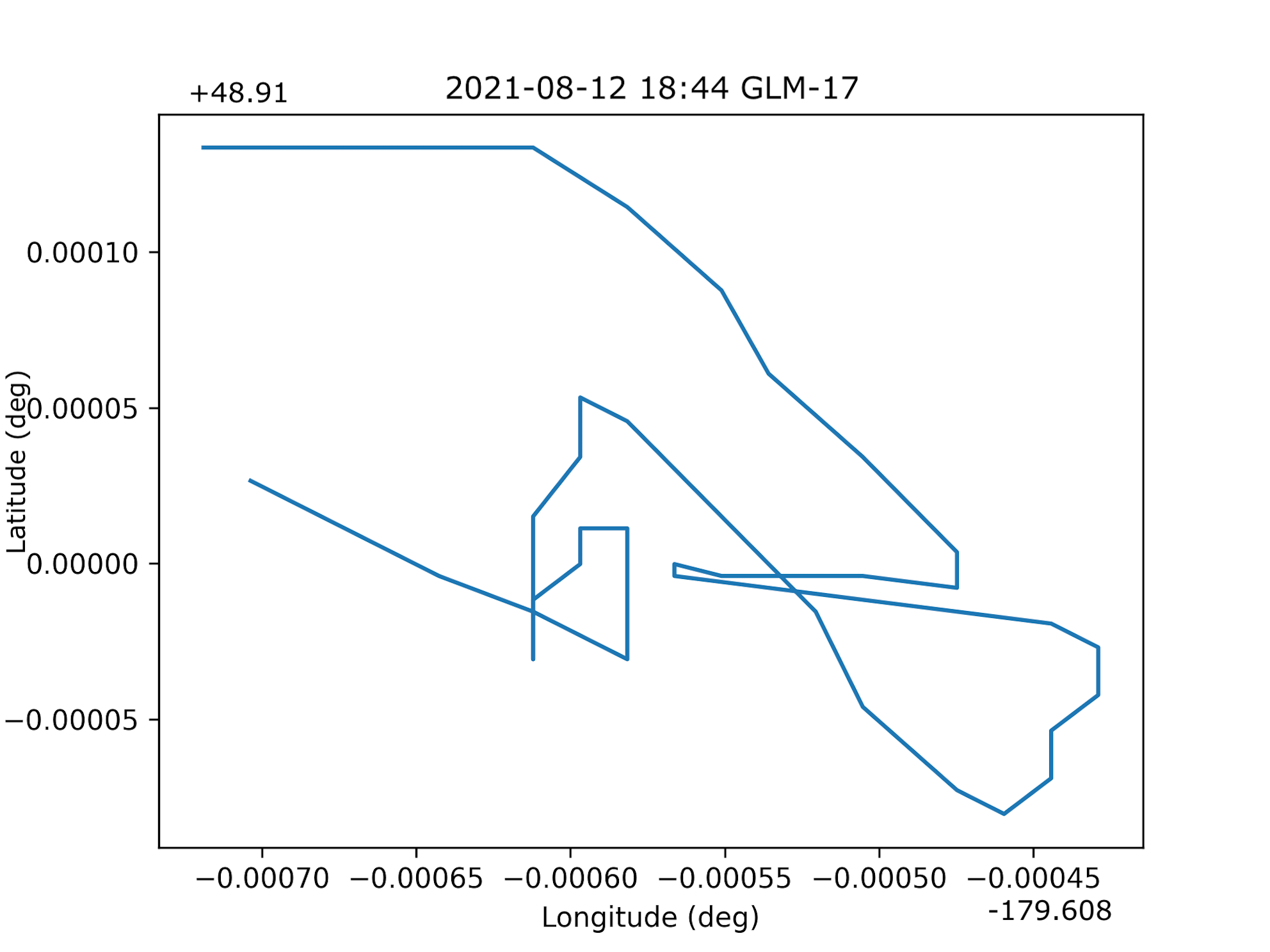}
  \caption{An example of a GLM-17 fireball detection with very bad ground track, spanning one pixel without any sense of direction of travel.}
  \label{fig:ground_track_bad}
\end{figure}

In addition to stereo cases where velocity can be measured, we calculate the anticipated transverse velocity from a single GLM detection by multiplying the cosine of the radiant altitude with the shower's velocity. This is cross-referenced with the velocity calculation per satellite as a validation measure. Theoretically, the transverse velocity is related to the look angle from the detector and absolute shower velocity. 

Comparing the apparent travel direction with the local radiant azimuth is an additional check on shower association suitable for observations with relatively linear progression. The bearing angle indicates the entry direction of the fireball into the atmosphere. For this filter, the first and last data points are used as an average to account for the pixel jump and potential outlying points. For some events with less linear paths, we applied a least squares fit to the ground track.  Some events have poor ground track, and thus a bearing angle or velocity estimate were not possible. We selected events with the better ground track even if there were other higher energy events passing the other filters. For the bearing angle, we accepted events where the observed azimuth was within $\pm$20$^{\circ}$ of the shower azimuth. 

As a final filter for shower membership we also check for the presence of a simultaneous USG event. As USG data has been shown to have no significant shower signature \citep{Brown2015} an accompanying USG detection we can usually take as indicating a non-shower origin. USG provides additional information to potentially eliminate events which are not shower-related, particularly through height at peak brightness and full velocity vector, information available for most USG detections. USG data is also used as a GLM calibrator (see Section~\ref{sec:glm_vs_usg}) as the USG total radiated energy can be compared with GLM  energy.

In addition to the foregoing, we make use of USG versus GLM energies as a discriminator for shower candidates. As we will show later, the shower velocity produces a significant correction to the total GLM energy using the approach of \cite{vojacek}. When there is disagreement between USG and GLM energies, we recalculate the GLM energy using an average velocity of 20 km/s \citep{brown_flux}, typical of asteroidal impact speeds for events where no USG velocity is reported. If this lower speed produces better agreement between GLM and USG energies, it suggests the shower velocity is not appropriate for the fireball and it is removed as a possible shower event.

\begin{figure*}
  \includegraphics[width=\linewidth]{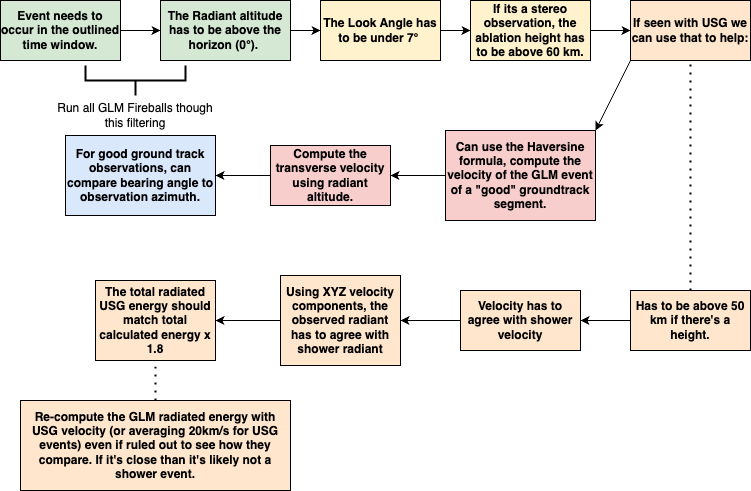}
  \caption{Summary of the filtering process to identify probable shower-related fireballs in GLM fireball data.}
  \label{fig:filtering}
\end{figure*}

The flowchart in Figure \ref{fig:filtering} outlines the process used to determine the shower membership of the most energetic events within our "possible fireballs" list.

To identify individual, large shower-fireballs, the main filters we consider are, the height of the observation, the altitude of the radiant and the time of the event relative to the peak of the shower. We also seek agreement between the observed and known shower radiant azimuth and transverse speed for the most energetic fireballs. 

\subsection{Collecting Area and Shower Flux}
\label{sec:CA_methods}

The collecting area in the Earth's atmosphere for a given shower is the overlap between the GLM field of view (FOV) and the meteor shower radiant projected onto the Earth's surface. The flux of a meteor shower is found by computing the number of events per time-area product (TAP) as a function of solar longitude for a specific year.

\begin{figure}
  \includegraphics[width=\linewidth]{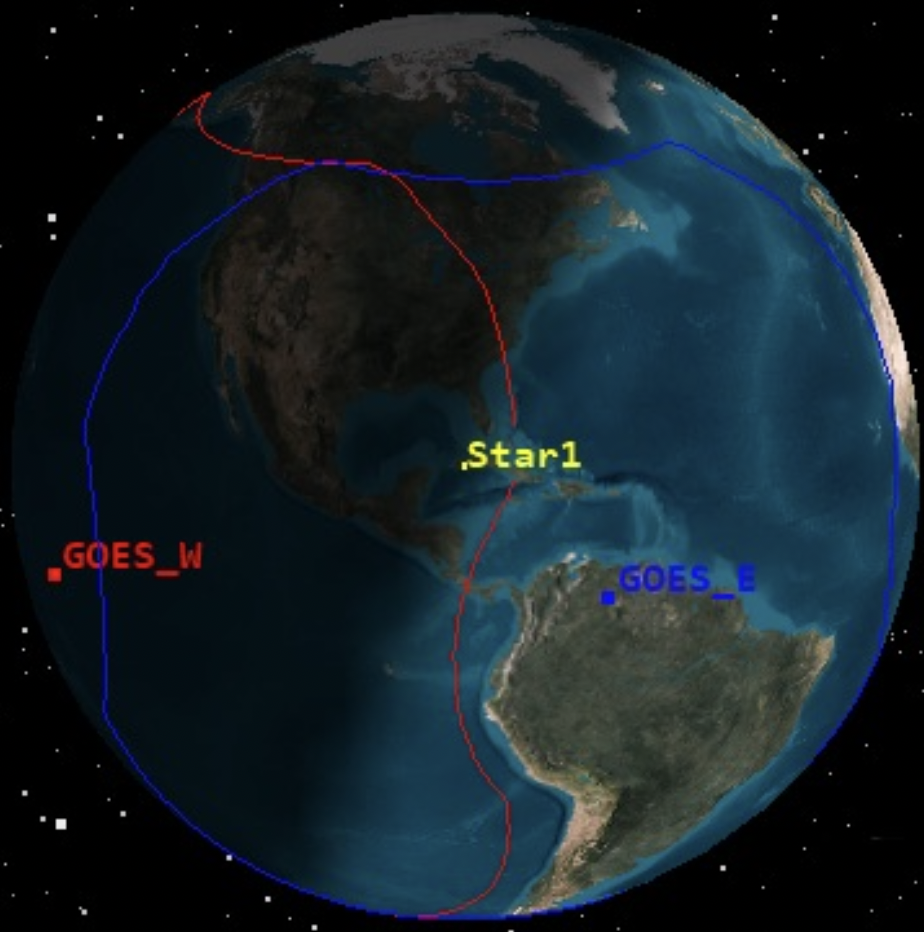}
  \caption{A screenshot from the STK program used to determine the atmospheric collecting area for the Leonid shower. Here the Earth is shown from the vantage point of the radiant with the "Star1" being the sub-radiant point of the Leonids at on 2021-11-17 12:00:00 UTC. The GOES East (blue region) and GOES West Satellite (red region) GLM fields of view are shown.}
  \label{fig:stk}
\end{figure}

\begin{figure}
  \includegraphics[width=\linewidth]{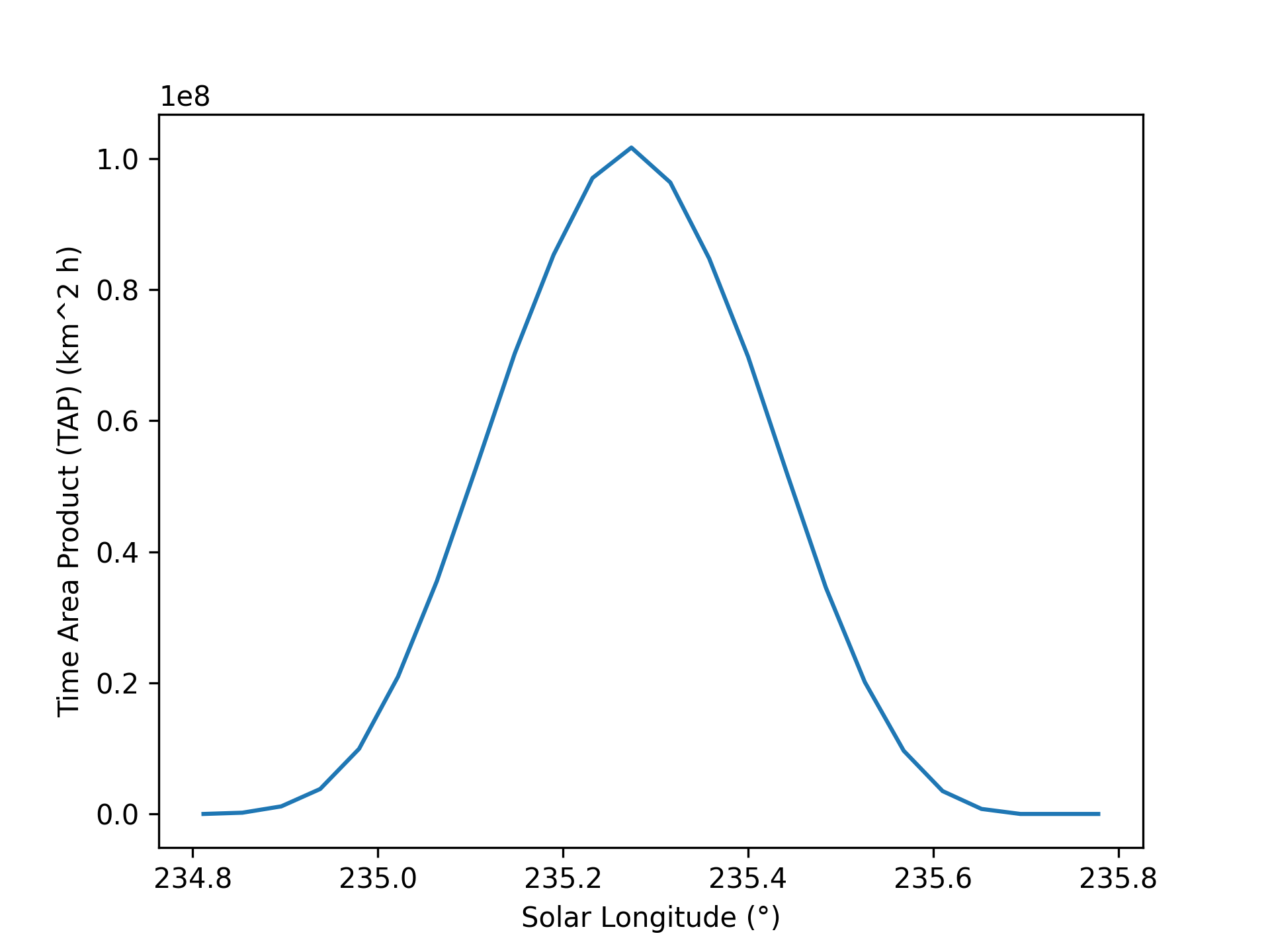}
  \caption{An example of the variation of the time-collecting area product for the Leonids in 2020 near the time of peak activity for GLM-16. The solar longitude is J2000.}
  \label{fig:leo_TAP}
\end{figure}

Determining the intersection between these two areas was intricate due to irregular GLM FOV shapes. The software "Systems Tool Kit" (STK)\footnote{Source: https://licensing.agi.com/stk/} helped visualize how shower radiants were projected from the vantage point of each of the satellites. The GLM FOV latitude/longitude boundaries were extracted for GOES-16 and GOES-17 via the fireballs package \footnote{Source: https://fireballs.readthedocs.io/en/latest/} and added as area targets in STK. The shower radiant projection to the Earth's disk is a time-dependent area, approximated as an Earth-radius disk. It was computed considering Earth's rotation and the varying radiant positions and radiant drift. By simulating the projection of the radiant to Earth, we observed how GLM FOVs intersected the disk over time, see Figure \ref{fig:stk}. Each hour, the GLM FOV's pixel percentage within the disk was measured, calculating the area for a 24-hour period. This provided the time-area product needed for an initial flux estimate. Figure \ref{fig:leo_TAP} shows an example of the time area product for the Leonids during the day of the peak of the shower in 2020.

To compute shower fluxes using the collecting area, GLM events were binned in 0.042 and 0.25 solar longitude degrees, corresponding to 1 and 6-hour bins. Fluxes were calculated per year by dividing the number of events per bin by the time-area integrated product for that bin width.

All GLM fireball events occurring within bins were filtered using the first two filter criteria described earlier for fireballs occurring within the GLM FOV and during the activity time of the shower as defined in Section \ref{sec:filter}. These are part of the "possible" shower related events list.

Having estimated a flux value, the effective limiting mass of the measurement needs to also be determined.

\subsection{Shower Meteoroid Mass Estimates and Calibrations}
\label{sec:cali}

One of the primary objectives of this study is to utilize data from an instrument initially designed for lightning detection and apply it to fireball detections of meteor showers. The GLM instrument collects spectral data within the oxygen line triplet (OI-1) at 777 nm, which represents a narrow band of the spectrum. The key challenge is to transform this narrow-band data into a full fireball spectral profile and ultimately determine an estimate of the meteoroid mass. We relied heavily on the GLM energy calibration procedure proposed by \cite{vojacek}. This calibration was the primary method used in our analysis and we therefore provide a brief summary of their approach and results.

\begin{figure}
  \includegraphics[width=\linewidth]{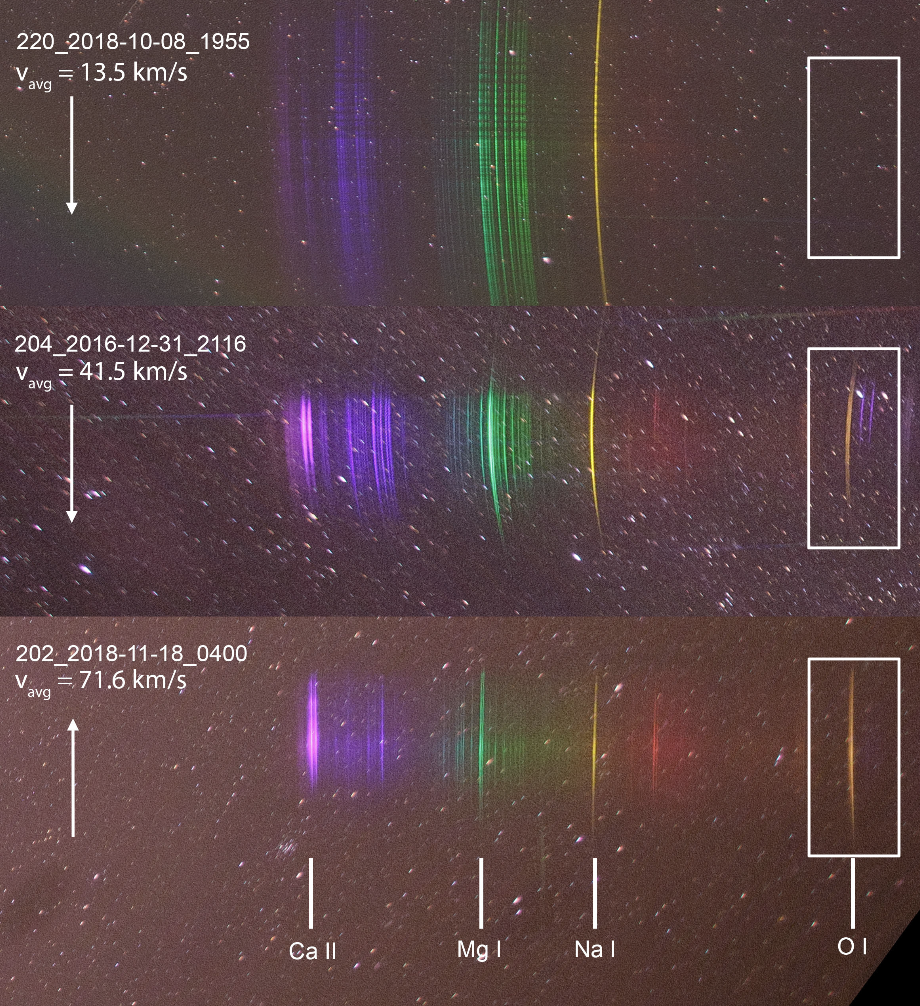}
  \caption{Taken from \cite{vojacek}. This shows as an example of the differing importance of the Oxygen line at 777 nm for different meteor speeds. These range from low (top) speeds to high (bottom) speeds. These observations were taken by the European Fireball Network. The arrow indicates the direction of travel of the meteor and the spectral lines were aligned  to show the difference in the oxygen line}
  \label{fig:spec_voj}
\end{figure}

For their study, \cite{vojacek} recorded spectra of fireballs from ground-based cameras with varying velocities, focusing on the 777 nm line emission strength. Figure \ref{fig:spec_voj} (taken from \cite{vojacek}) illustrates examples of low, medium, and fast velocity fireballs, highlighting the prominence of the 777 nm line, particularly in medium and high velocity fireballs.

\cite{vojacek} established a relationship between the velocity and the ratio of intensity from the 777 nm line to the measured total radiant intensity by fitting a least squares model to the data. This radiant intensity, denoted as $I_{777}$, is the same as calculated by the GLM detector.

However, it is important to note that the GLM data available on the NASA fireballs site is categorized as Level 2 data (L2), which includes a lightning correction. We discuss the implications of this correction later in Section \ref{sec:glm_lims}. The radiant intensity, as measured at the focal plane of the GLM, is expressed as follows: 

\begin{equation}
I_{777} = \frac{E_{GLM} \times R^2}{\Delta t \times A}
\label{eqn:i777}
\end{equation}

Here $E_{GLM}$ is the GLM reported energy in joules, ${\Delta t}$ represents the exposure time (2 ms), $A$ stands for the area of the effective lens aperture (0.0098 m$^2$), and $R$ denotes the distance from the satellite to the detection point in meters. With these units the final intensity is given in W/steradian (note that for later mass calculations using this energy, W/steradian needs to be in W so a factor of 4$\pi$ was removed). To calculate $R$, we employed the ground position for the fireball detection and computed the range to the satellite. However, it is worth noting that depending on the altitude at which the fireball occurs, this range may be short by a few tens of kilometers, an adjustment which has negligible impact on the final intensity.

Using the relationship they found in their ground-based fireball data between intensity and velocity, \cite{vojacek} was able to compute the total intensity of the fireball across the entire spectrum relative to that measured at 777 nm. Thus for $I_{Total}$ as per Eq. \ref{eqn:total_int} in \cite{vojacek}:

\begin{equation}
\label{eqn:total_int}
I_{Total} = \frac{I_{777}}{10^{0.026 \times v -3.294}}
\end{equation}

\noindent where $v$ is the in-atmosphere velocity of the fireball. \cite{vojacek} also investigated calibration using magnitude estimates derived from light curves; we will explore this alternate calibration in Section \ref{sec:corr_fac}.

Having estimated $I_{Total}$ it is possible to calculate the total fireball energy and ultimately the mass as the velocity for each shower is well known. However, as a final step to estimate mass, we require an estimate of the relationship between the luminous efficiency (the ratio between the visual light energy over the total energy) ($\tau$) and intensity:

\begin{equation}
I = \tau \frac{dE}{dt}
\label{equ:lumeff}
\end{equation}

\noindent where the differential of energy with respect to time can be determined by taking the derivative of the kinematic energy to obtain:

\begin{equation}
\frac{dE}{dt} = \frac{1}{2}\frac{dm}{dt}v^2 + v \frac{dv}{dt}
\label{eqn:diff_energy}
\end{equation}

\noindent The second term in equation \ref{eqn:diff_energy} is assumed to be negligible. Rearranging for $\frac{dm}{dt}$ and integrating, we can get the total meteoroid mass as:

\begin{equation}
m_{Total} = \frac{2}{\tau \times v^2} \int I dt
\label{eqn:mass}
\end{equation}

Here the integrated intensity with respect to time gives the total radiated fireball energy in joules. This then allows us to estimate the meteoroid mass with suitable choice for luminous efficiency. 

\subsection{Luminous Efficiencies}
\label{sec:lum}

Estimating the meteoroid mass once the fireball radiated energy is known, requires knowledge of the luminous efficiency. Luminous efficiency is a theoretically simple concept but in practice challenging to compute as it may depend on velocity, composition, height and meteoroid mass \citep{Subasinghe2017, Popova2019}. 

\begin{figure}
  \includegraphics[width=\linewidth]{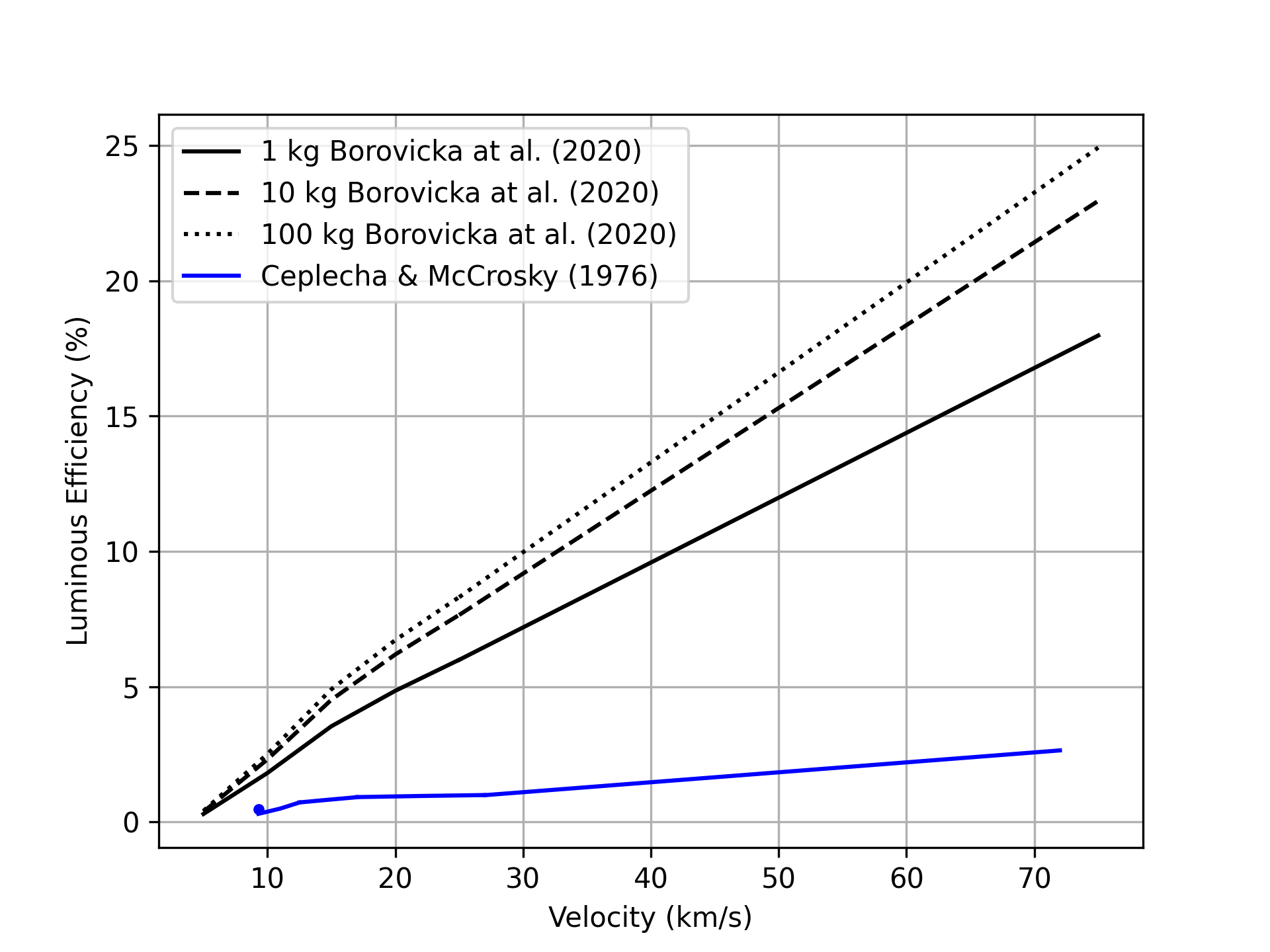}
  \caption{An estimate of the luminous efficiency ($\tau$) as a function of velocity from \cite{Ceplecha_McCrosky_1976} and also by mass following \cite{Borovicka_2020}.}
  \label{fig:taus}
\end{figure}

In this study, we explored the possibility of using the luminous efficiency relation between $\tau$ and velocity proposed by \cite{Borovicka_2020} and \cite{Ceplecha_McCrosky_1976} as shown in Figure \ref{fig:taus}. The luminous efficiency model adopted in \cite{Borovicka_2020} is based on data in \cite{Ceplecha1996} as analysed in \cite{ReVelle2001}. These are almost exclusively lower speed fireballs, the fastest being 40 km/s and hence the relation is extrapolated to higher speeds for our shower analysis where no data are present. 

The $\tau$ values from \cite{Ceplecha_McCrosky_1976} are based in part on empirical fits to a wider range of speeds, but make several additional assumptions/approximations. Foremost among these is the assumption (from \cite{opik1958}) that luminous efficiency increases proportionally with velocity. From comparison with meteorite-producing fireballs it is widely accepted \citep{Borovicka2019} that the values of $\tau$ from \cite{Ceplecha_McCrosky_1976} are too small at lower speeds for fireballs. However, as emphasized by the order of magnitude difference between the two models at higher speeds as shown in Figure \ref{fig:taus} there is uncertainty at our shower speeds of interest. 

For fireballs at high shower speeds, the only empirical estimates are from \cite{borovicka_1997} and \cite{Brown2007}. \cite{borovicka_1997} examined spectral data from two bright (-11 magnitude) Perseids and determined a luminous efficiency in the V-band of 1.5\% and 2.4\% respectively. \cite{Brown2007} compared acoustic energies from infrasound signals with light curves for a Leonid and two Perseids, finding panchromatic luminous efficiencies of 4.3\% for the Leonid, while the Perseids produced values of 2.6\% and 7.1\%. 

 In light of these measurements, we chose to adopt the relations derived by \cite{Ceplecha_McCrosky_1976} for high-speed showers as these produce values closer to the available empirical estimates, but emphasize that resulting masses could easily be in error by a factor of several.  

The following table includes the luminous efficiencies used for the showers in our study where \cite{Borovicka_2020} was adopted for the low-speed Taurids and only for the low speed showers as it was derived using data from events with lower speeds. and \cite{Ceplecha_McCrosky_1976} for the remaining higher speed showers. Again, we emphasize that the derived mass could differ based solely on this choice of $\tau$.

\begin{table}[!ht]
    \centering
    \begin{tabular}{|l|l|l|}
    \hline
        \textbf{Shower } & \textbf{Velocity (km/s)} & $\mathbf{\tau}$ \% \\ \hline
        \textbf{Leonids} & 70.2 & 2.6 \\ \hline
        \textbf{Perseids} & 59.1 & 2.2  \\ \hline
        \textbf{eta Aquariids} & 65.7 & 2.4   \\ \hline
        \textbf{Taurids} & 27 &5  \\ \hline
        \textbf{Orionids} & 66.3 & 2.4   \\ \hline
    \end{tabular}
       \label{tab:shower_tau}
    \caption{Luminous Efficiency ($\tau$) adopted for each shower in our study as a fraction of total initial kinetic energy.}
\end{table}

\section{Results}
\label{sec:results}

As noted earlier, examining GLM fireball rates near the time of the peak for each of the showers originally identified for analysis (see Table \ref{tab:whipple}), it became apparent that the GLM signal for the Geminids, Quadrantids, Southern delta Aquariids and Lyrids was very weak and significant sporadic contamination was expected. This is evident when examining Figure \ref{fig:all_events} as none of these showers show significant enhancement in GLM fireball rates more than one standard deviation above the average background at the time of their peaks. This lack of a clear signal and high likelihood of contamination led us to focus instead on the three showers for which conspicuous activity is apparent: the Leonids, Perseids and eta Aquariids, with the intermediate strength Orionids and Taurids summarized in the appendices.

\subsection{GLM Meteor Shower Flux}
\label{sec:CA_results}

As detailed in Section \ref{sec:methods}, we determined the atmospheric collection area for major meteor showers exhibiting clear temporal signals in GLM detections. This area was calculated by intersecting the GLM field of view with the projection of the meteor shower radiant onto Earth. The shower's collection area exhibits a 24-hour periodicity, which we then applied to the two-week shower observation window as mentioned in Section~\ref{sec:CA_methods}.

\begin{figure}
  \includegraphics[width=\linewidth]{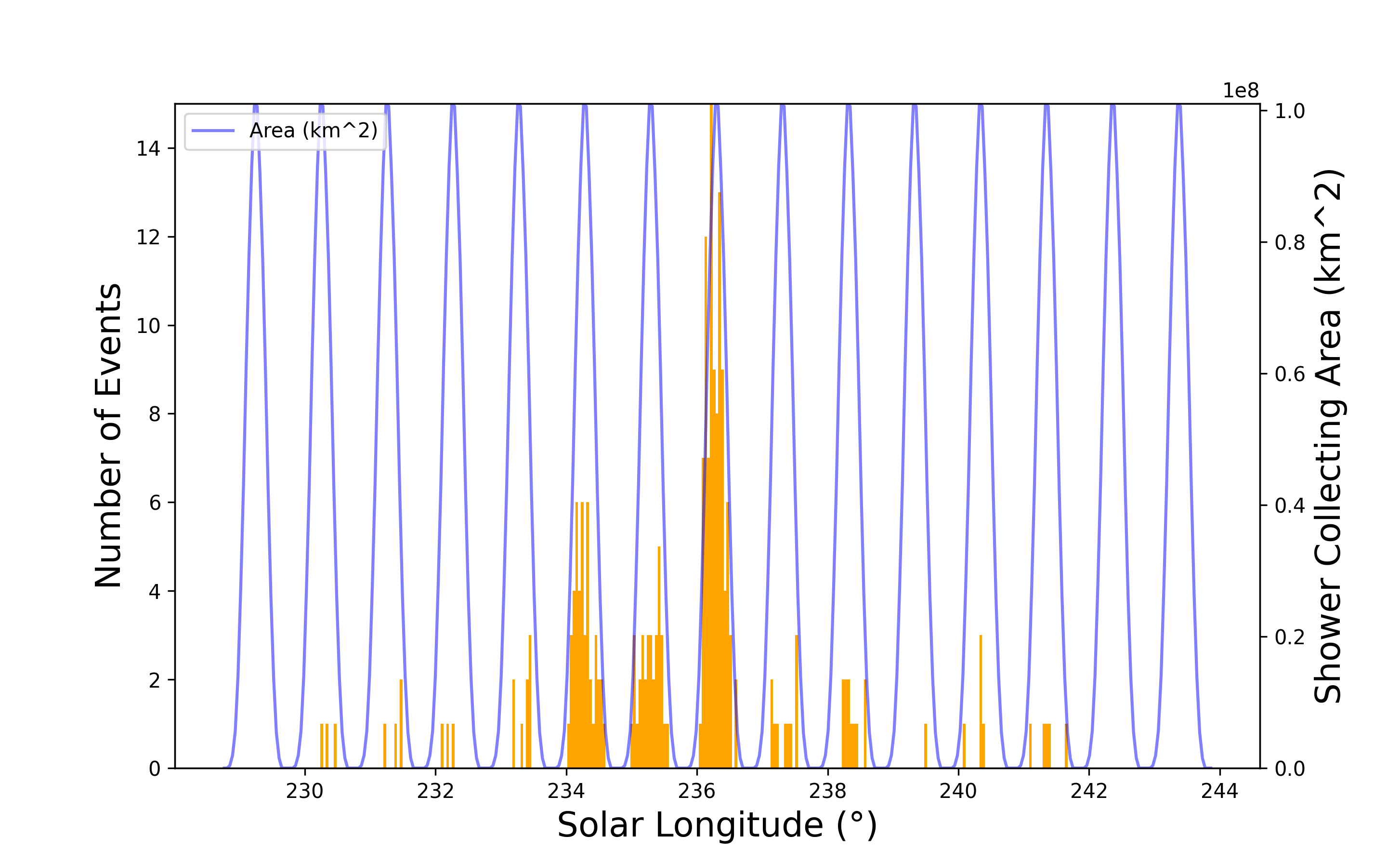}
  \caption{The number of fireballs detected per hour near the peak of the 2020 Leonids (orange bars) by GLM-16 in comparison to the shower collecting area (blue lines).}
  \label{fig:leo_2020_CA}
\end{figure}

To validate our approach, we conducted an hourly examination of the GLM detection rate for the most prominent showers and compared it to our collection area plot. An illustrative example is presented in Figure \ref{fig:leo_2020_CA} for the 2020 Leonids observed with GLM-16, considering all possible Leonid events that passed our first two filters outlined in section \ref{sec:filter}. As anticipated, the fireball rate closely aligns with the shower collection area, reinforcing our selection of these events as predominantly belonging to the shower.

Figure \ref{fig:leo_2020_CA} reveals distinct "gaps" in the frequency of fireball events when the collecting area is low, with a consistent increase as the collecting area rises. This trend is particularly noticeable on the day of the peak, near $\lambda$\textsubscript{\(\odot\)} = 236$^{\circ}$.

\begin{figure}
  \includegraphics[width=\linewidth]{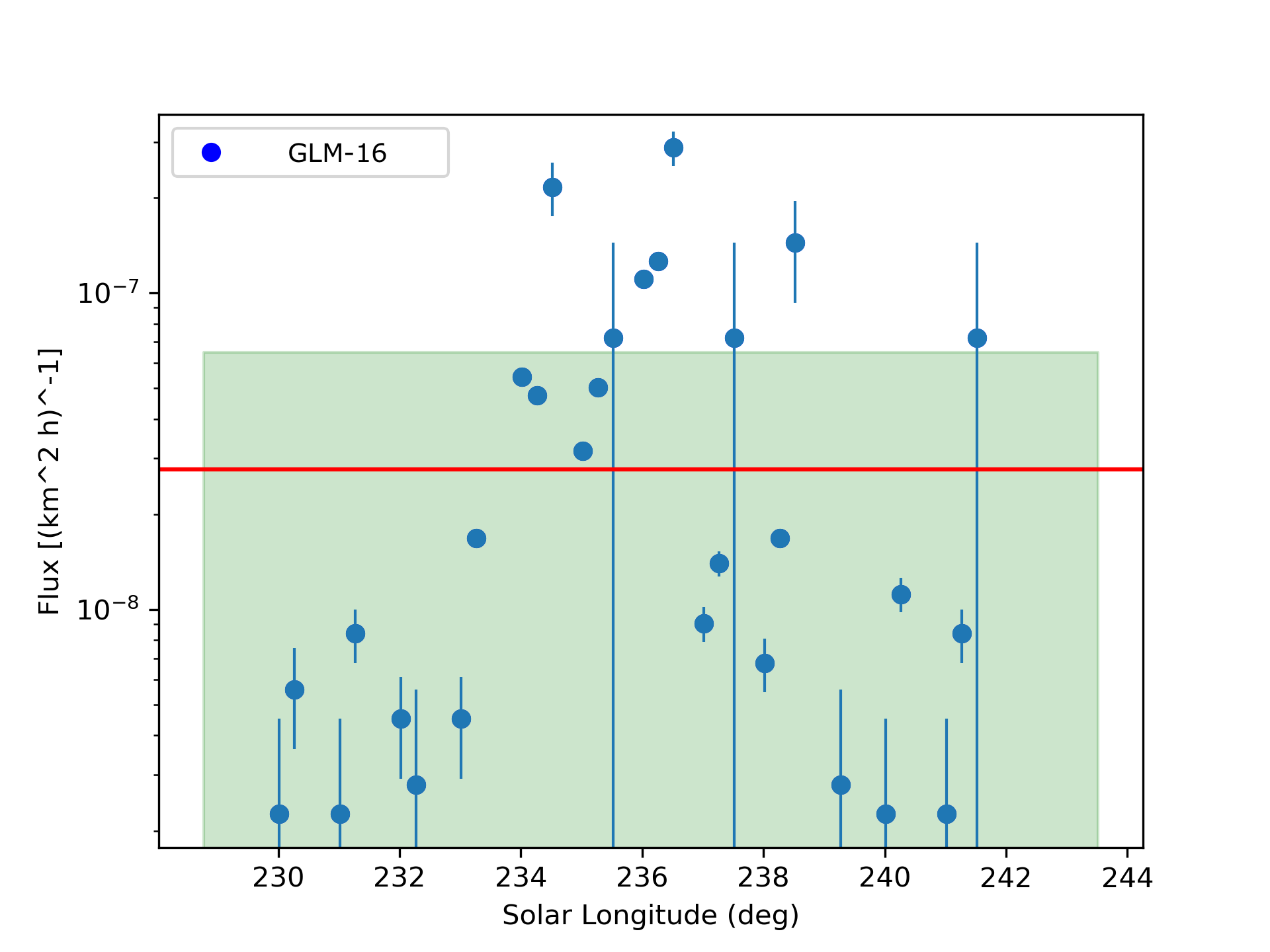}
  \caption{The 2020 Leonid flux as recorded by GLM-16 together with its variance and the background average (red line) and background variance (green shaded region). Here the peak Leonid fireball flux occurs near $\lambda$\textsubscript{\(\odot\)} = 236.5$^{\circ}$. }
  \label{fig:flux_plot_leo_2020_16}
\end{figure}

To obtain a shower time-area product (TAP), we integrated the area under the curve, yielding the number of events per integrated hour of data. To calculate flux, we determined the number of events per square kilometer per hour using both the count and the TAP. Figure \ref{fig:flux_plot_leo_2020_16} displays these 2020 Leonid flux values alongside their corresponding variances together with the background flux. Given the limited number statistics in many bins for hourly windows, we employed 0.25-degree binning, which corresponds to approximately 6 hours in Figure \ref{fig:flux_plot_leo_2020_16}.

Our highest fluxes occur around the peak of the Leonids, near $\lambda$\textsubscript{\(\odot\)} = 235-237$^{\circ}$. Similar fluxes were computed for the Perseids and eta Aquariids, the two additional showers showing strong signatures in GLM data, with similar plots in the Appendix for the Orionids and Taurids. We use the peak flux for each of these three showers to estimate the largest shower meteoroid expected in GLM data given our total cumulative number of shower events as a function of energy, which we assume is a power law. We also will use these fluxes to compare with our individually identified largest, single GLM shower fireballs. The events used for this flux estimation only consider the spatial and temporal initial filters, which still could contain sporadic events but as evidenced by the strong signal strength over the background near the shower peaks, we infer this contamination to be negligible. 

\subsection{Identifying the Largest GLM fireballs for each Major Shower: Single Event Approach}
\label{sec:biggest_fireballs}

To establish the largest meteoroids in a given stream we take two approaches. One is to measure the cumulative energy - frequency distribution of shower fireballs and determine the knee or roll-off in numbers at the largest sizes. This statistical approach establishes what size we should expect to detect, assuming a power-law holds over a range of fireball energies, given our known shower TAP.

The other approach is to identify the largest, individual GLM fireball from our list of potential shower fireballs that passed our first two spatial and temporal filters. In this approach we aim to isolate the most energetic single GLM fireball which is likely related to the shower by imposing additional filters outlined in Section \ref{sec:filter} This latter approach is what we describe in this section.

Upon compiling our initial list of GLM events identified as potential shower members, we proceeded to calculate their masses using the photometric mass calculation method outlined in section \ref{sec:lum}. For these mass calculations, we assumed a single shower velocity for all fireballs in a given stream.

Our analysis of the largest masses began at the individual event level. As discussed earlier, available USG data was first used to eliminate some candidates based on factors like velocity and height. Any fireballs with concurrent USG detections were excluded from our analysis. In most cases, their velocities were too low to be consistent with the showers under study, with the exception of one Taurid fireball. Upon closer examination of this particular detection, using the x-y-z components of velocity, we found that its radiant did not match the expected radiant for the Taurids, leading us to remove it as a possible Taurid member. Furthermore, we recalculated the GLM energies following Eq. \ref{eqn:total_int} for these fireballs but using an average USG velocity of 20 km/s, appropriate for the typical USG fireball population \citep{Brown2015} in place of the potential shower speed. In most instances, these recalculated energies better agreed with the USG energies, which we interpret to indicate that the assumed shower velocity was not correct for these events. These were also removed from further consideration.

\begin{figure}
  \includegraphics[width=\linewidth]{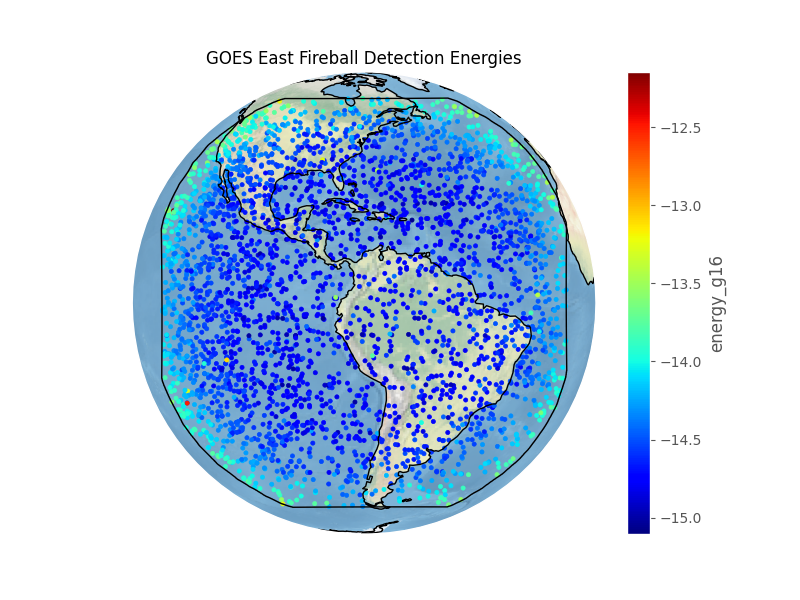}
  \caption{Fireballs detected by GLM-East as a function of the log of the total GLM fireball energy.}
  \label{fig:goes_east}
\end{figure}

\begin{figure}
  \includegraphics[width=\linewidth]{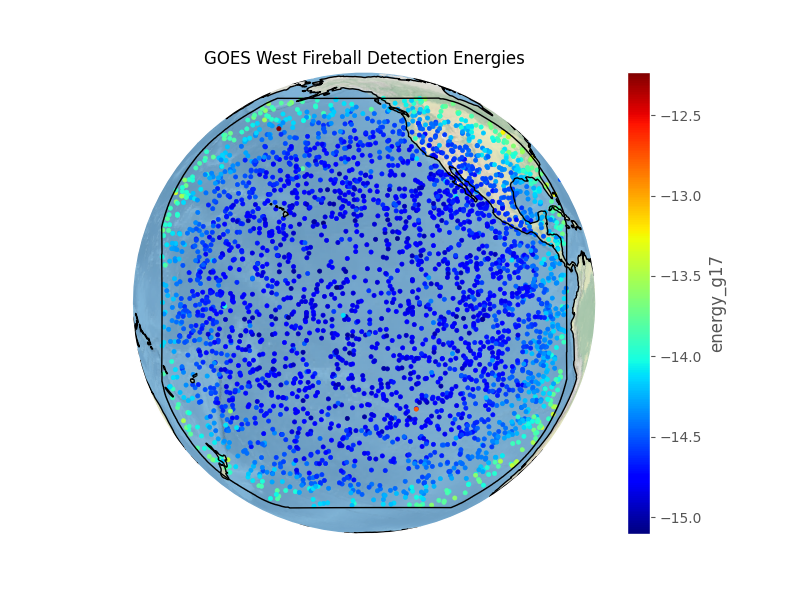}
  \caption{Fireballs detected by GLM-West as a function of the log of the total GLM fireball energy.}
  \label{fig:goes_west}
\end{figure}

At this stage of analysis one aspect that had to be considered was the look angle of the detection, which influence the energy calculations. Observations made by the GLM satellite to the nadir and a few degrees off nadir tend to be most sensitive, while those near GLM's field of view edge are least sensitive. The detection limit may vary by almost an order of magnitude in energy near the extreme edges \citep{jenniskens, smith_morris_rumpf_longenbaugh_mccurdy_henze_dotson_2021}. We see this effect when examining the angular distribution of GLM events as a function of the logarithm of their energies as shown in Figures \ref{fig:goes_east} and \ref{fig:goes_west}.  This effect is a combination of the detector being less sensitive near the edges (and hence only detecting the brightest events) due to the shift in the effective bandpass to shorter wavelengths and a probable over-correction to the energy when the standard lightning sensitivity correction is applied on L2 data as a similar effect is found for lightning \citep{smith_morris_rumpf_longenbaugh_mccurdy_henze_dotson_2021}.

In their study, \cite{jenniskens} discusses the impact of Sun-blocking and narrow band-filters on the shift in the amount of 777 nm emissions as a function of the angle from the nadir position (0$^{\circ}$). Examining their Figure 3, we inferred that observations with a look angle exceeding 7° would likely lead to energy overestimation. The optimal range for accurate energy estimation falls within 3$^{\circ}$ to 5$^{\circ}$, while observations between 0$^{\circ}$ to 2$^{\circ}$ tend to result in slight energy underestimation. 

Additional evidence for this overestimation bias is that we systematically observed that bright events with concurrent USG detections had an energy ratio of GLM divided by USG that significantly increased for observations with larger look angles. In contrast, the ratio remained relatively moderate for events with look angles under 7$^{\circ}$. Consequently, for our survey to establish the most energetic shower fireballs, we excluded GLM observations with look angles exceeding 7$^{\circ}$ to mitigate this energy overestimation bias. With this angular cut in place, we are justified in using the raw L2 energies for the remaining events as representative of the true I$_{777}$ for the event, as they have dominant 777 nm emission at high speeds, much like lightning. 

Having identified our list of most energetic possible shower members we then examined those with stereo measurements to extract velocities. We used this additional information to  compare to our expected shower velocity estimates.

In Table \ref{tab:Big_fireballs1} we summarize our individual, largest GLM shower fireball detections from the online GLM database between 2019-2023 for which we have confidence in association with the three major showers in our study.

\begin{table}[!ht]
    \centering
    \begin{tabular}{|l|l|l|l|}
    \hline
\textbf{Shower}  & \textbf{Leonids} & \textbf{Perseids} & \textbf{eta Aquariids } \\ \hline
        \textbf{Date (UTC)} &  2020-11-18 09:12 & 2019-08-14 08:54 & 2020-05-08 10:06  \\ \hline
        \textbf{Solar Longitude (°)}  & 236.21 & 141.03 & 48.00  \\ \hline
        \textbf{Latitude (°)}  & 11.9 & 34.3 & 21.6  \\ \hline
        \textbf{Longitude (°)}  & -103.1 & -118 & -93.1  \\ \hline
        \textbf{Sensor }  & Stereo - GLM-17 & Stereo - GLM-17 & Stereo - GLM-16  \\ \hline
        \textbf{Look Angle (°)}  & 2.23 & 5.73 & 4.63  \\ \hline
        \textbf{Height (km)}  & 105& 81 & 81  \\ \hline
        \textbf{Radiant Altitude (°)}  & 32.97 & 38.38 & 32.38  \\ \hline
        \textbf{Total Radiated Energy (J)}  & 4.62E+08 & 1.28E+08 & 1.51E+08  \\ \hline
        \textbf{Photometric Mass (kg)}  & 7.3 & 3.4 & 2.9  \\ \hline
        \textbf{Magnitude}  & -13.5 & -13.4 & -12.7  \\ \hline
        \textbf{Shower Velocity (km/s)}  & 70.20 & 59.1 & 65.7  \\ \hline
        \textbf{Transverse Velocity (km/s)}  & 58.9 & 46.32 & 55.5  \\ \hline
        \textbf{GLM Velocity (km/s)}  & 89.9 & 100.1 & 38.1  \\ \hline
        \textbf{Radiant Azimuth (°)}  & 72.15 & 40.45 & 104.5  \\ \hline
        \textbf{GLM Bearing Angle (°)}  & 73.62 & 36.31 & 91.4 \\ \hline
    \end{tabular}
    \caption{The most energetic GLM fireballs associated with our three major showers found in our survey using the methodology described in the text. Here the shower radiant altitude is given at the time (given in UT) and location of the fireball and the total radiated energy is found from the GLM light curve following the procedure described in section \ref{sec:methods}. Here height refers to the height above the Earth surface of the detection. The transverse velocity is calculated using the radiant altitude. Note that the GLM Velocity is the velocity calculated using the Haversine formula, which is also used to compute the GLM bearing angle.}
    \label{tab:Big_fireballs1}
\end{table}

\subsubsection{Leonids}
\label{sec:biggest_leo}

The Leonids were the most robustly observed shower by GLM, with many events occurring during the peak of 2020. Out of several hundred GLM detected Leonids, the single largest  was of order 6-7 kg, compatible with the Whipple gas-drag limit and similar to the largest detected by ground-based cameras. The largest Leonid event was observed in the stereo region, where both masses calculated from the light curves are within a kilogram of each other. While the GLM-16 estimate of mass is 6.55 kg,  we included the larger of the two in the table. The observed GLM-17 velocity was slightly higher than the shower but was in our adopted range. The GLM-16 velocity was closer to the expected velocity, with a speed of 50 km/s. We see good agreement (to within a degree) between the shower radiant azimuth and the GLM bearing angle from the GLM-17 observation.

\subsubsection{Perseids}
\label{sec:biggest_per}

The GLM mass estimates for the largest Perseids were several orders of magnitude higher than the Whipple mass limit. While there were even larger Perseids with some (but not all) metric characteristics consistent with the shower, the most energetic fireball for which we have confidence of a Perseid association had a mass of order 3 kilograms. Thus GLM results for the Perseids suggest strongly that multi-kilogram to $\sim$5 kg Perseids are present in the stream. Here the radiant azimuth differed by four degrees and the speed agreed within a factor of two. The larger speed disparity reflects the short path length for this event. 

\subsubsection{eta Aquariids}
\label{sec:biggest_eta}

The eta Aquariids showed good agreement between the GLM mass values and the Whipple limit. Our most energetic GLM eta Aquariid fireball, for which we have confidence in the shower association, has a mass of order ~3 kg. This event was a stereo observation; however the GLM-17 observation angle was just outside our limits. It had an apparent mass of ~8 kg, but this is likely an overestimate. The GLM-16 detection better met the velocity and bearing angle estimates expected for the shower and was not outside our 7 degree acceptance limit. 

\subsubsection{Other Showers}
In the appendix we summarize a similar analysis of the largest fireballs likely associated with the Orionids and the Taurids. These showers were well observed by GLM, but had much smaller number statistics than the three principal showers.  As such, the confidence in the identification of the two most energetic showers fireballs is less so then with our three principle showers. For the Orionids we found the single most energetic event had a mass of 2.4 kg while for the Taurids the largest event had a mass of 150 kg.  

\subsection{Identifying the Largest GLM fireballs for each Major Shower: Cumulative Mass Distribution.}
\label{sec:method2}

The second approach we take to estimate the largest shower meteoroid is statistical. In this method we construct the cumulative mass distribution for each shower by including only GLM events that passed the first two primary filters for shower membership. Additionally, we restricted our selection to events with look angles under 7° to avoid overestimating energy at the tail of the distribution. To maintain self-consistency in the distributions, we excluded the most energetic events that were further investigated for shower membership but ultimately ruled out in section \ref{sec:biggest_fireballs}.

By taking all possible shower events occurring within the time window of the activity of each shower (where it is more than 1$\sigma$ above the background fireball rate) between 2019-2022, we then compute the mass for each fireball as described in section \ref{sec:methods}. The cumulative mass distributions consistently displayed a primary power-law behavior in the middle portion of the mass range, with a roll off in numbers dominating at the low mass end and a knee or drop off near the high mass end which typically showed a steeper power-law. The power-law trend in the cumulative mass distribution is characteristic of shower populations when observed using ground-based instruments \citep{Ceplecha1998}. 

To determine the mass index of the shower, we applied a "shallow fit" to the primary (middle portion) of the distribution. We also applied a fit to the steep end of the curve, referred to as the "steep fit," to constrain our flux as we extrapolate down the curve to larger masses. The steep fit is our focus as this is the best estimate of the flux of larger shower members; it will also provide a lower limit to the estimate of largest shower member. These fits provided us with the ability to estimate the expected number of events for any given mass. When combined with the integrated area obtained from our collection area analysis, this formed the basis for our flux calculations for GLM.

\begin{figure}
  \includegraphics[width=\linewidth]{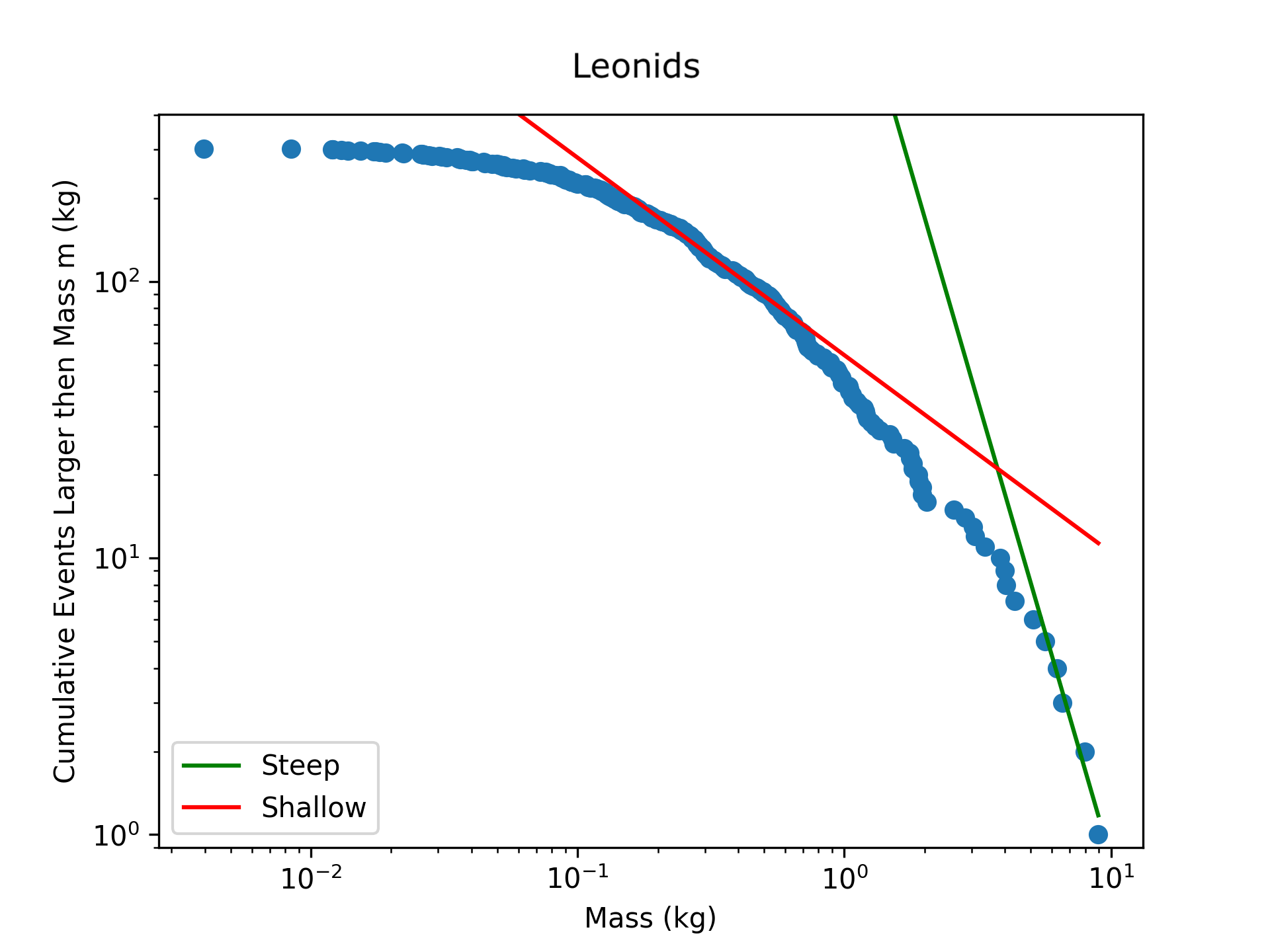}
  \caption{The cumulative mass distribution for the Leonids observed by GLM-16 between 2019-2022.}
  \label{fig:cm_leo}
\end{figure}

\begin{figure}
  \includegraphics[width=\linewidth]{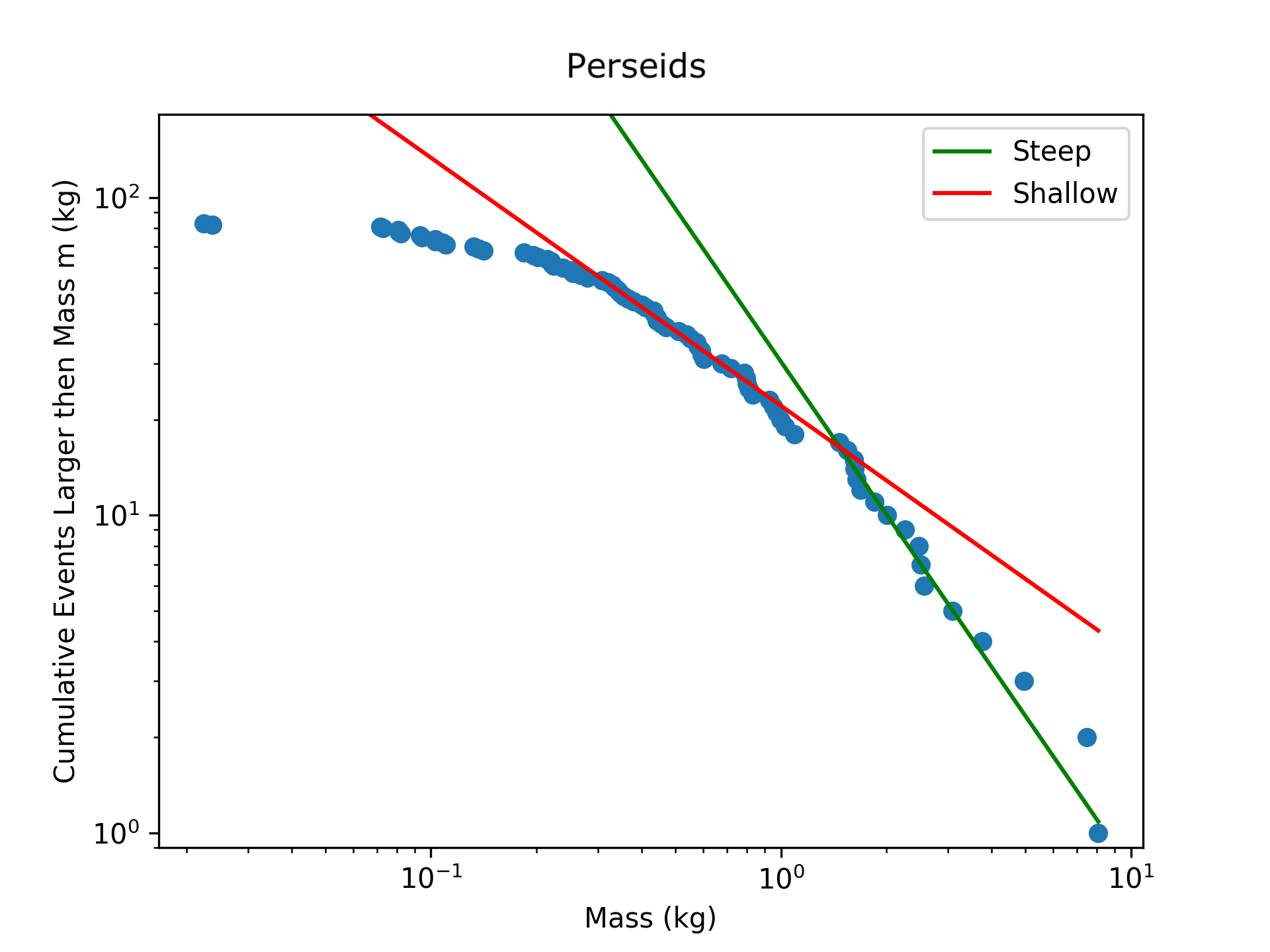}
  \caption{The cumulative mass distribution for the Perseids observed by GLM-16 between 2019-2022.}
  \label{fig:cm_per}
\end{figure}

\begin{figure}
  \includegraphics[width=\linewidth]{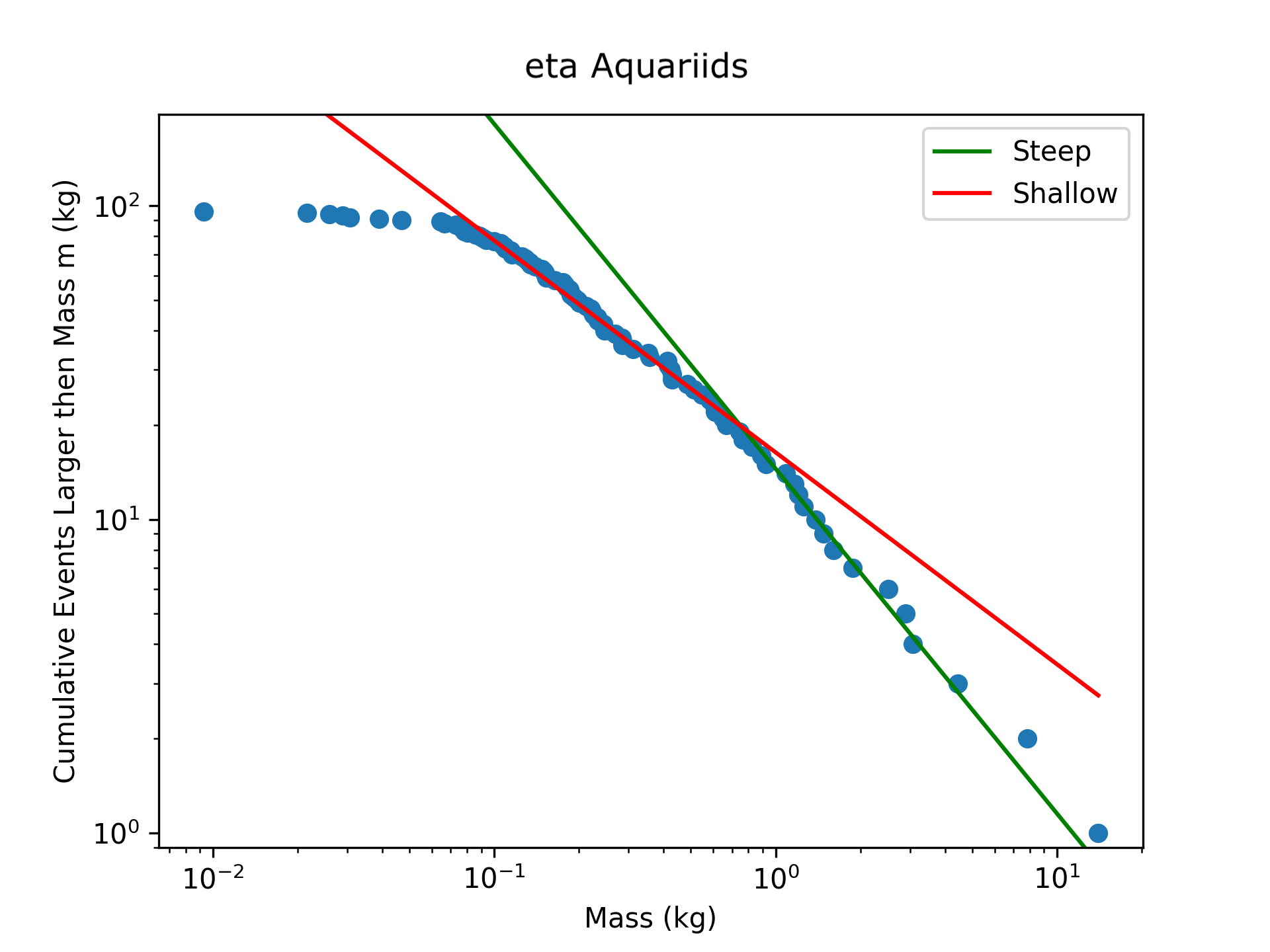}
  \caption{The cumulative mass distribution for the eta Aquariids observed by GLM-16 between 2019-2022.}
  \label{fig:cm_eta}
\end{figure}

Only the Leonids, Perseids, and eta Aquariids showed a sufficiently strong shower signature as detected by GLM to enable us to construct reliable mass distributions. The cumulative distributions for these three showers are shown in Figures \ref{fig:cm_leo}, \ref{fig:cm_per} and \ref{fig:cm_eta} respectively.

A common feature of all three cumulative distributions is a knee where the distribution steepens, in all cases at masses of order a few kilograms. We can determine this point by fitting a tangent line to the main "bump" of the curve. We initially fit a line to the beginning of the whole curve as well, and where the two fits intersect can give us where this knee occurs to determine a limiting mass. 

To utilize the cumulative mass distribution fits to make predictions regarding the largest shower mass that could be expected, we use the total TAP per shower and extrapolate the steep power fit. We choose to define the likely largest fireball which could reasonably be detected by GLM given the steep power-law distribution with a 50 percent chance of detectability, given the total collecting area-time product of the survey from 2019-2022.

To determine this, we employ the Percent Point Function available in the Poisson Python package \footnote{Source: https://docs.scipy.org/doc/scipy-0.13.0/reference/generated/scipy.stats.poisson.html}. We use the steep power law fit to then compute the mass we would have a 50\% chance of detection given our integrated TAP. Table \ref{tab:poisson_tab} provides this 50\% "Poisson mass" for each sensor and shower.

\begin{table}[H]
    \centering
    \begin{tabular}{|l|l|l|l|}
    \hline
        \textbf{Shower} & \textbf{Sensor}  & \textbf{Largest Mass (kg)} \\ \hline
        Leonids & GLM-16  & 10.5 \\ \hline
        Leonids & GLM-17  & 12.8 \\ \hline
        Perseids & GLM-16  & 10.7 \\ \hline
        Perseids & GLM-17  & 6.0 \\ \hline
        ETA & GLM-16  & 15.9 \\ \hline
        ETA & GLM-17  & 3.8 \\\hline
    \end{tabular}
    \caption{The Poisson mass for which there is a 50 \% detection probability using the cumulative mass fits for the steep portion of each shower flux curve for each sensor for masses above the "knee" in the distribution. For reference, the largest observed fireball for each shower was 7.3 kg for the Leonids, 3.4 kg for the Perseids and 2.9 kg for the eta Aquariids.}
    \label{tab:poisson_tab}
\end{table}

From Table \ref{tab:poisson_tab} we see the largest masses estimated this way are consistent, within a factor of 2, with the individual largest shower masses found in section \ref{sec:biggest_fireballs}. Thus our power-law fit is broadly consistent with our single largest identified shower member for our three primary showers.

\begin{figure*}
  \includegraphics[width=\linewidth]{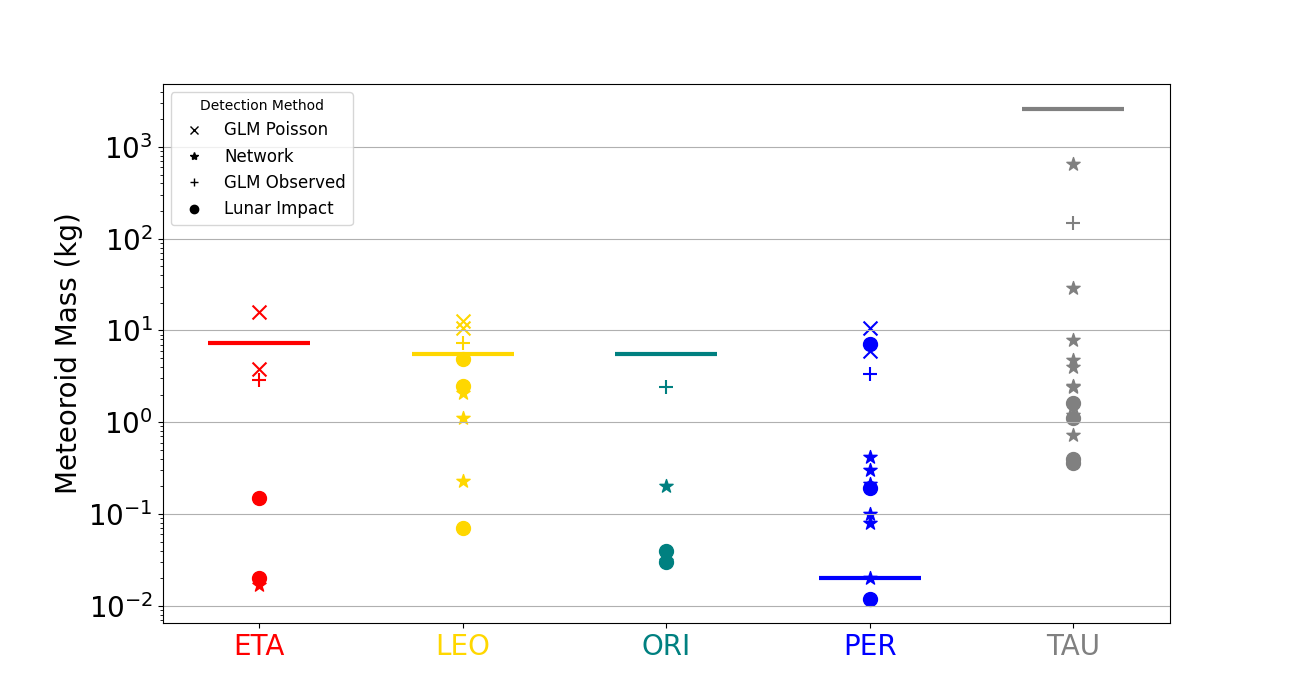}
  \caption{Largest meteoroid masses reported per stream from literature sources, both ground-based surveys and lunar impacts, and derived from GLM. The GLM limit estimated from extrapolating the steep power-law flux per shower where applicable is shown as a cross (GLM Poisson) while the GLM estimate of the largest observed shower fireball is shown by a plus (GLM observed). The networks used for each shower are outlined in section \ref{sec:lit_rev_sum} and refer to the European Fireball Network, Southern Ontario Meteor Network, the Meteorite Observation and Recovery Project, and the Prairie Network.}
  \label{fig:lit_mass_with_GLM}
\end{figure*}

We now compare our GLM fireball results using both the statistical approach and individual event identification to compare with literature values. This is shown in Figure \ref{fig:lit_mass_with_GLM}, which also includes the estimated Whipple gas-drag limited mass (as a horizontal line). The major outlier here are the Perseids, which contain much larger meteoroids than expected from gas-drag ejection.

In Table \ref{tab:flux_summary} we also provide a summary of the equivalent flux for each of the three major showers based on the shallow power-law fit, appropriate to the middle mass range of the shower. Here our completeness to a given limiting mass is defined as the intersection of the shallow fit power-law and the total number of shower events. Note the Poisson Fluxes are calculated for the total 4 years of data from 2019 to 2022, and thus the time-area products include this entire time-span. 

\begin{table}[!ht]
    \centering
    \begin{tabular}{|p{40pt}|p{40pt}|l|l|l|l|}
    \hline
        \textbf{Shower} & \textbf{Time Area Product } & \textbf{Poisson Mass } & \textbf{Poisson Flux } & \textbf{Limiting Mass} & \textbf{Limiting mass Flux} \\ 
       &$km^2$ h&(kg)&$km^{-2} h^{-1}$&(kg)&$km^{-2} h{-1}$\\ \hline
        \textbf{Leonids GLM-16} & 5.2E+10 & 10.5 & 1.34E-11 & 0.066 & 7.16E-08 \\ \hline
        \textbf{Leonids GLM-17} & 5.2E+10 & 12.8 & 1.32E-11 & 0.069 & 5.42E-08 \\ \hline
        \textbf{Perseids GLM-16} & 3.8E+10 & 10.7 & 1.82E-11 & 0.27 & 7.56E-08 \\ \hline
        \textbf{Perseids GLM-17} & 3.8E+10 & 6.0 & 1.79E-11 & 0.19 & 1.96E-07 \\ \hline
        \textbf{eta Aquariids GLM-16} & 5.2E+10 & 15.9 & 1.34E-11 & 0.076 & 2.01E-08 \\ \hline
        \textbf{eta Aquariids GLM-17} & 5.2E+10 & 3.8 & 1.19E-11 & 0.066 & 1.52E-07 \\ \hline
    \end{tabular}
    \caption{A summary of the total time-area product in units of $km^2$ h for shower collection for each sensor for the three primary showers in our study. The Poisson mass refers to the largest shower event we expect to see based on extrapolation of the steep sloped cumulative mass law and our TAP. The Poisson flux represents the equivalent flux to this mass using the steep power-law distribution. The limiting mass reflects the smallest shower meteoroid mass for which we judge GLM detection to be complete using the shallow power-law and the total cumulative number of events per shower per sensor.}
    \label{tab:flux_summary}
\end{table}

\section{Discussion}
\label{sec:diss}

\subsection{Limitations and basis of GLM L2 Data}
\label{sec:glm_lims}

The original purpose of GLM was lightning detection, but its adaptation for fireball detection presents several challenges. Specifically, we must consider the disparities between Level 0 (L0) "raw" data and readily accessible processed Level 2 (L2) data. \cite{ACM_Thom_2023}, conducted an analysis comparing the light curves of GLM fireballs in both L0 and L2 data.

Intermediate data processing involves filtering out lightning events and conducting post-processing for these events. The specifics of processing can be somewhat opaque, with some events showing differences in light curves when comparing L0 and L2 data, as reported in \cite{ACM_Thom_2023}.

For instance, they considered an event that occurred on 2022-07-07 01:49 UTC. It was observed that the L2 light curve for this event lacked the peak energy plateau in the light curve visible with L0, leading them to conclude that the L0 data was the more reliable dataset for this specific event. Furthermore, when comparing this event to a simultaneous USG event, the total integrated USG energy closely matched the L0 GLM data while the L2 GLM data was approximately half the value.

Most fireball examples examined by \cite{ACM_Thom_2023} showed small differences in energy between L0 and L2 data, but where disparities were noted, it was always the case that the L0 energy was higher than that derived from L2.

While these discrepancies may not significantly impact broader analyses, the precision of data becomes increasingly critical when assessing individual events, especially for the largest fireballs related to a meteor shower. For our purposes, the L2 data will suffice, but for future work, the release of L0 data could prove useful for detailed fireball analysis. We also emphasize that use of the L2 data in the present analysis implies that our energies may be underestimates - the true energies may be somewhat bigger for the largest shower fireballs. 

\subsection{GLM Energy Calibrations}

To correct from the narrow GLM 777nm bandpass to total energy, the effects of optics, geometry and the change in bandpass with look angle can be modelled assuming the fireball behaves as a blackbody. Such a continuum calibration (assuming a 6000K blackbody) has been produced for GLM-16 and GLM-17 by Lockheed Martin (Tillier, private communication). This calibration aims to rectify overestimated energies when the sensor is at positions away from the nadir and accounts for the small GLM bandpass sampled. Note that the absolute correction from these tables does not need to be applied to shower fireballs, which, like lightning, have a significant fraction of their spectral emission at 777nm. However, we still expect an overestimation of L2 energies near the edge of the GLM field of view as is also observed for lightning energy.

To explore this potential edge effect, we examined the GLM-16 fireball energies as a function of observation angle. We divided the look angle into 0.5-degree bins and averaged the fireball energies for observations falling within each bin.

\begin{figure*}
  \includegraphics[width=\linewidth]{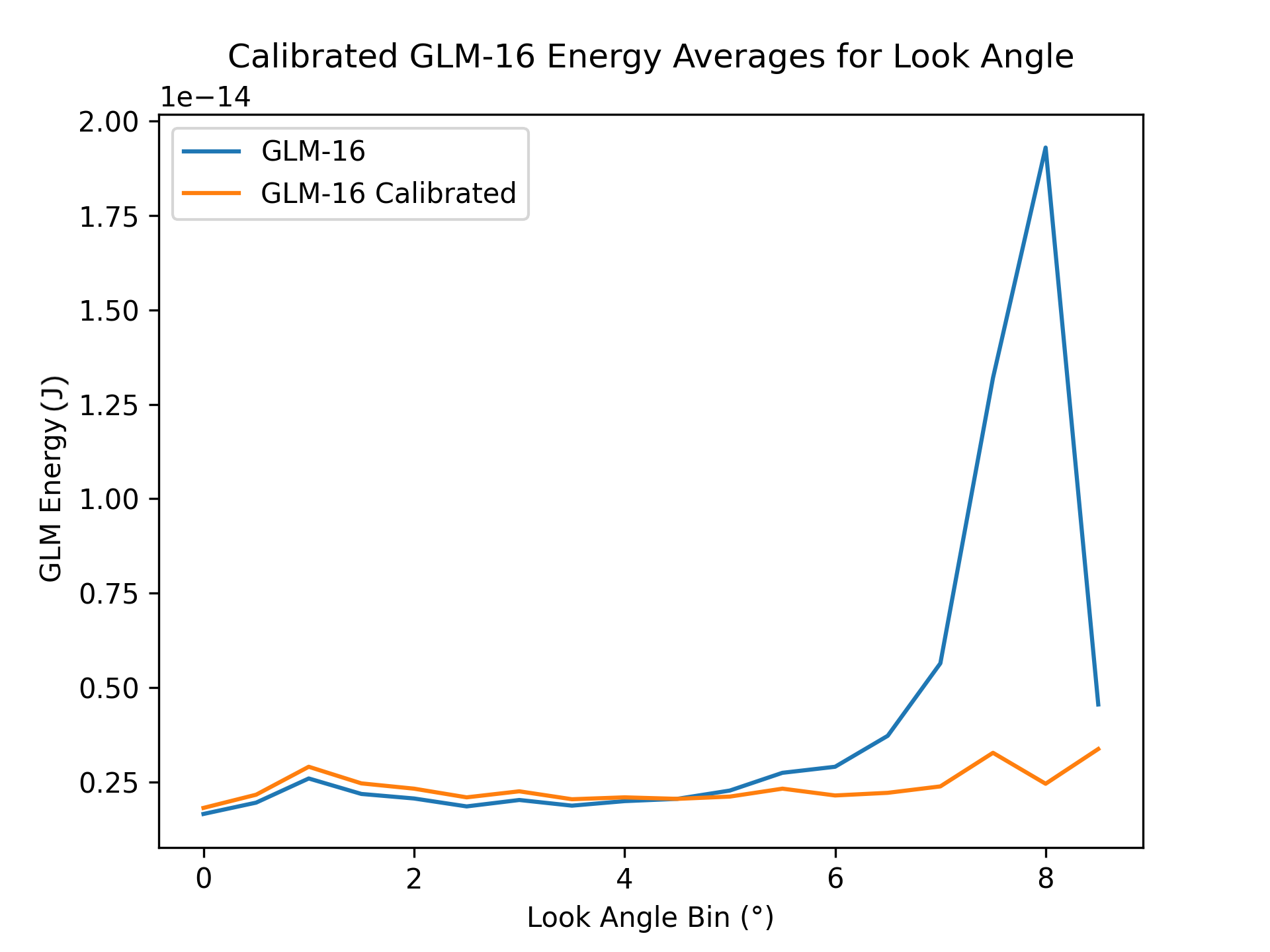}
  \caption{The average corrected L2 total fireball energy per 0.5 degree bin, where a correction equal to the inverse of the continuum calibration is applied to GLM-16 energy values as a function of look angle. The distribution of raw L2 energies is shown for comparison.}
  \label{fig:energy_cali16}
\end{figure*}

This gives a sense of how the apparent raw L2-based energies change with observation angle. As shown in Figure \ref{fig:energy_cali16} GLM-16 shows a large spike near the edge of the FOV for angles over 7° without any corrections (the blue line). There are about 3000 events for the GLM-16 sensor so it appears there is a significant overestimation in energy for GLM-16 near the edge of the detection threshold. We can see in both sensors a trend of increasing energy with look angle, implying that event energies near the edges are overestimated.

Using the Lockheed Martin calibration tables, we found the relative correction value using each events latitude and longitude. Similar to the methods for the energies, we divided the events into 0.5° bins and then averaged the look up table value for each look angle. The angle for potentially the least under/over corrected energies is at 4-5° \citep{Ozerov2023}; we took that to be our normalization point. We normalized all the bin calibration values to the 4.5° bin by dividing the averaged continuum by the value in the 4.5° bin, such that each bin had now a normalization factor that range from above 1 to below 1. We then multiply the continuum calibration normalization factor referenced to 4.5° to all the L2 energies. The result is shown in Figure \ref{fig:energy_cali16}. This procedure acts to isolate the relative effects of changes in L2 energy due to geometry, but does not apply an absolute correction as would be appropriate to slower fireballs with spectral energy distributions approximating a blackbody.

We can see the normalization for GLM-16 helps the overestimated energies significantly. Our analysis suggests that energies derived from GLM-16, in particular, at larger look angles should be used with extreme caution.

In our analysis to isolate shower fireballs we remove all events beyond a 7° look angle. Examination of Figure\ref{fig:energy_cali16} shows that this cut-off should limit systematic overestimation near the edges to less than a factor of $\sim$2 for both systems.  Note however, this continuum calibration applies to slower speed fireballs, when the spectral continuum is dominant. When we look at the high speed events, the 777nm spectral line is dominant and as such is similar to the lightning detection; thus it doesn't need to be applied to our shower related events. The simple cut off at a look angle beyond 7° produces accurate L2 energies. 

\subsection{Different Correction Factors to use for Total Intensity}
\label{sec:corr_fac}

As previously discussed in Section \ref{sec:cali}, we have adopted the relation between total intensity and GLM intensity from \cite{vojacek}. To extrapolate this relation beyond the brightness range where it was derived, \cite{vojacek} conducted a series of visual magnitude calibrations, incorporating various corrections that can be applied to GLM light curves. These corrections include a brightness correction and consideration of the presence of a flare in the light curve.

The flare correction is actually an indirect means of gauging the typical ablation height as \cite{vojacek} notes that changes in the ablation height affects the I$_{777}$ line. However, it's important to note that ablation height information is only available for stereo events in GLM and is not available for the majority of detections. Consequently, a flare/height correction could not be universally applied to all GLM events. We therefore assume that all our shower fireballs have similar ablation heights, noting that for most cases, following the results of \cite{vojacek}, this will introduce at most of order one stellar magnitude in brightness uncertainty. 

The brightness correction, on the other hand, is rooted in the division of EN fireballs into velocity bins. It was chosen to explore variations in the strength of the oxygen line to account for fireballs that may have magnitudes outside of the observed range. However, this approach does not account for the influence of velocity on radiation. To address this, \cite{vojacek} employed a least squares fit to derive magnitude correction constants based on velocity. The non-brightness corrected magnitude is derived from the line intensity and velocity as:

\begin{equation}
\label{eqn:voj_vis_mag}
m_v =  - 2.5 \times \log_{10}{I_{777}} + 0.0948 \times v - 3.45
\end{equation}

whereas the brightness corrected magnitude is given as: 

\begin{equation}
\label{eqn:voj_bc_mag}
m_v = cm(v) \times \log_{10}{I_{777}} +dm(v) - 2.5 \times \log_{10}{I_{777}} + 0.0948 \times v - 3.45
\end{equation}

with correction constants that depend on velocity given by:

\begin{equation}
\label{eqn:voj_bc_correction_const_c}
cm(v) = -0.022 \times v + 0.79
\end{equation}
and 

\begin{equation}
\label{eqn:voj_bc_correction_const_d}
dm(v) = 0.102 \times v - 3.31
\end{equation}

The fit is given by equation \ref{eqn:voj_bc_mag} with the parameters $cm(v)$ and $dm(v)$.

To assess the robustness of this correction we can compare the brightness for USG observations with GLM as was also done by \cite{vojacek}, though in our cases we have many additional events compared to what was available to \cite{vojacek}.

\subsection{USG and GLM Calibration Comparison}
\label{sec:glm_vs_usg}

To assess the accuracy of using L2 energies ideally we want to compare to an independent technique which measures energy for the same fireballs. As this requires large area coverage, the only currently available dataset which meets this requirement is USG data. The fireballs detected by USG are very bright and are not shower-related. This implies that the 777nm emission will be comparatively weak. As shown by \citet{jenniskens} for USG events the 777nm emission is part of the broad, continuum spectral energy distribution, often assumed to be a blackbody.

We now compare concurrent observations of fireballs captured by USG and GLM to estimate the accuracy of the brightness correction. In our dataset, which encompasses the GLM operational years 2020-2022, we have a total of 8 stereo events, 7 GLM-17-only observed events, and 18 GLM-16-only observed events also detected by USG. Detailed information about these events can be found in the summarized table presented in the Appendix.

Intensity light curves for some USG data were converted to equivalent magnitude to first determine which calibration best suits the GLM data from \cite{vojacek}. USG light curves were converted to magnitudes  using the following \citep{Brown1996}:

\begin{equation}
\label{eqn:peter_mag_usg}
m_v = 6 - 2.5\log_{10}I
\end{equation}

With some of the USG light curves, we can compute magnitudes using Eq.\ref{eqn:peter_mag_usg} to compare to the GLM magnitude curving using the two \cite{vojacek} relations: the uncorrected visual magnitude given by Eq.\ref{eqn:voj_vis_mag} and by the brightness corrected magnitude give by Eq.\ref{eqn:voj_bc_mag}. 

\begin{figure}
  \includegraphics[width=\linewidth]{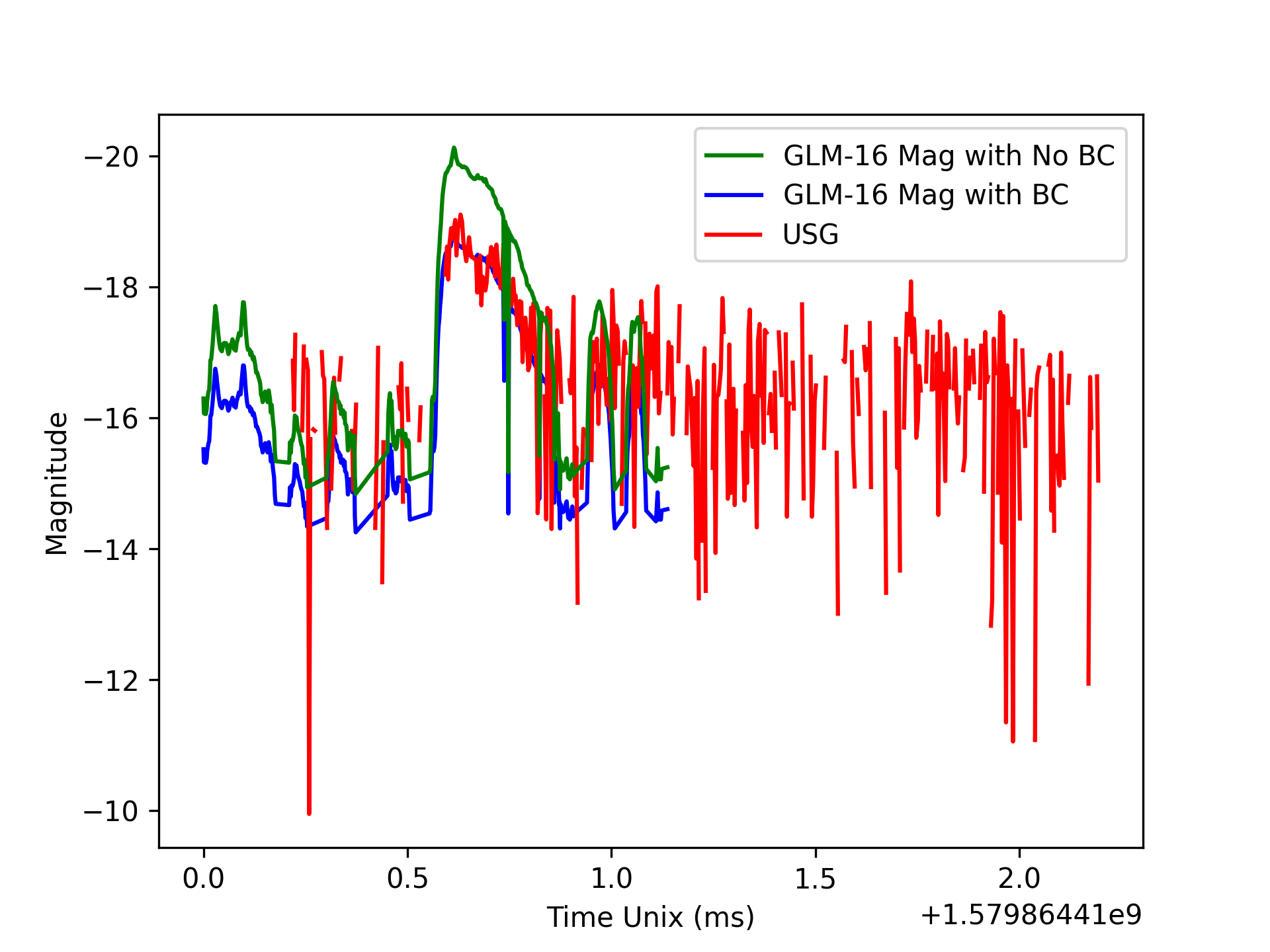}
  \caption{A comparison between the USG and GLM-16 magnitude curve for GLM-16, showing magnitude with (and without) brightness correction for GLM-16 and the USG magnitude for comparison. This fireball occurred on 2020-01-24 11:13:31 UTC at 28.0N 35.8W.}
  \label{fig:usg_glm_mag1}
\end{figure}

\begin{figure}
  \includegraphics[width=\linewidth]{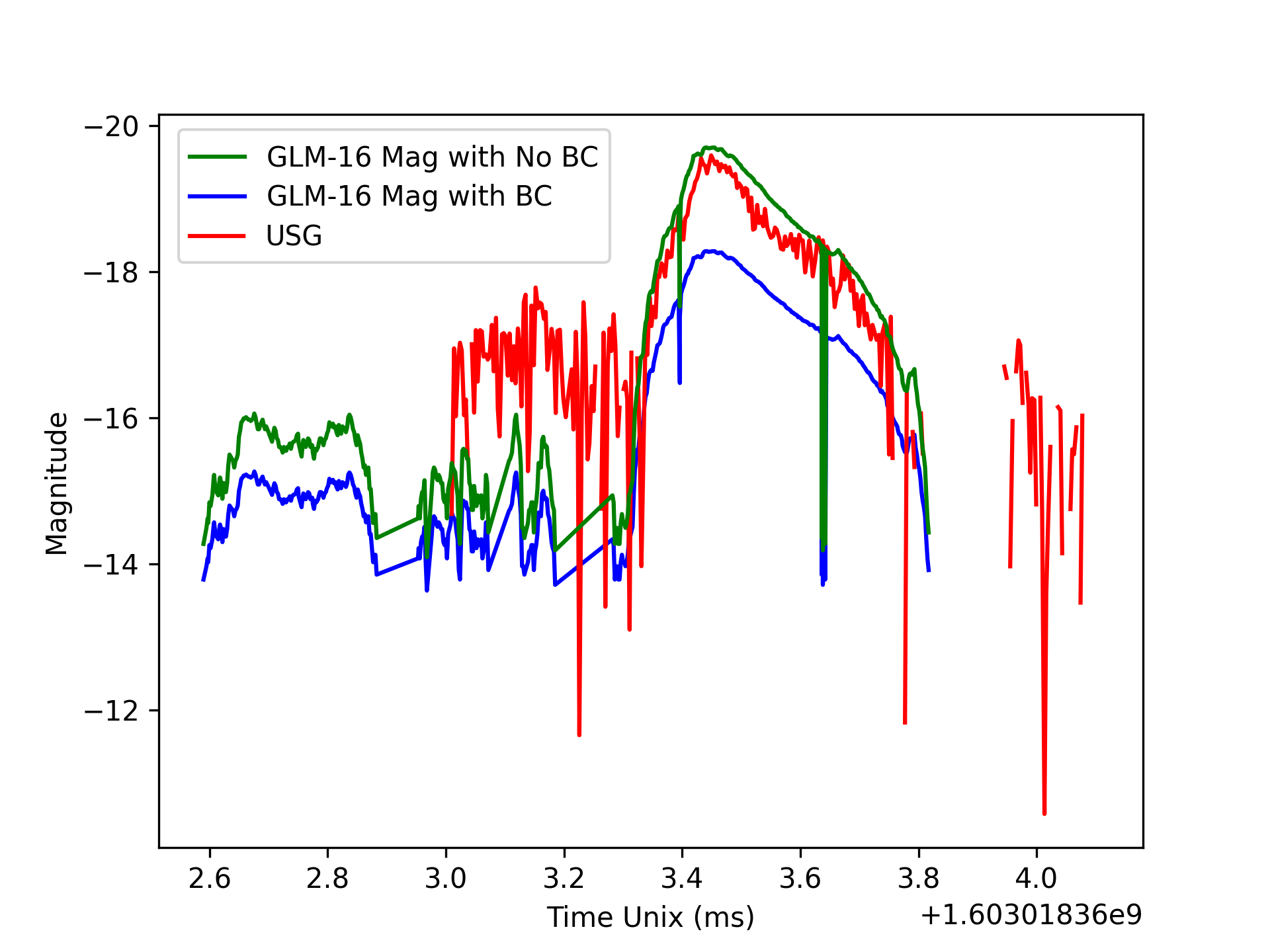}
  \caption{A comparison between the USG and GLM-17 magnitudes for GLM-17, brightness correction magnitude for GLM-17 and the USG magnitude. This fireball occurred on 2020-10-18 10:52:43 UTC at 11.5S 135.9W.}
  \label{fig:usg_glm_mag2}
\end{figure}

An example of a comparison between the GLM and USG magnitude curves is shown in Figure \ref{fig:usg_glm_mag1} for a GLM-16 event, with a USG velocity used of 21.1 km/s and for a GLM-17 event with a USG velocity of 16.2 km/s in Figure \ref{fig:usg_glm_mag2}.

Additional events shown in the appendix are \ref{fig:usg_glm_magapp1}, \ref{fig:usg_glm_magapp2}, and \ref{fig:usg_glm_magapp3}. One limitation in these comparisons for our work arises from the limited velocity range for USG-detected events, with the majority falling within the range of 10 km/s to 20 km/s, except for one event at 42.3 km/s. Even among events with relatively similar velocities, it's not immediately clear which correction (brightness corrected or standard) best matches the USG magnitude curve. For example, in the case of the GLM-16 event, the brightness corrected magnitude curve seems to be a better fit, while for the GLM-17 event, the curve with no brightness correction provides a better match to the USG curve.

Given this ambiguity, we chose to use the standard relation without the brightness correction for our work, but note that this may lead to more than a magnitude of systematic offset in some cases.

Before proceeding with mass calculations between USG and GLM events, we conducted a replication of the work presented in \cite{vojacek} as a verification step. We compared the total fireball energies determined from GLM, calibrated as per their method, against the USG energy, using a division factor of 1.85. This factor accounts for the fact that USG measurements cover the entire spectrum, while the calibration in \cite{vojacek} considers only the 380-850 nm bandpass.

\begin{figure}
  \includegraphics[width=\linewidth]{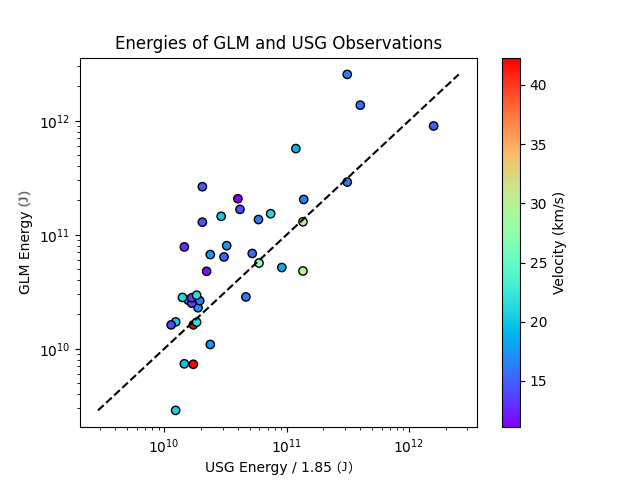}
  \caption{A comparison of total fireball energies for GLM (without brightness correction) with those determined from USG light curves as a function of velocity (color bar).}
  \label{fig:usg_glm_rep}
\end{figure}

Figure \ref{fig:usg_glm_rep} shows this comparison as a function of USG speed. Overall, this shows agreement with the results obtained in \cite{vojacek}. However, it's worth noting some differences, such as the larger dataset used in our analysis and the fact that the light curves in \cite{vojacek} for GLM include certain corrections to fill in data gaps, which were not applied in our work. Nevertheless, the overall trend, namely that there is a systematic overestimate (on average) between GLM event energies and those from USG is consistent with the findings of \cite{vojacek}. This overcorrection is likely due to those observations having high look angles and consequently overestimated energies.

To properly compare the energies between the USG and GLM detections,  we make use of the Lockheed Martin continuum calibrations. To facilitate this comparison, we utilized the USG velocity as the accepted velocity for the GLM calculations.

We also examine the effect of the look angle on the raw GLM energies see if all observations with a higher look angle from GLM tend to be overestimated when compared to USG. 

\begin{figure}
  \includegraphics[width=\linewidth]{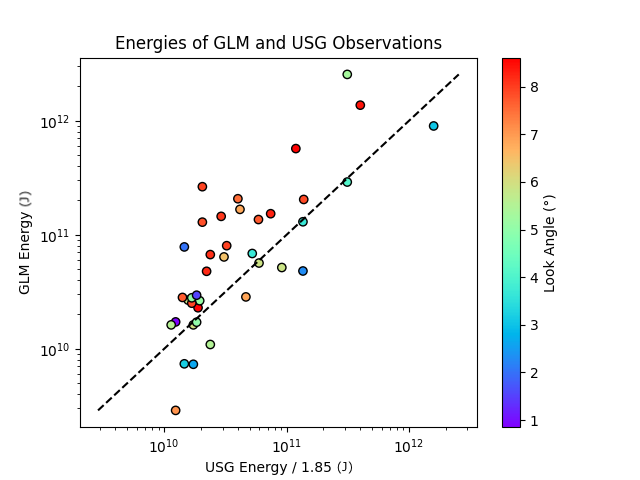}
  \caption{Concurrent fireballs recorded by GLM and USG showing energy correlation as a function of GLM look angle.}
  \label{fig:usg_glm_lookangle}
\end{figure}

\begin{figure}
  \includegraphics[width=\linewidth]{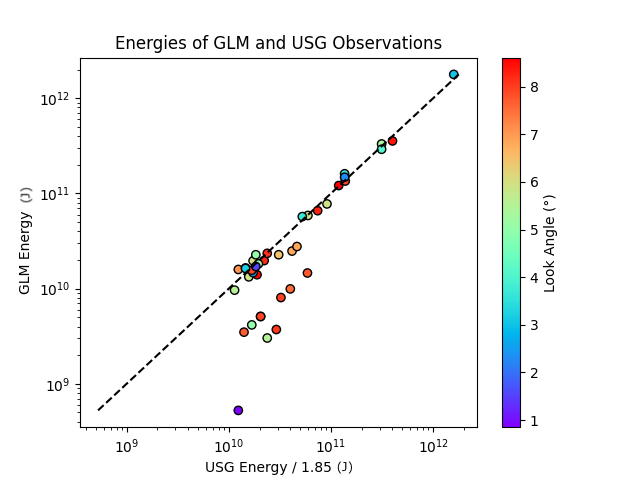}
  \caption{Concurrent fireballs recorded by GLM and USG showing energy correlation as a function of GLM look angle with the continuum calibration.}
  \label{fig:usg_glm_lookangle_cont}
\end{figure}

It is clear from Figure \ref{fig:usg_glm_lookangle} that there is a trend of GLM events at high look angles showing the greatest overestimate compared to USG.  Applying our earlier continuum correction to the GLM energies produces a revised comparison plot shown in Figure \ref{fig:usg_glm_lookangle_cont}.

It is immediately apparent that the continuum correction greatly improves the agreement between USG and GLM energies. The cluster of events paralleling the 1:1 line at energies between 10$^{10}$ and 10$^{11}$ J we believe to be an artifact of the processing which occurs in transforming data from L0 to L2 which has the net effect of removing some pixel intensity. This comparison suggests that in using L2 data we may, in some cases, produce underestimates by of order a factor of two in total energy, but are unlikely to overestimate the total fireball energy.

\subsection{Upper mass cutoff in streams}
\label{sec:upper_mass_cut}

Among the three streams showing the strongest signal in GLM data (LEO, PER and ETA) we find that observations for the largest mass meteoroids in the LEO and ETA are consistent with that expected from gas-drag formation alone (see Figure \ref{fig:lit_mass_with_GLM}. This is not surprising as the parent bodies (55P/Tempel-Tuttle and 1P/Halley respectively) of these streams have been well observed and have long been presumed to have formed from gas-drag sublimation \cite[e.g.][]{Egal2020}. Moreover, both of these streams have been successfully modelled as having been produced from gas-drag sublimation \citep{vaubaillon2003, Egal2020a}, showing observations consistent with this mode of production from the parent on essentially its current orbit.

From Whipple's criterion (Eq. \ref{eqn:whip}) it is clear that the perihelion distance of the comet makes a substantial difference to the expected maximum meteoroid mass which can be ejected. Hence, if the stream has been formed when the parent comet has a similar perihelion distance to its current value, we expect the upper meteoroid size to be close to the predicted value. 

Backward integrations of the Halleyid streams \citep{Egal2020a} show that the age of the ETAs are ~5000 years. During this time 1P/Halley's perihelion distance has actually decreased slightly, so the lack of meteoroids larger than the Whipple limit is consistent both with gas-drag limited mass ejection, the estimated age/dynamical evolution of the parent and its known nuclear size.

The Leonids show the strongest shower signal among all GLM fireball detections, with a daily rate at the time of peak more than 15$\sigma$ above the average GLM fireball background rate (see Figure \ref{fig:all_events}). The larger meteoroids in the LEO stream are quite young, with most having an age less than 2000 years (and probably much younger) \citep{Asher1999, Brown2000}, a consequence of the concentration effect of the 5:14 mean motion resonance with Jupiter. During the last 2000 years, the perihelion distance of 55P/Tempel-Tuttle has remained nearly constant \citep{Yeomans1999}, explaining the consistency between the Whipple prediction of the largest stream meteoroid and observations.

The Perseids show different behaviour than the ETA and LEO - the largest observed Perseids are more than three orders of magnitude in mass above the gas-drag limit. The Perseids are much older than either the ETA or LEO, with estimates suggesting an age of order 10$^{4-5}$ years \citep{Brown1998, jewitt1996}. During this time, the perihelion of 109P/Swift-Tuttle has not changed significantly \citep{Neslusan2022}, assuming no significant non-gravitational forces. Thus it does not appear the observation of meteoroids more than a factor of 10 in size larger than predicted by the Whipple formula can be due to older ejections having occurred at much lower perihelion distances.

A possible explanation is that 109P/Swift-Tuttle had higher gas production rate in the past than the average value assumed in Whipple's formula. As summarized in \cite{Harmon2004} the maximum meteoroid radius which can be drag-force lifted from a nucleus is proportional to both the gas production rate and the thermal gas expansion velocity. 

The latter could play a role if many meteoroids are ejected from 109P from jets where the expansion velocity would be expected to be higher, allowing larger meteoroids to escape. This is consistent with the presence of jets observed both at the 1992 return of the comet \citep{Jorda1994} and the 1862 return \citep{Sekanina1981}.

Alternatively, large Perseid meteoroids may be low density/fluffy and therefore experience greater drag than assumed in the Whipple formulation, again leading to larger ejected particles.

From our observations, it is clear that decimeter-sized meteoroids are still present in all three streams and that these have survived for long periods. Radar observations of cometary comae have long suggested the presence of centimeter particles near many active cometary nuclei \citep{Harmon2004}. Similarly, models of comet formation and observations from Rosetta suggest decimeter-sized particles of water ice and refractory grains should be abundant \citep{blum2017} consistent with our observations.

\subsection{GLM Shower Fireball Flux:  Comparison to Shower Fluxes at Smaller Masses}
\label{sec:flux}

To place into context our GLM shower-fireball flux we compare to shower fluxes at smaller masses obtained through other observation methods. For this we employ two distinct approaches:

\textbf{Collecting Area (CA) Flux:} This method utilizes the collecting area discussed in Section \ref{sec:CA_methods}. We select the highest flux value around the peak of each shower. To determine the equivalent limiting mass appropriate for this flux, we use the shallow-sloped cumulative mass distribution from GLM for each shower and identify the curve's turn-off at low masses. We then find the mass where the shallow fit intersects the total count and take this to be the effective limiting mass. In the case of the Leonids observed by GLM-16, for example, we find a limiting mass of 0.066 kg and an associated magnitude of -10.5. We interpret this to mean that our flux is complete to this mass (and larger).

\textbf{Poisson Flux:} This approach involves calculating flux by using the steep power-law fit from the cumulative mass distribution. We select the Poisson mass (corresponding to detection at 50\% probability) and find the number of associated events with that mass or greater using the steep fit to the curve. We then calculate the time-area product for the shower to determine an average flux. These fluxes are appropriate to the largest events based on Poisson statistics and are labeled as 'Poisson' in the accompanying figures.

\begin{figure}
  \includegraphics[width=\linewidth]{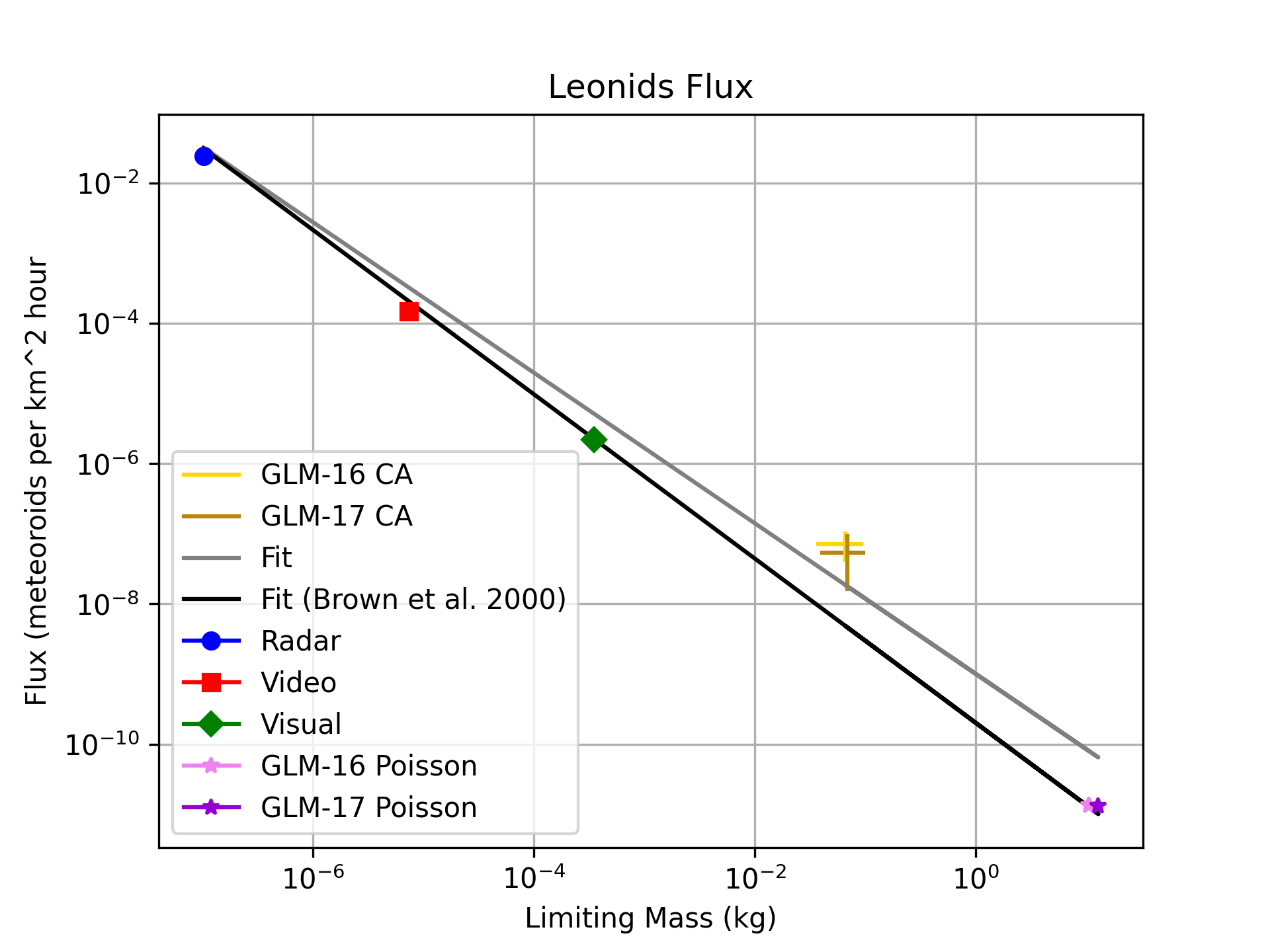}
  \caption{Meteoroid flux for the Leonids across eight orders of magnitude in mass using fluxes from \cite{brown_flux} at small sizes and GLM values at the large end. The black line represents the fit from \cite{brown_flux} of s=1.95 $\pm$ 0.05 compared to the new fit (grey line) with s=2.08 $\pm$ 0.08.}
  \label{fig:flux_brown}
\end{figure}

\begin{figure}
  \includegraphics[width=\linewidth]{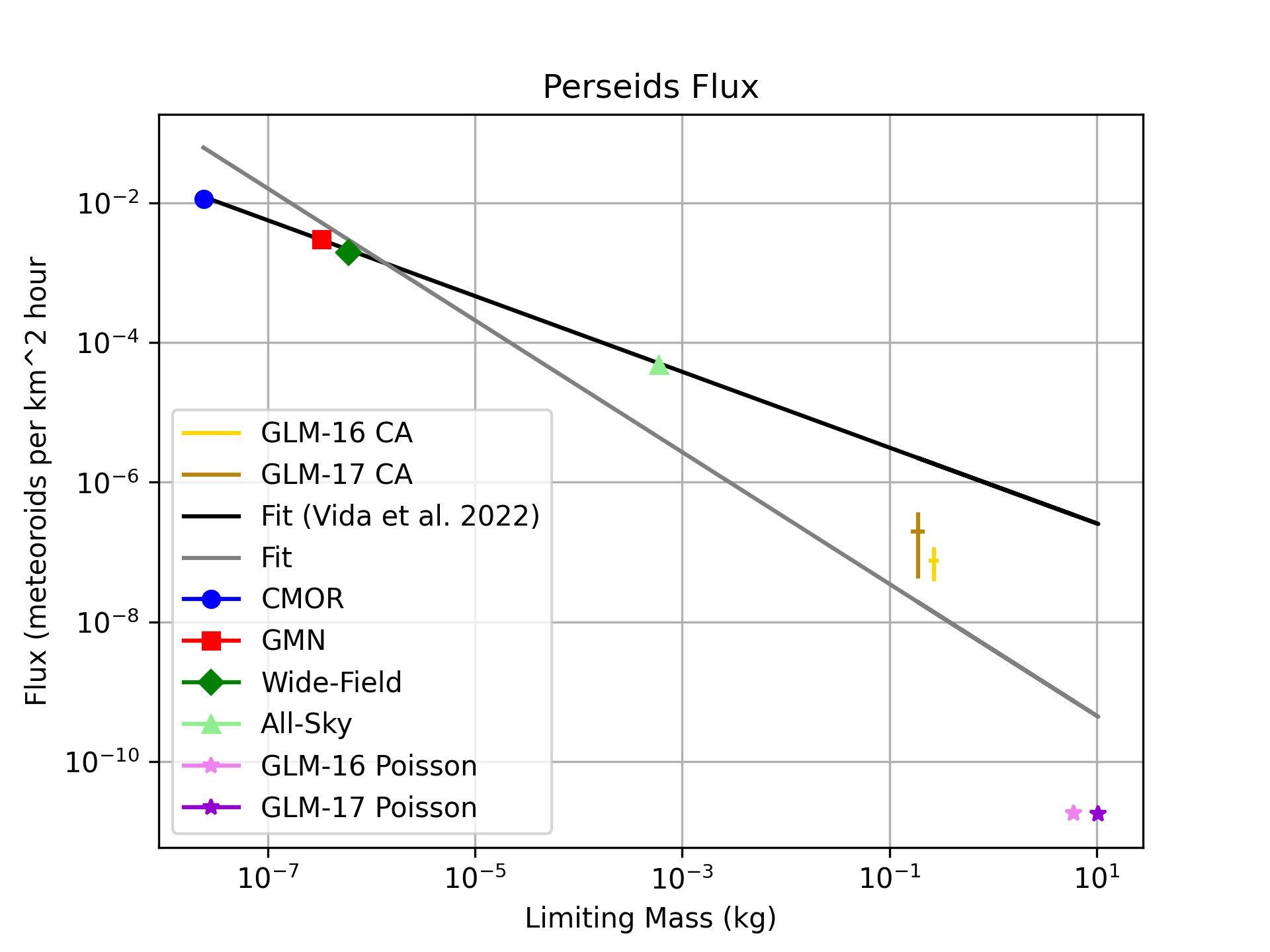}
  \caption{Meteoroid flux for the Perseids employing values from \cite{vida_flux} at small Perseid masses and GLM at larger masses. The black line fit has s=1.83 $\pm$ 0.10 from \cite{vida_flux}. The grey fit shows the GLM data with s=1.94 $\pm$ 0.10. }
  \label{fig:flux_vida}
\end{figure}

\begin{figure}
  \includegraphics[width=\linewidth]{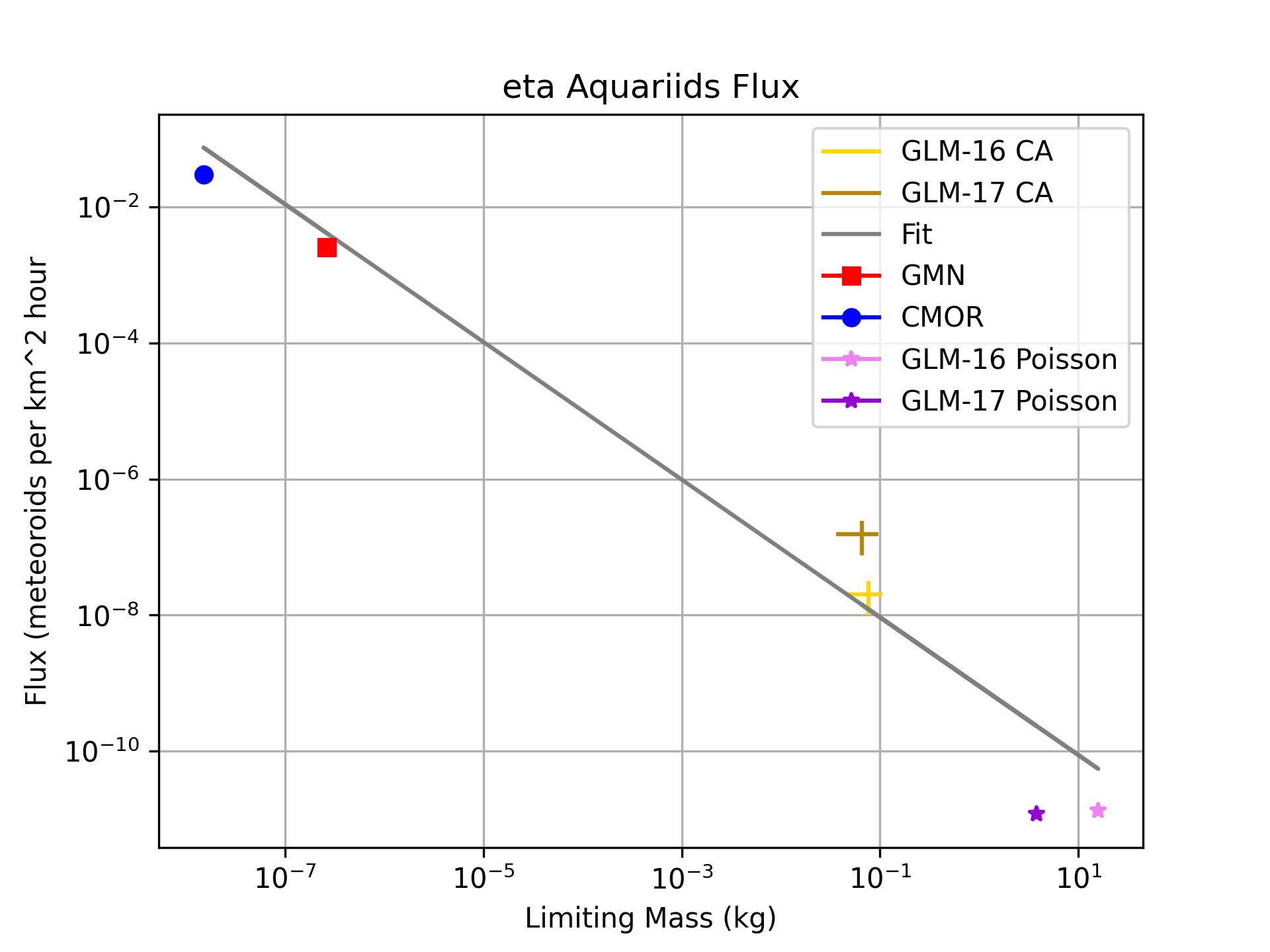}
  \caption{Meteoroid Flux for the eta Aquariids using optical, radar and GLM flux estimates. The grey solid line represents a mass index of s=2.00 $\pm 0.09$}
  \label{fig:eta_flux}
\end{figure}

The Poisson and Collecting-area based GLM flux estimates are compared to fluxes computed at smaller masses using radar and optical instruments in Figures \ref{fig:flux_brown} - \ref{fig:eta_flux}. Here we perform a regression fit (solid black line in all plots) to the flux values at lower masses and assume a constant power law applies across all mass ranges. 

For the Leonids, we use the radar and optical fluxes appropriate to the 1999 storm from \cite{brown_flux}. Figure \ref{fig:flux_brown} is a concatenation of the fluxes reported in \cite{brown_flux} and our GLM fluxes. Our values reflect the average Leonid flux of the 4 years (2019-2022) of GLM measurements for the Leonids in the solar longitude interval from 231-241. Noting that the Leonids in these years did not produce any significant outbursts, we assume returns were near the normal ZHR level of 20-30 \citep{Brown1994}. As such we adjusted the references fluxes from \cite{brown_flux} by a factor of 100, as the peak flux from 1999 corresponded to a ZHR of 3700 \citep{Arlt_Rubio_Brown_Gyssens_1999}. 

The resulting regression fit between 10 and 10$^{-7}$ kg produces a differential mass index of 2.08$\pm$0.08. This is in remarkably good agreement with our Poisson flux estimate for the largest Leonid meteoroid detected by GLM. Our peak flux using our collecting area approach is a factor of several above this constant power law fit, but also has significantly larger uncertainty as shown by the error bars which represent only one standard deviation.

A similar analysis for the Perseids is accomplished using the values at smaller masses for the shower as presented in \cite{vida_flux}. In contrast to the Leonids, here the power law curve, equal to an s=1.8, when extrapolated to large sizes produces expected fluxes much higher than GLM estimates. 

The flux profile here may suffer from several biases at smaller sizes. For the radar measurements, it is known that backscatter systems (like the Canadian Meteor Orbit Radar, the source of the flux number here) detect fewer Perseids than expected, potentially due to larger effects from initial radius and/or fragmentation than found in the general radar echo background \citep{Campbell2002}. This would tend to make the power law distribution shallower. 

As well, the all-sky (bright meteor) Perseid flux data point in this figure is from a single night at the peak of the shower (Aug 11/12, 2016) as reported in \cite{Ehlert2020}. They note that the resulting differential mass index is much smaller than typical literature values (which are near 1.8) and suggest that the shower was unusually rich in large meteoroids due to an outburst in this year. Moreover, restricting the all-sky measured fluxes to the peak night of this outburst may further enhance the flux value relative to the interval of solar longitude from 136-142 we use for the GLM flux estimation.   

If we take the optical fluxes as likely most reliable and use a fixed power law of s=1.8 we see in Figure \ref{fig:flux_vida} reasonable  agreement with GLM fluxes using the collecting area method (GLM-16/17 CA) at larger masses. Alternatively, there may be a real turnover in the flux of Perseids at the largest sizes deviating significantly from our assumed constant power law.

Finally we consider the GLM flux of the eta Aquariids. Here we use fluxes derived from the Global Meteor Network (GMN) using the methodology of \cite{vida_flux} averaged from 2021-2023 and from the Canadian Meteor Orbit Radar (CMOR) averaged from 2002-2014 as described by \cite{Campbell-Brown2014} at small sizes. The results, shown in Figure \ref{fig:eta_flux}, show reasonable agreement between the GLM flux estimates from the collecting area method and smaller particle sizes for a fixed mass index of 2 (solid line). The flux estimates for the Poisson approach are slightly lower than this curve, potentially reflecting a real deficit of very large eta Aquariids. 

Using the cumulative mass distributions from the shallow fit of GLM data alone, we can estimate the mass index of the population of large meteoroids in each stream. This is summarized in Table \ref{tab:mass-index-fireball}. 

\begin{table}[!ht]
    \centering
    \begin{tabular}{|l|l|l|l|}
    \hline
        \textbf{Shower} & \textbf{Limiting Mass (kg)} & \textbf{Sensor} & \textbf{Mass Index} \\ \hline
        \textbf{Leonids } & 0.07 & GLM-16 & 1.72 $\pm$ 0.01\\ 
        \textbf{}  ~ & ~ & GLM-17 & 1.70 $\pm$ 0.01\\ \hline
        \textbf{}  ~ & / & Flux from Full Mass Range & 2.08 $\pm$ 0.08\\ \hline
        \textbf{Perseids} &  0.25 & GLM-16 & 1.78 $\pm$ 0.005 \\ 
        \textbf{}  ~ & ~ & GLM-17 & 1.77 $\pm$ 0.005 \\ \hline
        \textbf{}  ~ & / & Flux from Full Mass Range & 1.94 $\pm$ 0.1 \\ \hline
        \textbf{eta Aquariids} & 0.07 & GLM-16 & 1.68 $\pm$ 0.05\\ 
        \textbf{} ~ & ~ & GLM-17 & 1.79 $\pm$ 0.05\\ \hline
        \textbf{} ~ & / & Flux from Full Mass Range & 2.00 $\pm$ 0.09\\ 
        \hline
    \end{tabular}
    \caption{The mass indices of derived GLM cumulative distributions for shallow fit of the Leonids, Perseids and eta Aquariids. The Flux from Full Mass Range row is a summary from the comparison spanning the full mass range per shower, anchored at small masses using other observation methods as summarized in section \ref{sec:flux}.}
    \label{tab:mass-index-fireball}
\end{table}

\section{Conclusions}
\label{sec:conclusions}

In this study, we demonstrate the utility of GLM instruments to analyze specific meteor showers. Employing temporal and spatial filtering, we provide estimates of the largest meteoroids present in several major showers by measuring the power-law of the cumulative mass distribution at large sizes and isolating the individual most energetic event per shower.  Among the showers examined in this paper—Leonids, Perseids, and eta Aquariids—all displayed strong signals above the background GLM fireball rate average. Evidence for some signal from the Orionids and Taurids was also found; all other major showers showed no significant enhancements above the background GLM fireball rate.

Among our main conclusions are:
\begin{enumerate}
    \item The LEO, PER and ETA all contain multi-kilogram meteoroids, with estimated maximum masses from GLM of approximately 7.3 kg, 3.4 kg, and 2.9 kg, respectively.
    \item The GLM estimated largest masses for the LEO and PER are comparable to the largest lunar impacts which had estimated masses of 5 and 7 kg respectively.
    \item Comparison with gas-drag models shows that the LEO and ETA are consistent with the classical Whipple limit suggesting gas-drag release from the parent comet in more or less its current orbit captures the formation process.
    \item The Perseids show clear evidence for much larger meteoroids than the simple Whipple gas-drag formulation predicts. We suggest that this is indicative of ejection processes more complicated than captured by the Whipple formula, possibly including mantle release, jet-driven ejection or if large Perseids had very low bulk densities at ejection.
    \item The LEO and ETA mass distributions are well represented by a single differential mass index of s=2.08$\pm$0.08 and s=2.00$\pm$0.09 respectively from 10$^{-7}$ kg < m < 1 kg. 
    \item The luminous efficiency relation of \cite{Ceplecha_McCrosky_1976} appears to be most appropriate to high speed shower fireballs based on agreement with empirical measurements of $\tau$ from \cite{borovicka_1997, Brown2007}.
    \item The Taurids show evidence for very large meteoroids, though not beyond the nominal Whipple gas-drag limit, which is large due to 2P/Encke's small perihlion distance.
    \item From GLM measurements, the Orionids appear to have large meteoroids comparable to the ETA with maximum masses of order 3-5 kg. However, this estimate has greater uncertainty than for the ETA due to low number statistics.
    \item GLM fireball energies are overestimated using L2 data alone, particularly at larger satellite look angles. We also demonstrated the importance of the continuum correction for slower fireballs through comparison of USG and GLM energies.
\end{enumerate}

As expected, we find that GLM is most sensitive to fast shower fireballs and showers with high fluxes, with the former condition being most important. The apparent difference in detectability of the ETA compared to the ORI reflects this effect as the Orionids have a smaller flux than the eta Aquariids. 

The current integrated time-area-product for GLM is just sufficient to have a significant probability of detection of multi-kilogram shower fireballs for the LEO, ETA and PER. As the GLM TAP increases, more stringent values on the extreme upper mass limits in these streams will become possible. 

 Our study also demonstrates the potential use of GLM in estimating meteoroid flux, complementing other observational methods. For the Leonids, our Poisson flux estimation showed good agreement with established fits in \cite{brown_flux}, although our collection area flux was about an order of magnitude above a simple power-law fit. In the case of the Perseids, both our Poisson and collection area fluxes were below the fit proposed by \cite{vida_flux}. This may be due to higher than average all-sky fluxes and underestimation by radar. 

Additionally, we investigated the eta Aquariids, using flux estimates from the Global Meteor Network (GMN) and the Canadian Meteor Orbit Radar (CMOR) to compare with out GLM fluxes. Our collection area flux agreed with a simple power law fit, while the Poisson flux was well below the extrapolated power law.

Analysis of GLM data in the current study, highlights limitations of the data. These include:

1. The processing pipelines for GLM data, particularly those specific to meteor detections, are not entirely disclosed, potentially affecting the accuracy of GLM energies. Obtaining unprocessed L0 data could mitigate this uncertainty for future events.

2. Limitations exist in the bandpass for reported energy from lightning detection sensors. While efforts have been made to calibrate energies in the 777 nm line, further work on calibrating reported GLM energies is warranted.

3. The determination of photometric mass heavily relies on the luminous efficiency factor, $\tau$, which remains uncertain in meteor research, potentially impacting our mass estimates.

4. Our filtering methods provide an estimation of shower membership, lacking precision. Incorporating velocity and radiant data from additional Stereo events can significantly enhance the probability assessment of shower membership.

\section{Acknowledgements}

\label{section:acknow}
This work was funded by the NASA Meteoroid Environment Office under cooperative agreement 80NSSC21M0073. We thank Jeff Smith and Anthony Ozerov for their work on estimating GLM velocities for stereo observations that was kindly shared with us, as well for their great input and help using the fireballs python package. We also thank Thomas Edwards and Clem Tiller for their continuous input and help in analysing GLM data. Clem Tillier also kindly provided GLM continuum calibration tables used in our work. We also thank Vlastimil Vojacek for feedback insight to the energy calibrations work that was extensively used in this work and Zbyszek Krzeminski for his fireball reductions work and Jana Wisniewski for assistance in computing collecting area.

% Import the bibliography file bibliography.bib
\bibliography{references}

\newpage

\appendix
\section{Collecting Area and Fluxes for ETA and PER } \label{app:ca}

Included here are additional examples of fluxes for the PER and ETA which were summarized in the main paper. Here we have the individual yearly examples of the GLM fluxes for the Perseids and the eta Aquariids:

 \begin{figure}[ht!]
  \includegraphics[width=\linewidth]{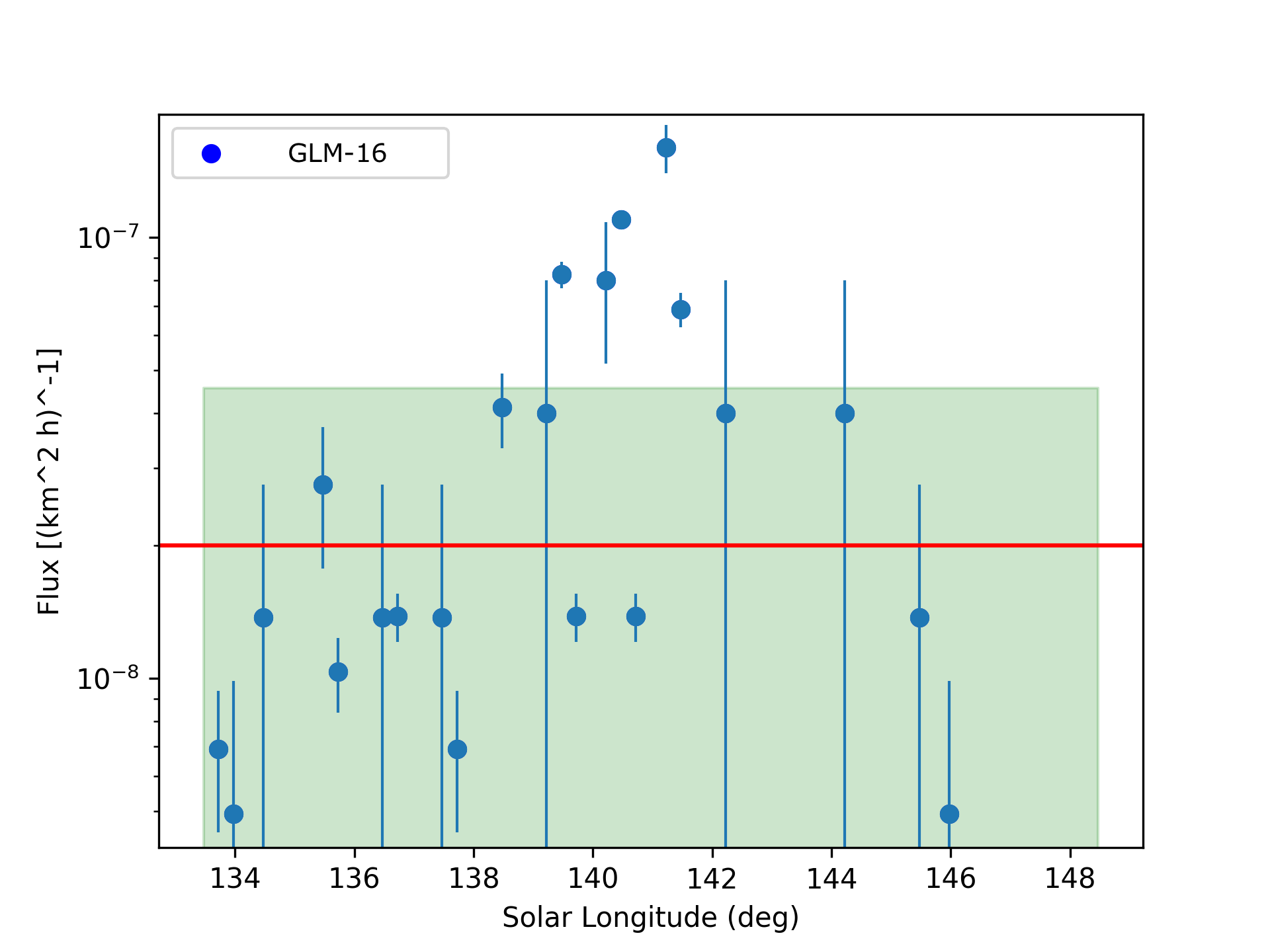}
  \caption{The 2020 Perseid flux as recorded by GLM-16 together with its variance and the background average (red line) and background variance (green shaded region).}
  \label{fig:flux_plot_per_2020_16}
\end{figure}

 \begin{figure}[ht!]
  \includegraphics[width=\linewidth]{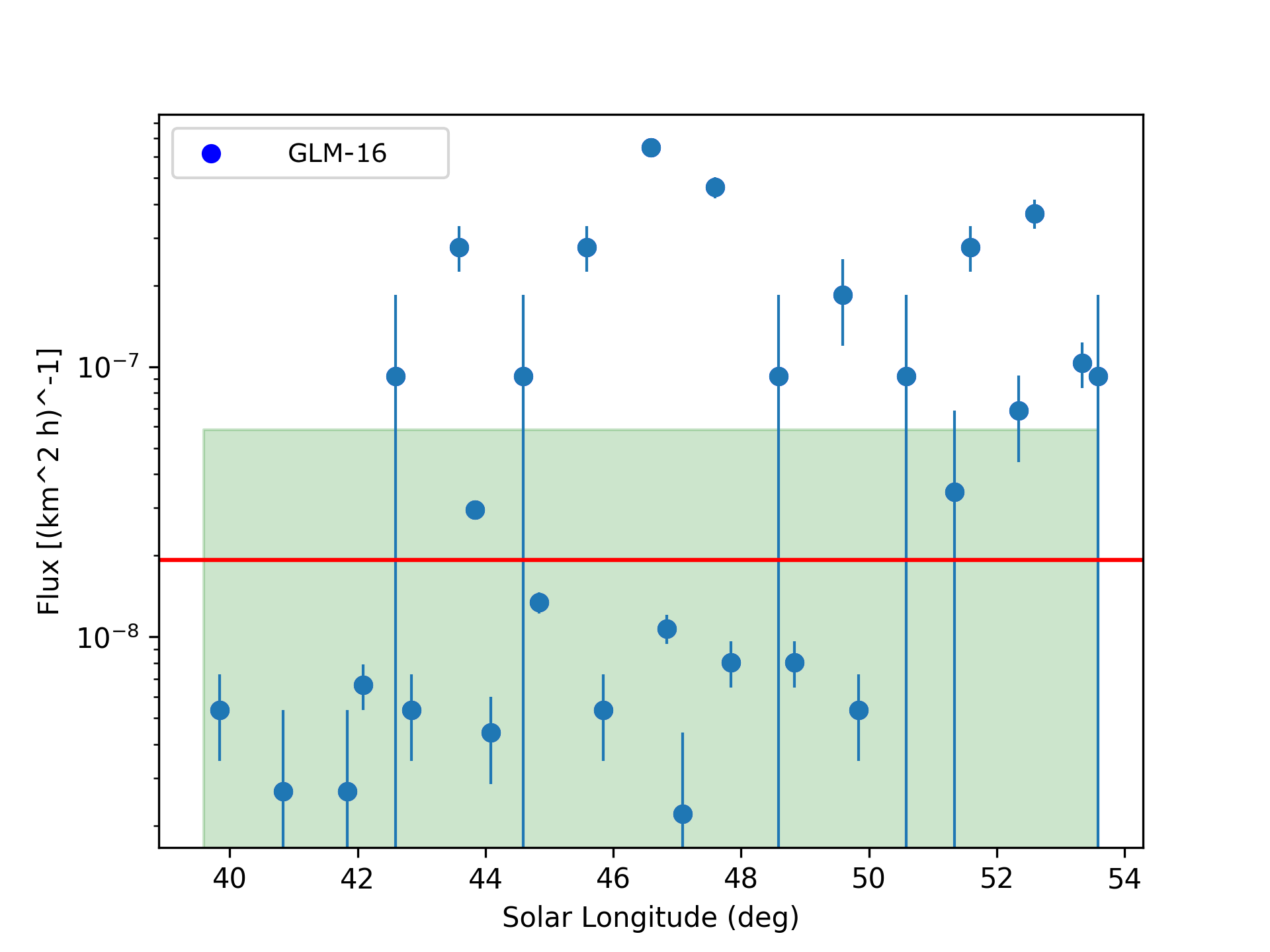}
  \caption{The 2021 Eta Aquariid flux as recorded by GLM-16 together with its variance and the background average (red line) and background variance (green shaded region).}
  \label{fig:flux_plot_eta_2021_16}
\end{figure}
\newpage

\section{Detailed Data for Largest PER, LEO and ETA Fireballs} \label{app:largest}
The following figures are the raw L2 light curves and positional information for the most energetic LEO, PER and ETA fireballs:

 \begin{figure}[ht!]
  \includegraphics[width=\linewidth]{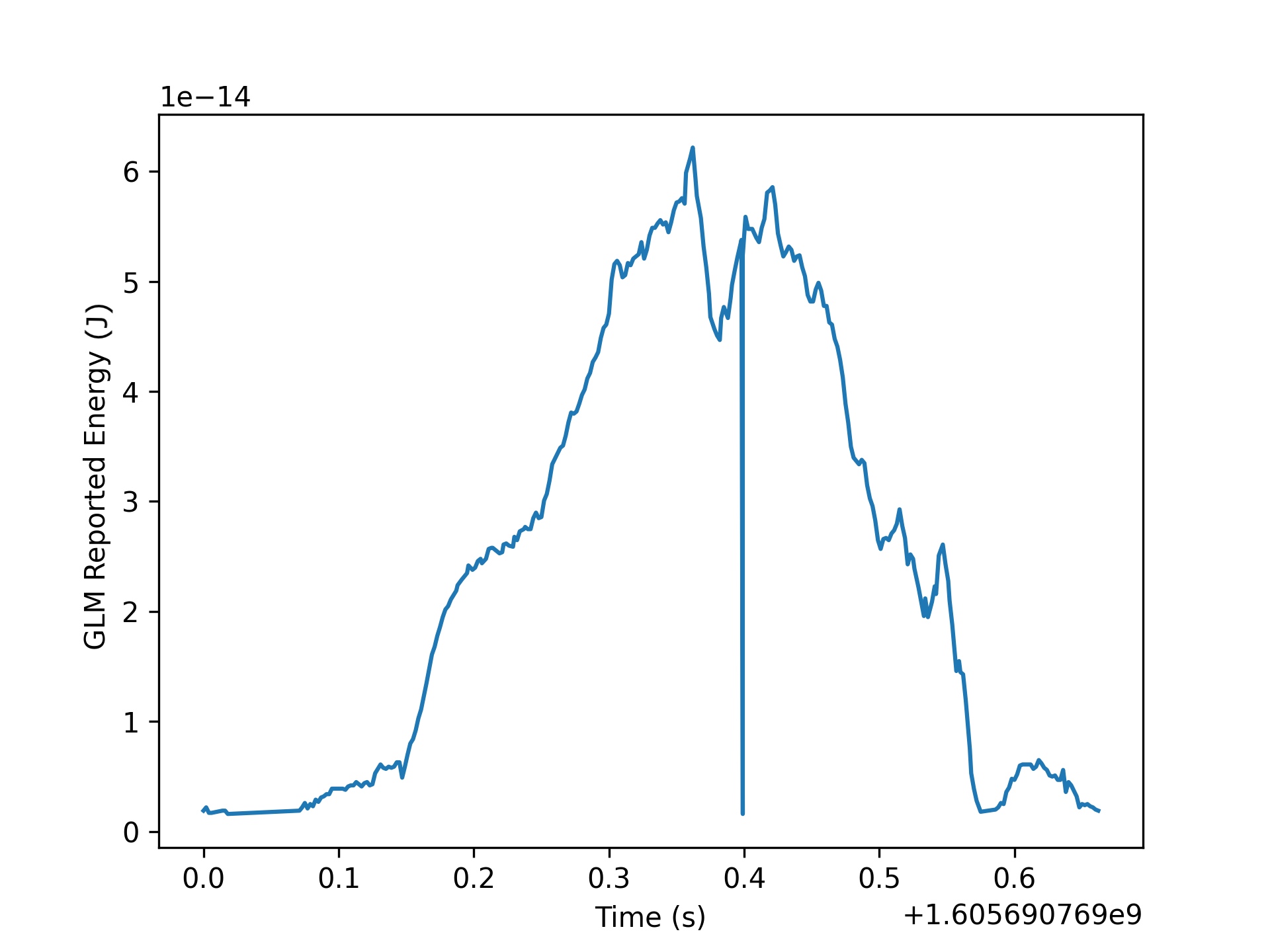}
  \caption{The light curve for the most energetic Leonid fireball as measured by GLM-16 recorded on 2020-11-20 09:12:00 with the timescale starting at T0=09:12:00 UT. See Table \ref{tab:Big_fireballs1} for more detail.}
  \label{fig:leo_light}
\end{figure}

 \begin{figure}[ht!]
  \includegraphics[width=\linewidth]{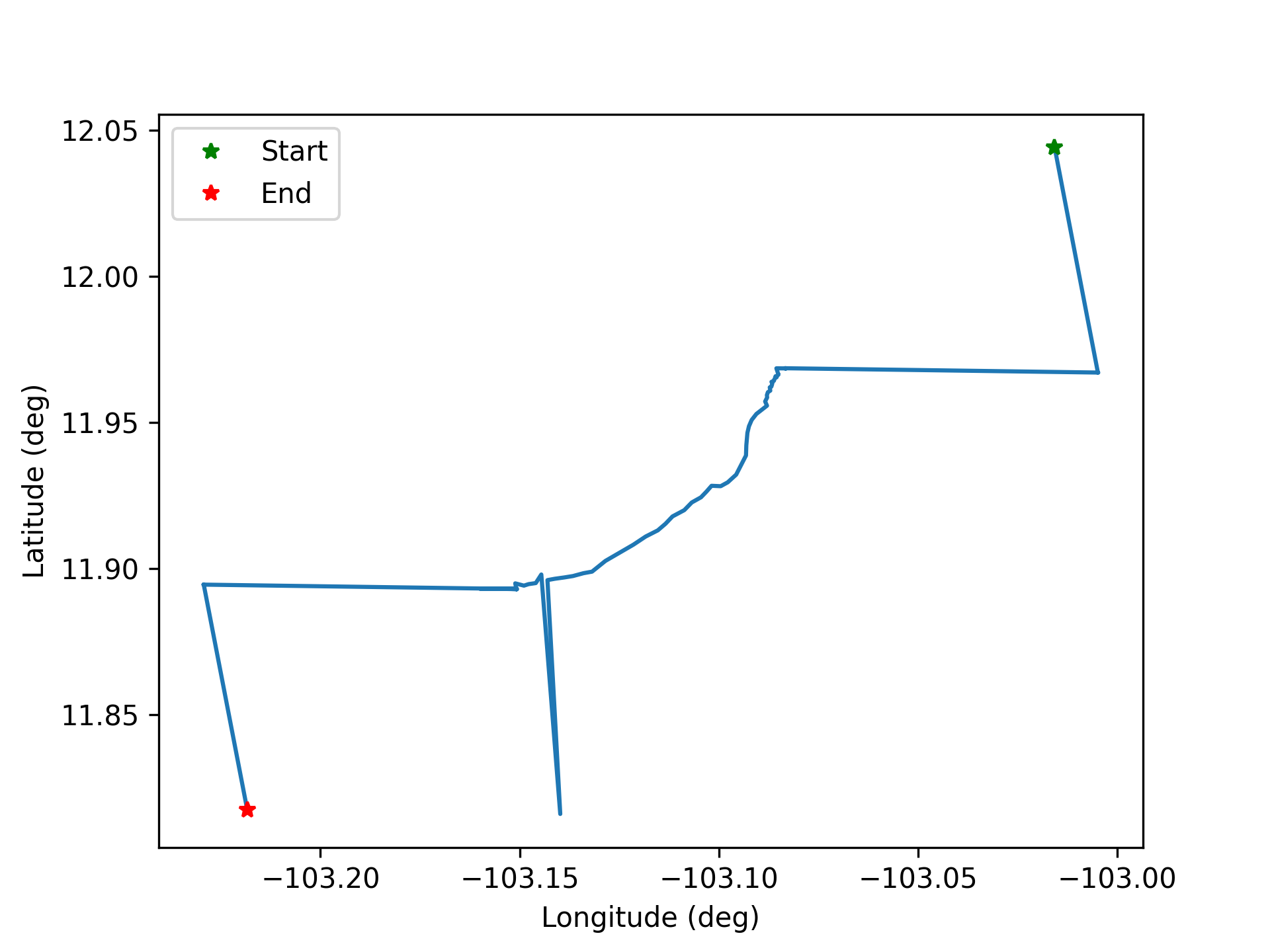}
  \caption{The ground track for the most energetic Leonid fireball as measured by GLM-16 recorded on 2020-11-20 09:12:00. See Table \ref{tab:Big_fireballs1} for more detail.}
  \label{fig:leo_pos}
\end{figure}

 \begin{figure}[ht!]
  \includegraphics[width=\linewidth]{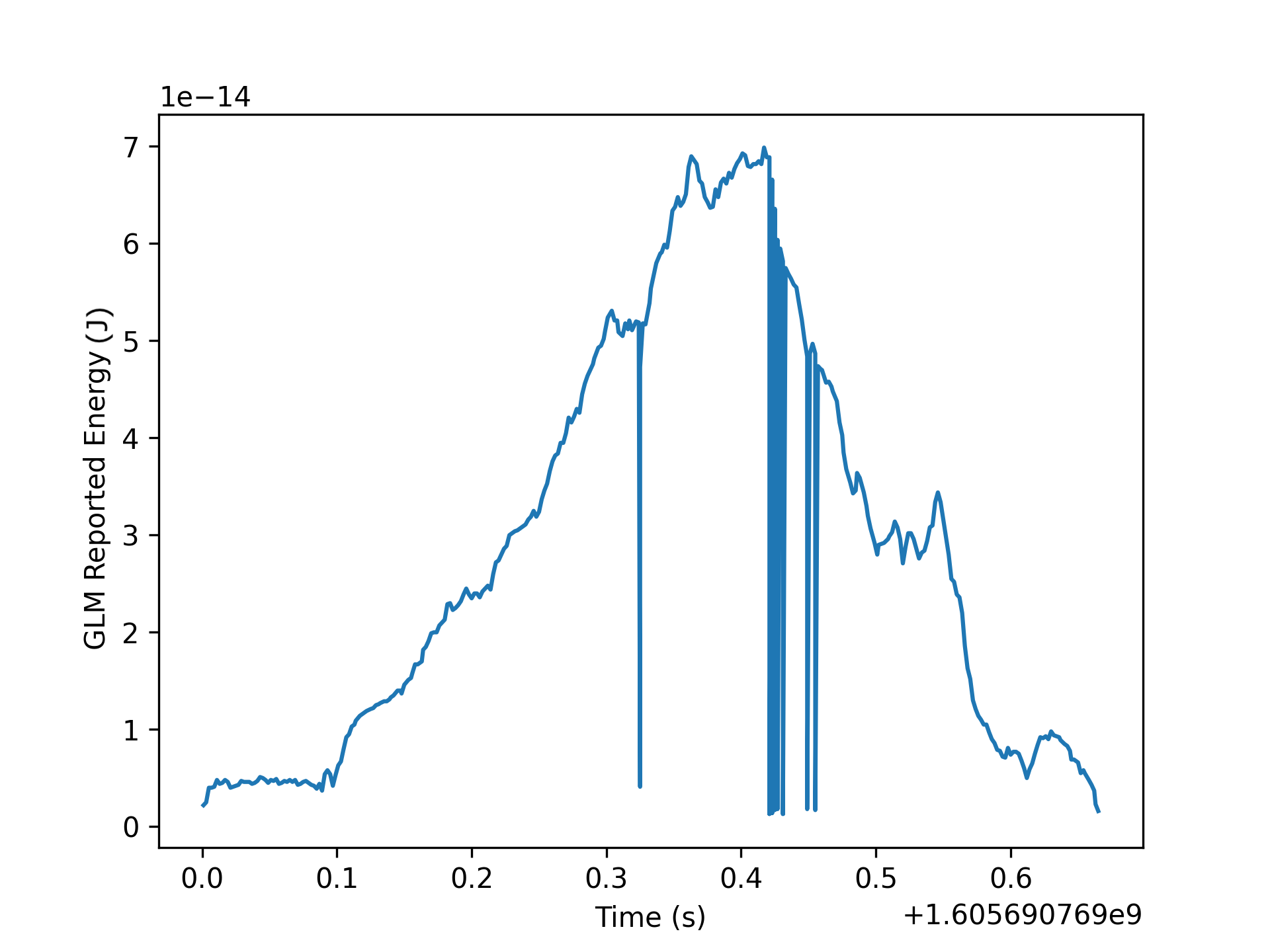}
  \caption{The light curve for the most energetic Leonid fireball as measured by GLM-17 recorded on 2020-11-18 09:12:00 with the timescale starting at T0=09:12:00 UT. Drops in the light curve can be artefacts from the pixel ground track. See Table \ref{tab:Big_fireballs1} for more detail.}
  \label{fig:leo_light2}
\end{figure}

 \begin{figure}[ht!]
  \includegraphics[width=\linewidth]{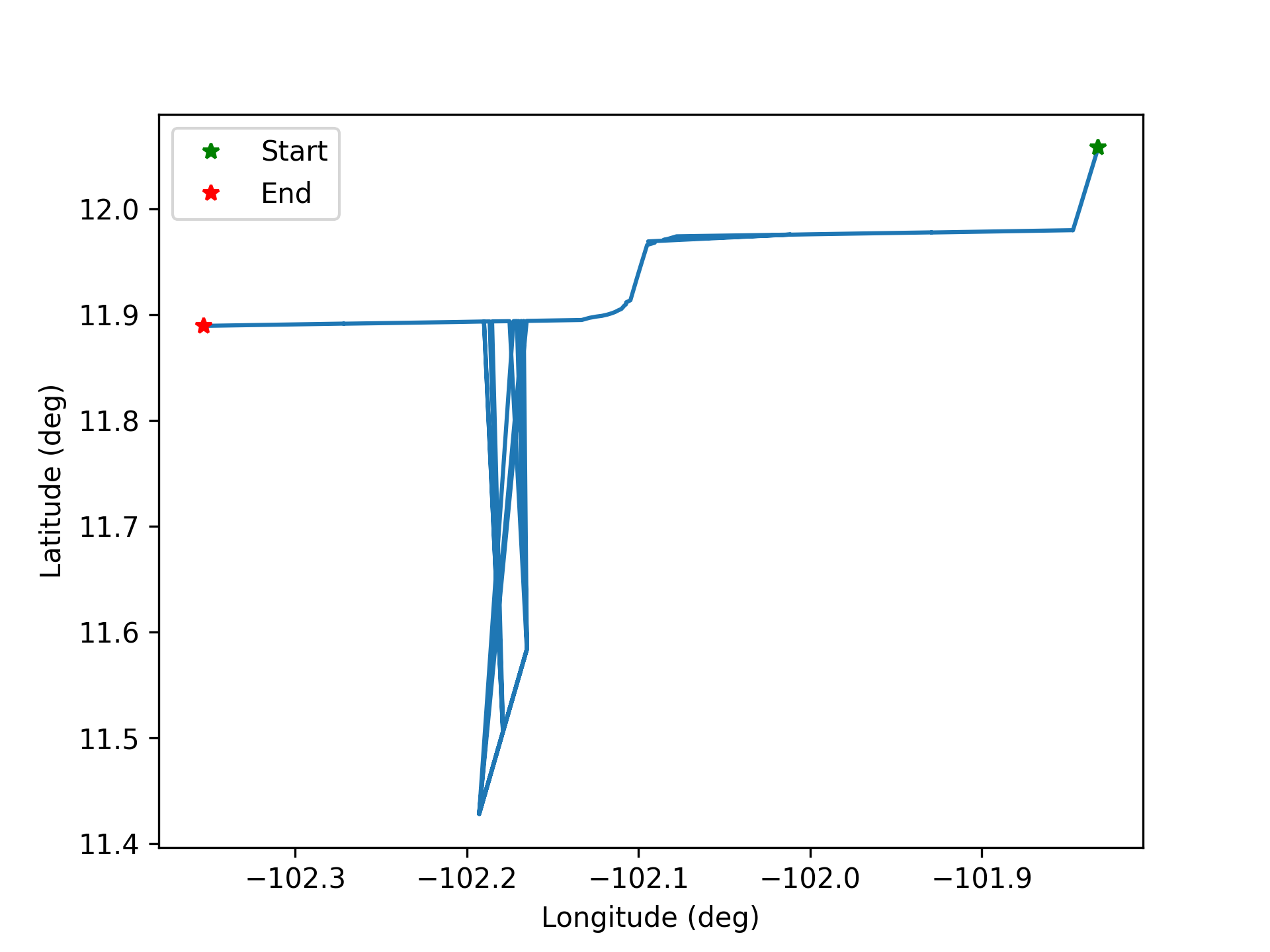}
  \caption{The ground track for the most energetic Leonid fireball as measured by GLM-17 recorded on 2020-11-20 09:12:00. Artefacts in the ground track are likely due to pixel coverage. See Table \ref{tab:Big_fireballs1} for more detail.}
  \label{fig:leo_pos2}
\end{figure}

 \begin{figure}[ht!]
  \includegraphics[width=\linewidth]{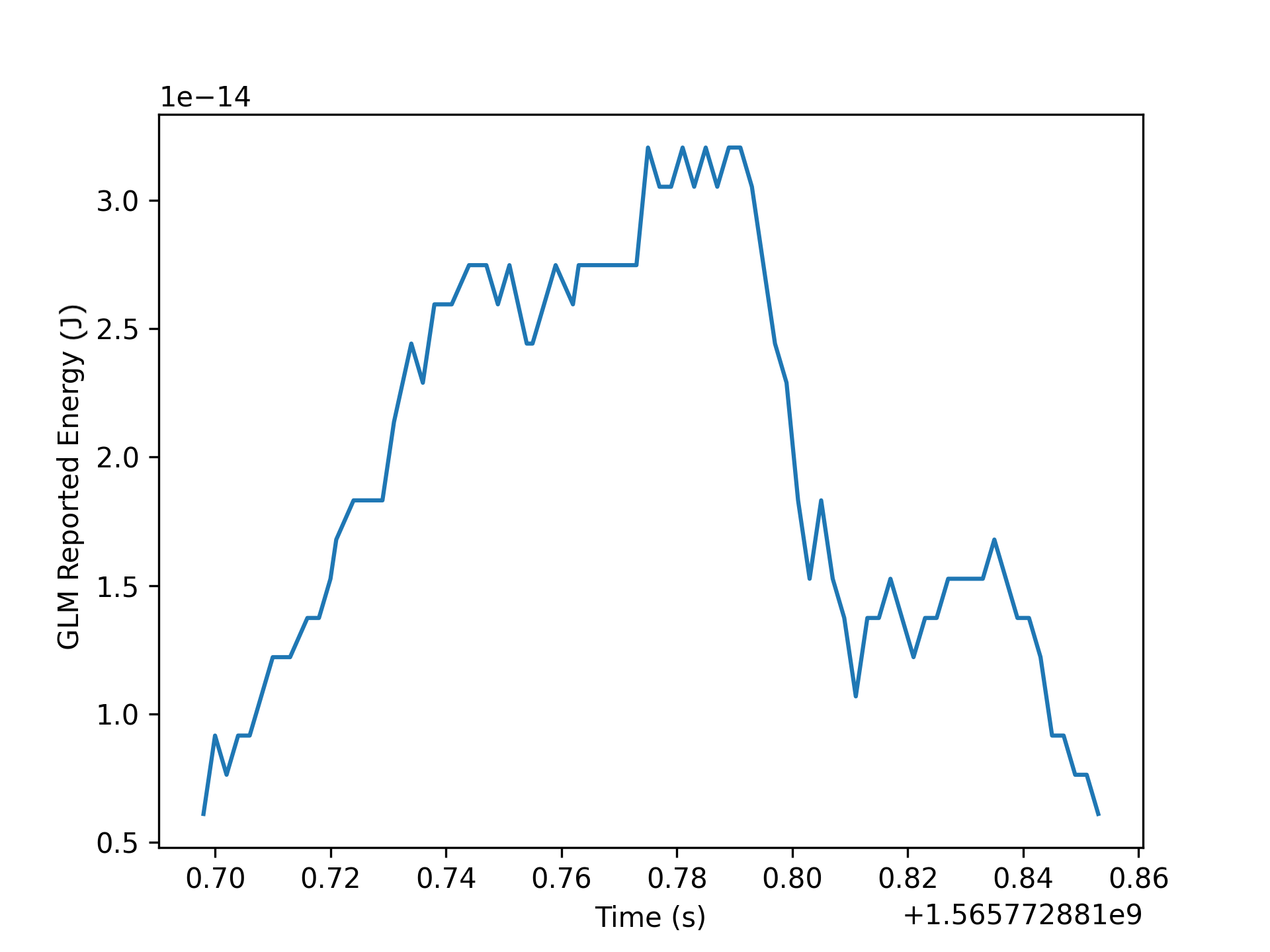}
  \caption{The light curve for the most energetic Perseid fireball as measured by GLM-16 recorded on 2019-08-42 08:54:00 with the timescale starting at T0=08:54:00 UT. See Table \ref{tab:Big_fireballs1} for more detail.}
  \label{fig:per_light}
\end{figure}

 \begin{figure}[ht!]
  \includegraphics[width=\linewidth]{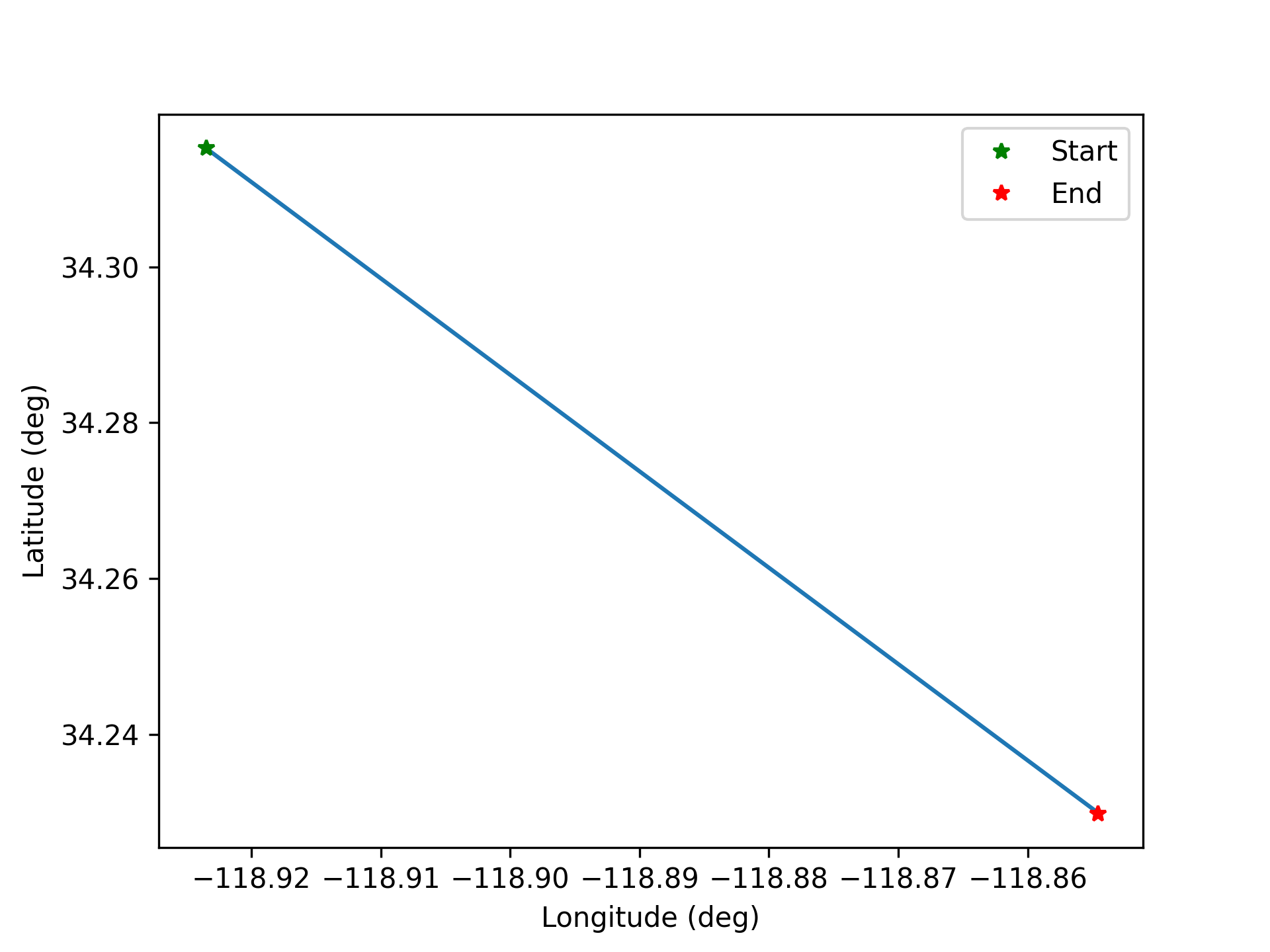}
  \caption{The ground track for the most energetic Perseid fireball event as measured by GLM-16 detected on 2019-08-42 08:54:00. See Table \ref{tab:Big_fireballs1} for more detail.}
  \label{fig:per_pos}
\end{figure}

 \begin{figure}[ht!]
  \includegraphics[width=\linewidth]{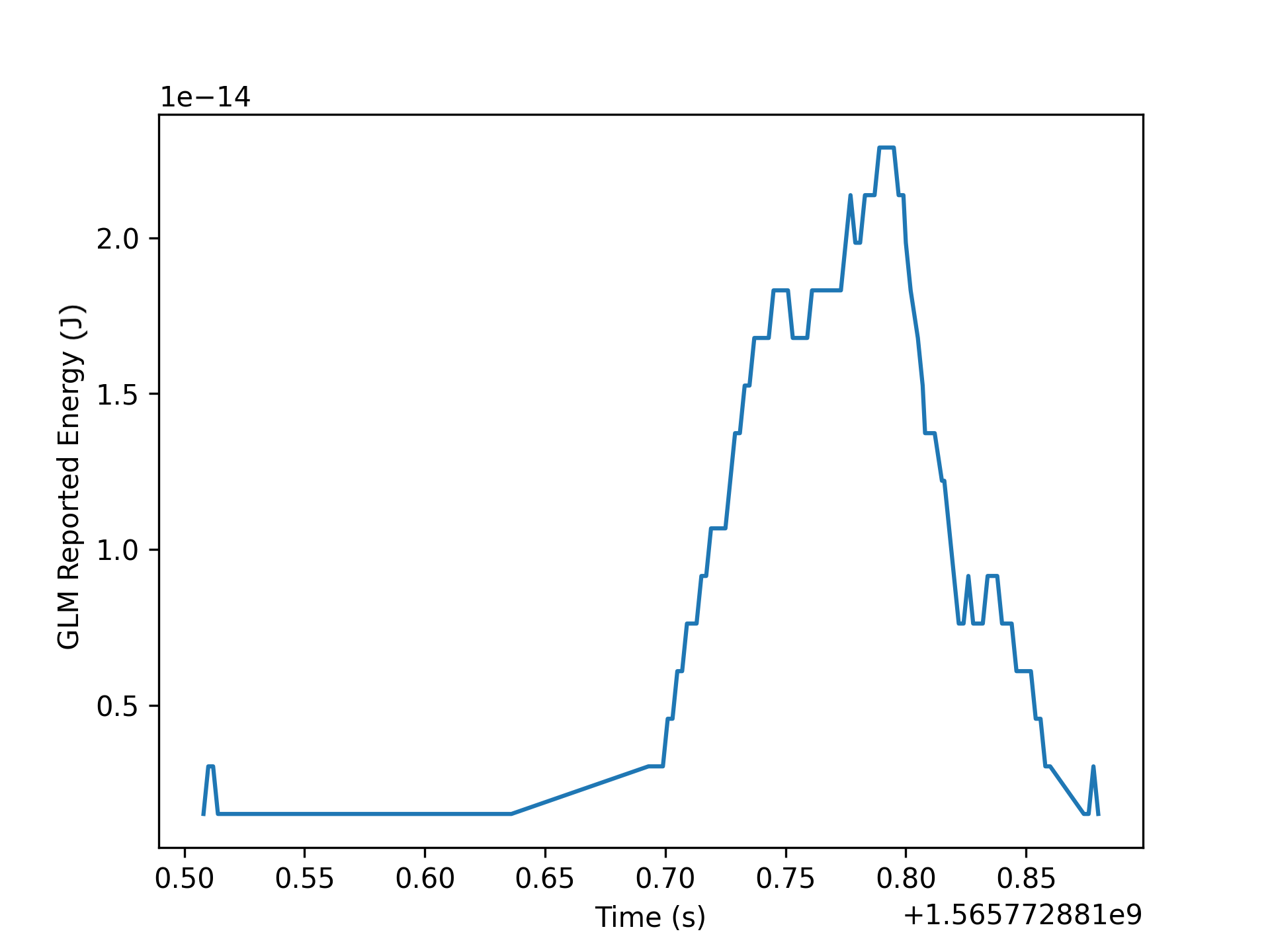}
  \caption{The light curve for the most energetic Perseid fireball as measured by GLM-17 recorded on 2019-08-42 08:54:00 with the timescale starting at T0=08:54:00 UT. See Table \ref{tab:Big_fireballs1} for more detail.}
  \label{fig:per_light2}
\end{figure}

 \begin{figure}[ht!]
  \includegraphics[width=\linewidth]{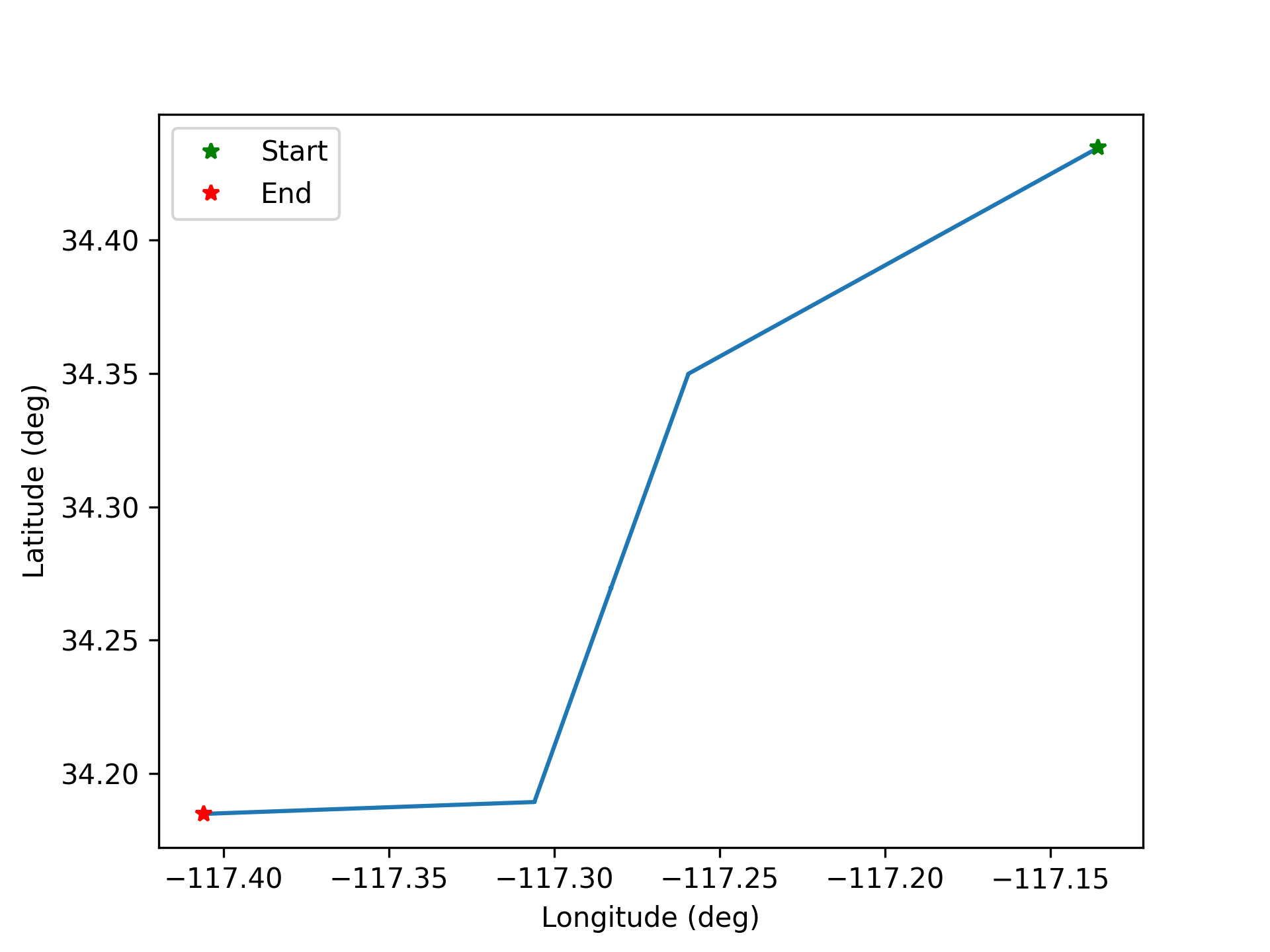}
  \caption{The ground track for the most energetic Perseid fireball event as measured by GLM-17 detected on 2019-08-42 08:54:00. See Table \ref{tab:Big_fireballs1} for more detail.}
  \label{fig:per_pos2}
\end{figure}

 \begin{figure}[ht!]
  \includegraphics[width=\linewidth]{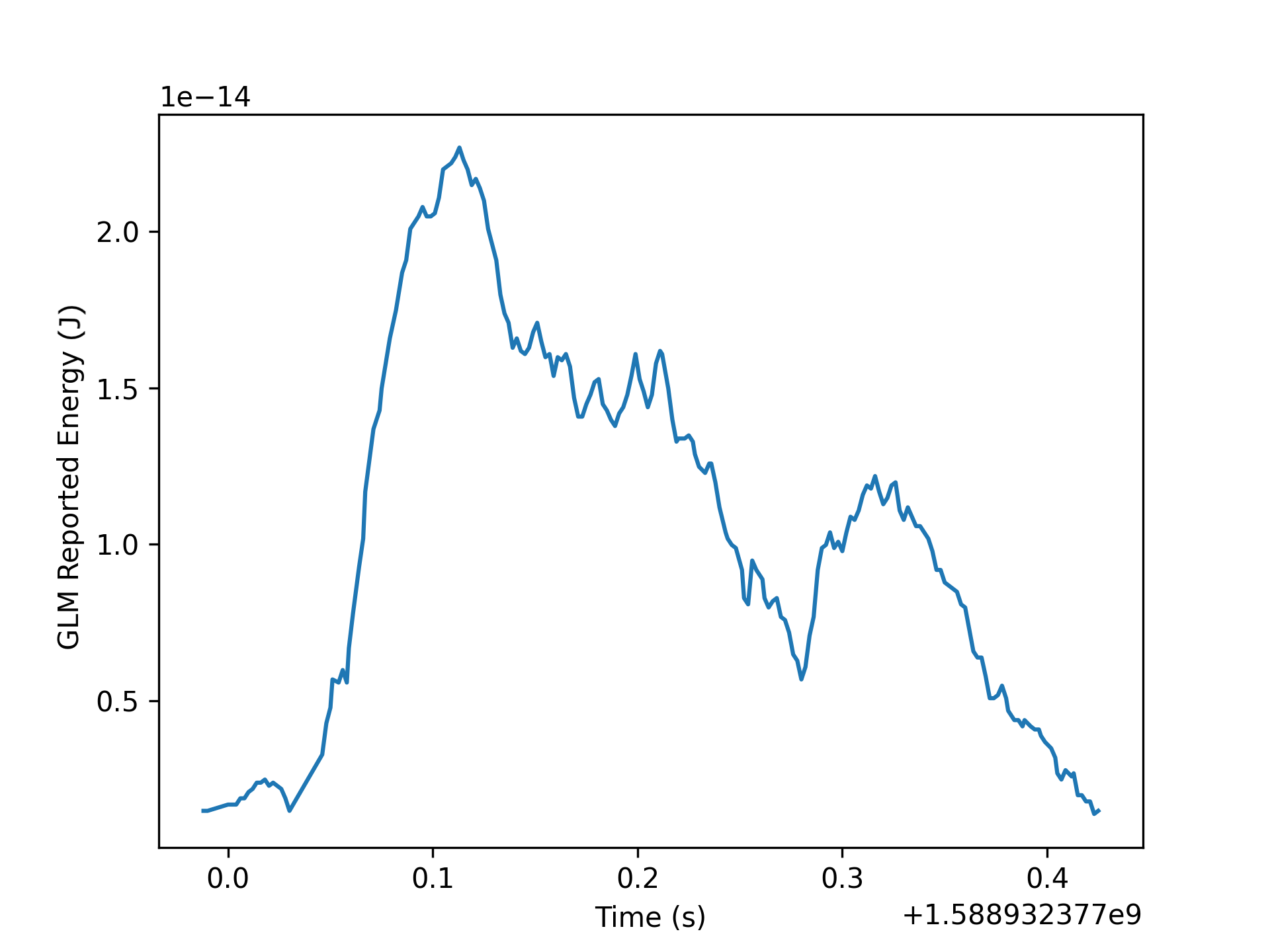}
  \caption{The light curve for the most energetic Eta Aquariid fireball as measured by GLM-16 detected on 2020-05-08 10:06:00 with the timescale starting at T0=10:06:00 UT. See Table \ref{tab:Big_fireballs1} for more detail.}
  \label{fig:eta_light}
\end{figure}

 \begin{figure}[ht!]
  \includegraphics[width=\linewidth]{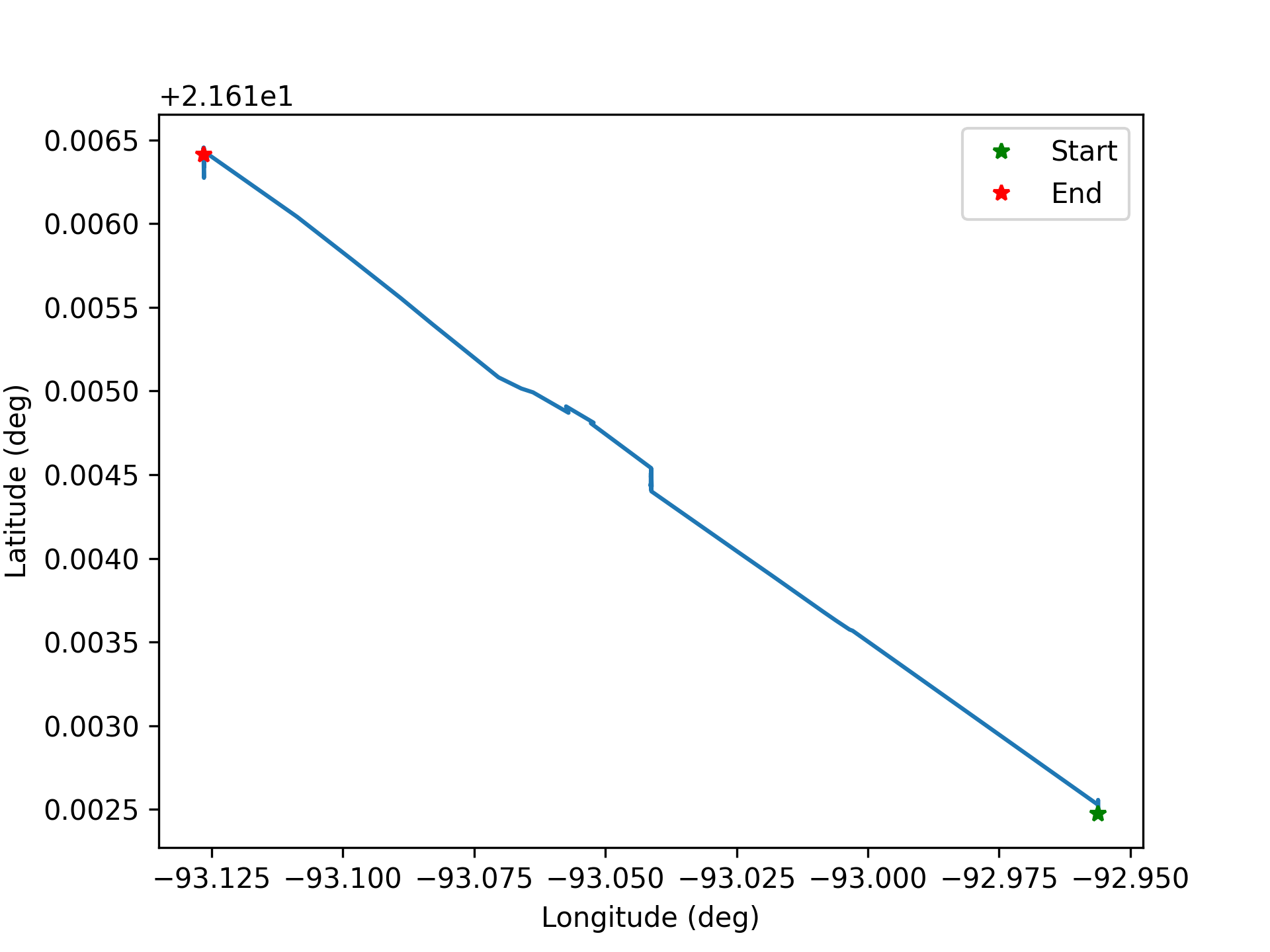}
  \caption{The ground track most energetic Eta Aquariid fireball as measured by GLM-16 detected on 2021-05-07 10:20:00. See Table \ref{tab:Big_fireballs1} for more detail.}
  \label{fig:eta_pos}
\end{figure}

 \begin{figure}[ht!]
  \includegraphics[width=\linewidth]{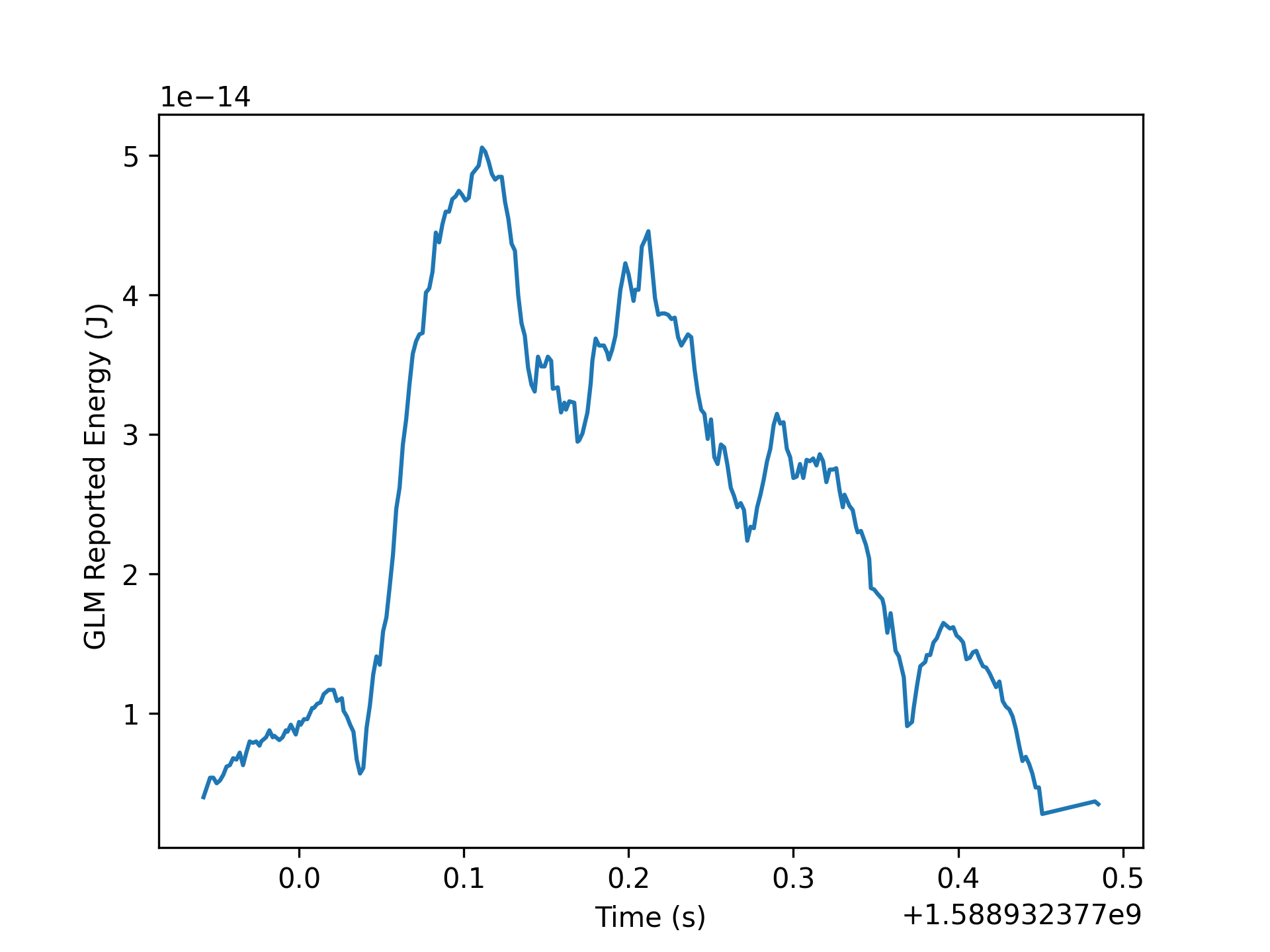}
  \caption{The light curve for the most energetic Eta Aquariid fireball as measured by GLM-17 detected on 2020-05-08 10:06:00 with the timescale starting at T0=10:06:00 UT. See Table \ref{tab:Big_fireballs1} for more detail.}
  \label{fig:eta_light2}
\end{figure}

 \begin{figure}[ht!]
  \includegraphics[width=\linewidth]{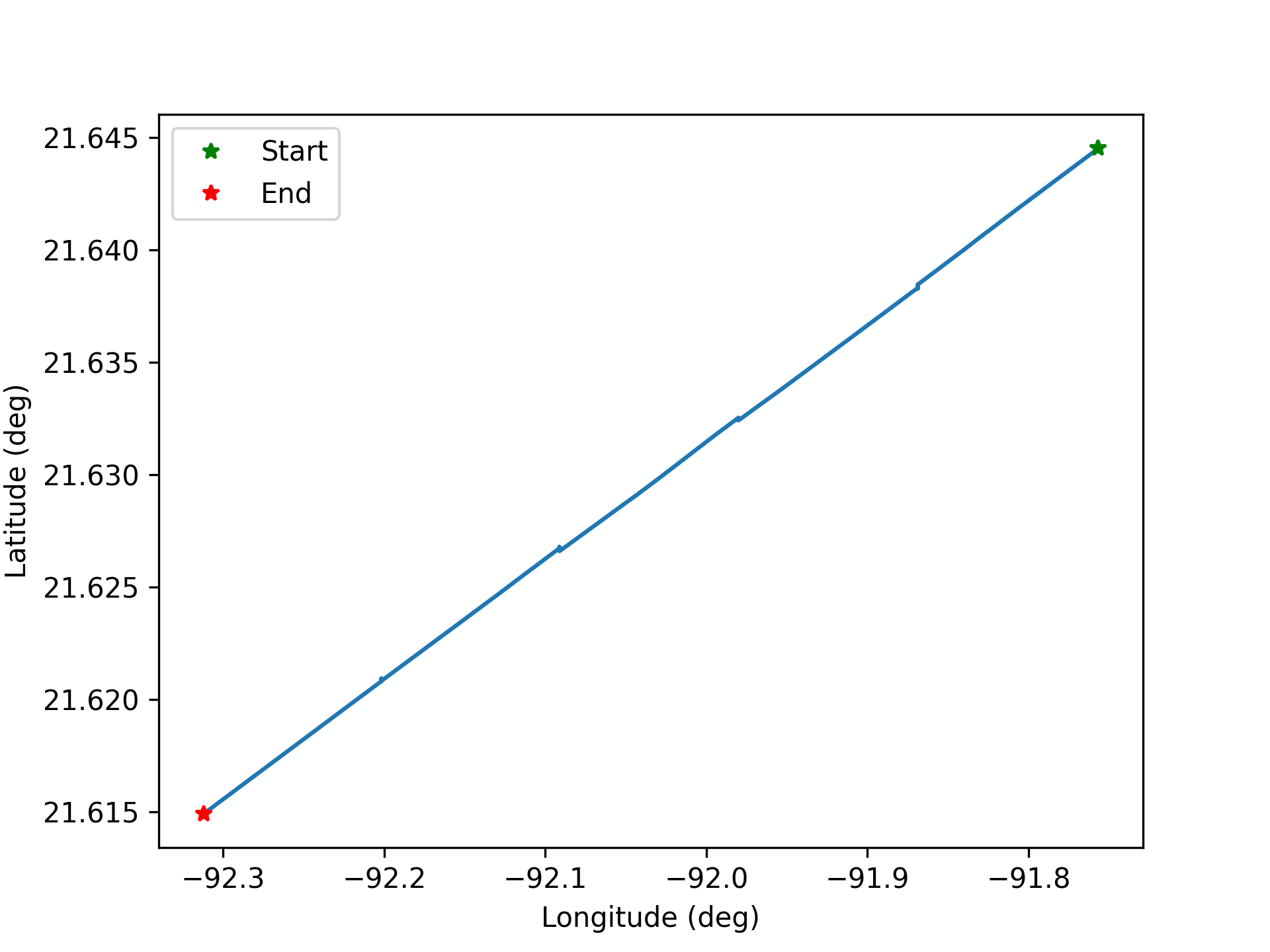}
  \caption{The ground track most energetic Eta Aquariid fireball as measured by GLM-17 detected on 2020-05-08 10:06:00. See Table \ref{tab:Big_fireballs1} for more detail.}
  \label{fig:eta_pos2}
\end{figure}

\clearpage
\newpage
\section{Cumulative Mass Distributions as Measured by GLM-17} 
\label{app:massindex}
The following figures show the cumulative mass distributions for all three of our most significant showers (LEO, PER and ETA) as measured by the GLM-17. These complement Figures \ref{fig:cm_leo} to \ref{fig:cm_eta} in the main text which shows the cumulative distributions as observed by GLM-16.

 \begin{figure}[ht!]
  \includegraphics[width=\linewidth]{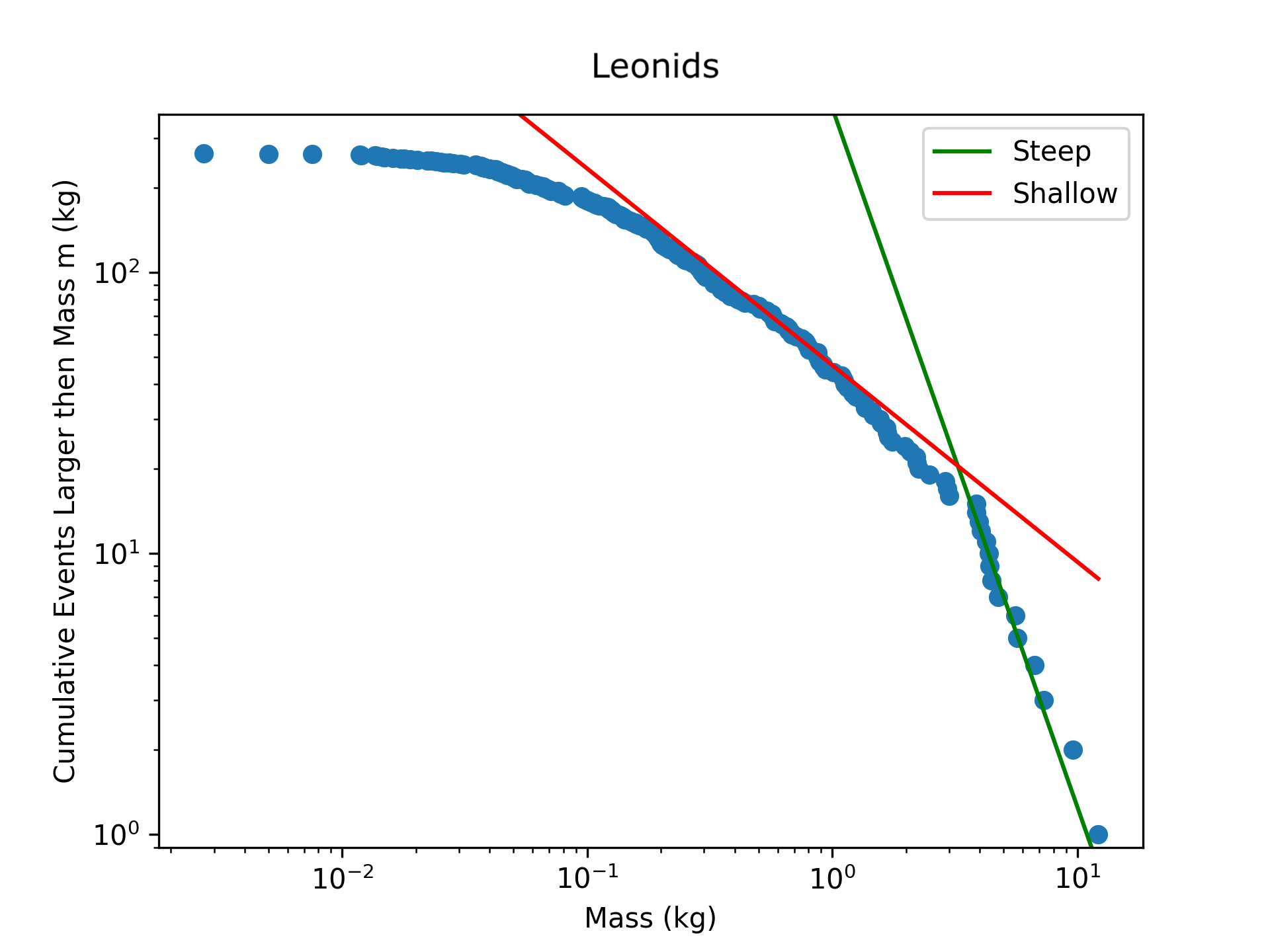}
  \caption{Cumulative Mass Distribution for Leonids for the GLM-17 Sensor for data collected between 2019-2022.}
  \label{fig:cm_leo17}
\end{figure}

 \begin{figure}[ht!]
  \includegraphics[width=\linewidth]{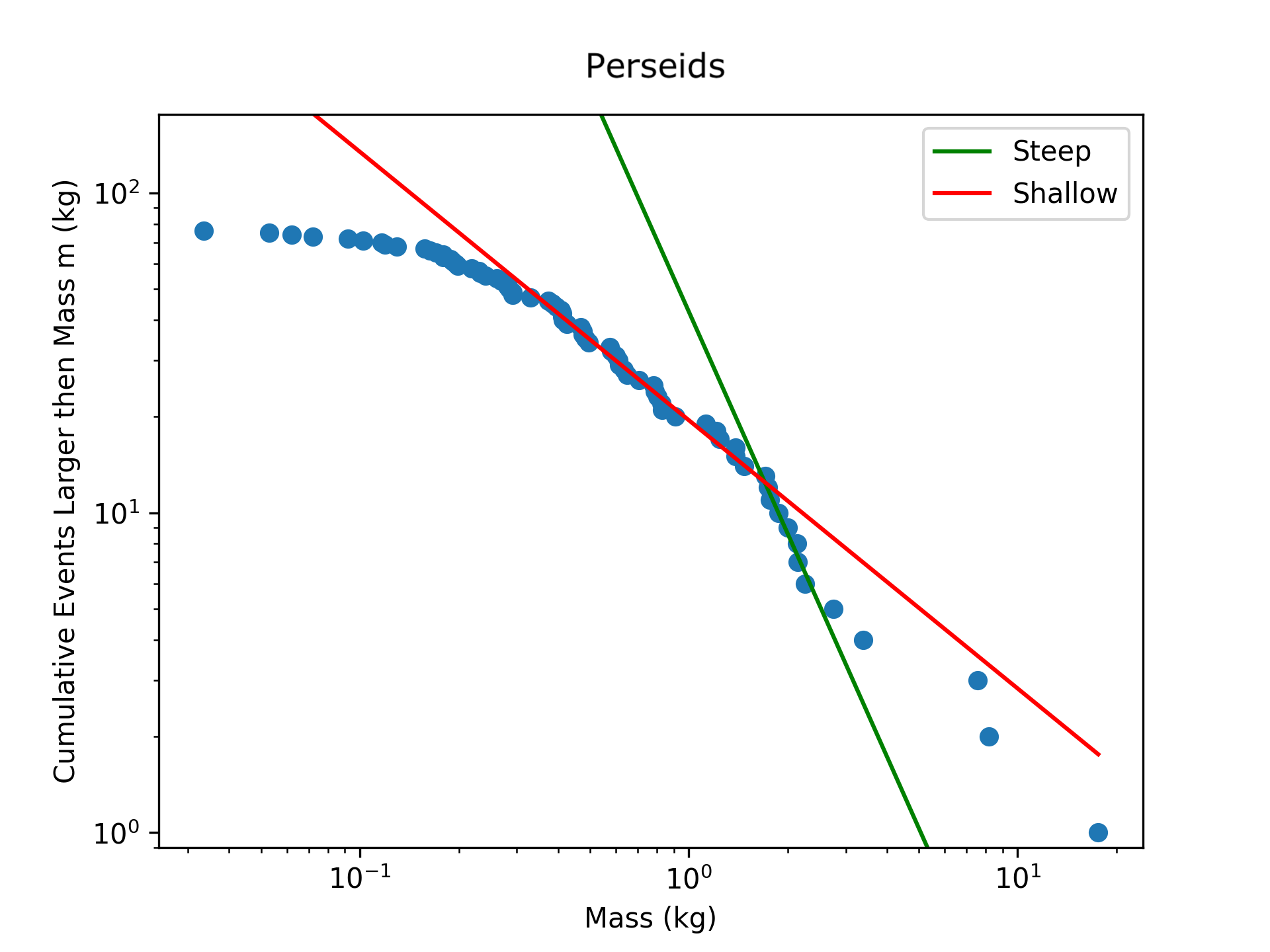}
  \caption{Cumulative Mass Distribution for Perseids for the GLM-17 Sensor for data collected between 2019-2022.}
  \label{fig:cm_per17}
\end{figure}

 \begin{figure}[ht!]
  \includegraphics[width=\linewidth]{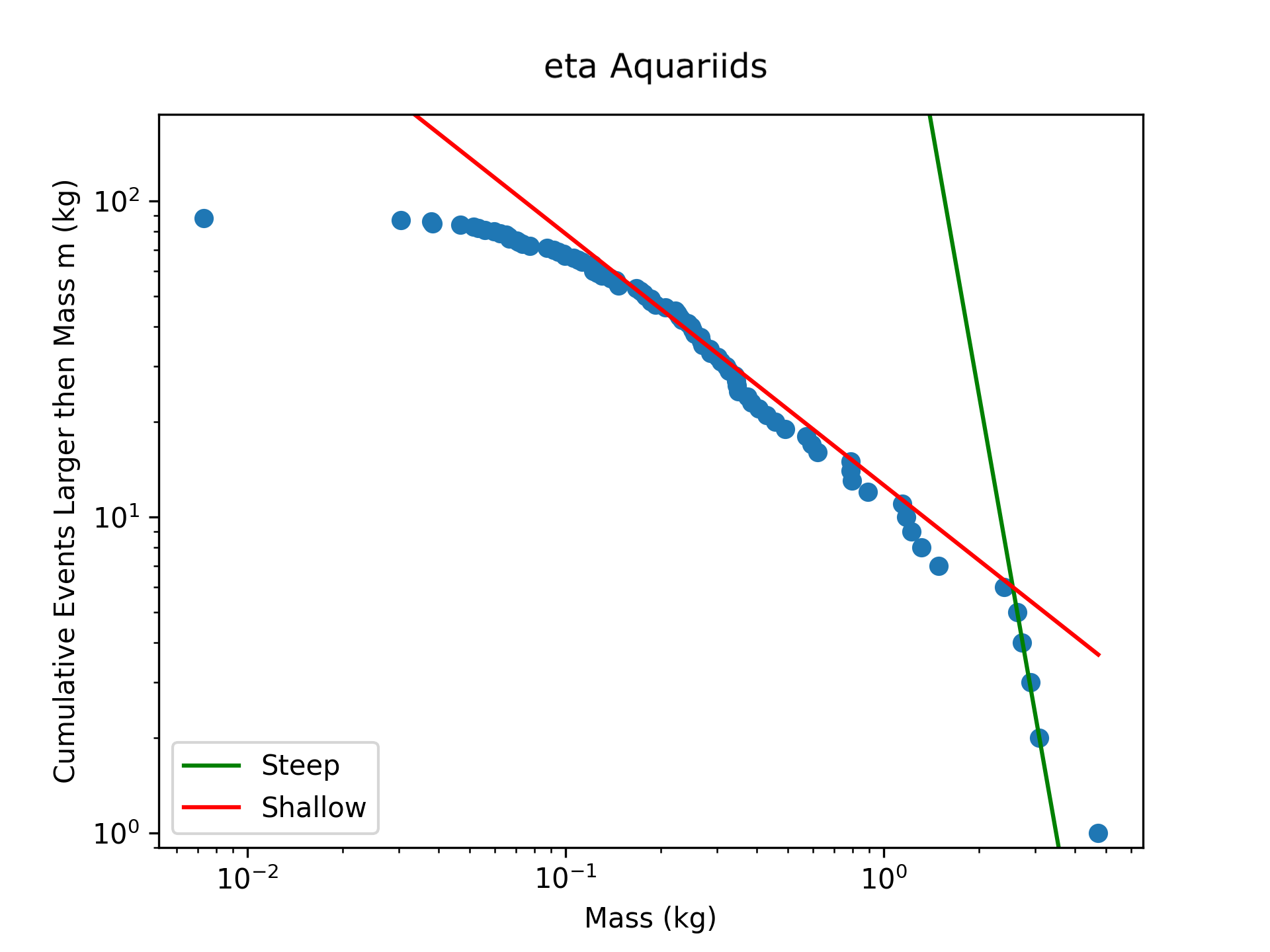}
  \caption{Cumulative Mass Distribution for eta Aquariids for the GLM-17 Sensor for data collected between 2019-2022.}
  \label{fig:cm_eta17}
\end{figure}

\clearpage
\newpage
\section{Details for all Fireballs Detected Concurrently by GLM and USG} 
\label{app:tab_usg}
In the following section, the table lists detection and metric information for all fireballs captured with both the USG sensor and the GLM sensor that were used for validation of the GLM energy approach as summarized in Figures \ref{fig:usg_glm_rep} to \ref{fig:usg_glm_lookangle_cont}. The figures  are additional comparisons between USG and GLM light curves from \ref{sec:glm_vs_usg}.

\newpage
\begin{sidewaystable}
    \centering
    \small
    \begin{tabular}{|l|l|l|l|l|l|l|l|l|l|l|l|}
    \hline
        Date & Sat & Vel (km/s) & Lat (°) & Long (°) & LA GLM (°) & USG E (J) & USG $\tau$  & $E_{GLM}$ (J)& LUT Value & C & C GLM Total E (J)  \\ \hline
        2017-07-23 6:12:36 & GLM-16 & 17.2 & -6.6 & -69.7 & 8.57 & 1.89E+10 & 0.069693 & 2.29E+10 & 4.17266E+13 & 0.7402 & 1.40E+10  \\ \hline
        2017-09-05 5:11:25 & GLM-16 & 14.7 & 49.3 & -166.9 & 7.80 & 2.05E+10 & 0.069847 & 1.29E+11 & 2.94053E+12 & 0.2485 & 5.10E+09  \\ \hline
        2018-10-05 0:27:01 & GLM-16 & 14.7 & -40 & -32.1 & 7.92 & 2.05E+10 & 0.069847 & 2.64E+11 & 8.29908E+12 & 0.2485 & 5.10E+09  \\ \hline
        2019-02-01 18:17:09 & GLM-16 & 16.3 & 22.6 & -83.5 & 4.08 & 3.13E+11 & 0.098822 & 2.90E+11 & 8.5376E+12 & 0.9313 & 2.91E+11  \\ \hline
        2019-04-14 17:54:32 & GLM-16 & 15.9 & 18.5 & -74.3 & 5.87 & 1.57E+10 & 0.069295 & 2.67E+10 & 4.11682E+13 & 0.8493 & 1.33E+10  \\ \hline
        2019-06-22 21:25:45 & GLM-16 & 14.9 & 14.9 & -65.8 & 3.06 & 1.59E+12 & 0.117364 & 9.00E+11 & 4.18371E+13 & 1.1147 & 1.78E+12  \\ \hline
        2019-06-30 16:52:47 & GLM-16 & 42.3 & 21.5 & -130 & 5.96 & 1.73E+10 & 0.069512 & 1.62E+10 & 7.14328E+12 & 1.134 & 1.96E+10  \\ \hline
        2019-09-12 2:34:57 & GLM-16 & 18.5 & 25.2 & -47.6 & 5.88 & 9.14E+10 & 0.084130 & 5.16E+10 & 3.09983E+13 & 0.8493 & 7.76E+10  \\ \hline
        2019-09-14 12:39:32 & GLM-16 & 15.9 & -38.9 & -32.9 & 7.71 & 5.89E+10 & 0.078925 & 1.36E+11 & 9.34746E+12 & 0.2485 & 1.46E+10  \\ \hline
        2019-09-28 10:40:18 & GLM-16 & 20.4 & -12.6 & -107.5 & 7.07 & 1.24E+10 & 0.067022 & 2.88E+09 & 3.34225E+13 & 1.282 & 1.59E+10  \\ \hline
        2019-11-28 20:30:52 & GLM-16 & 13 & 35.9 & -31.1 & 2.01 & 1.46E+10 & 0.067912 & 7.82E+10 & 1.01411E+13 & 1.1297 & 1.65E+10  \\ \hline
        2020-01-17 21:29:48 & GLM-16 & 15.5 & 19.4 & -65.9 & 3.69 & 5.24E+10 & 0.079924 & 6.85E+10 & 4.06174E+13 & 1.0892 & 5.71E+10  \\ \hline
        2020-01-24 11:13:30 & GLM-16 & 21.2 & 28.1 & -35.4 & 7.68 & 1.41E+10 & 0.067529 & 2.82E+10 & 1.85587E+13 & 0.2485 & 3.49E+09  \\ \hline
        2020-08-02 16:36:24 & GLM-16 & 11.1 & -35.3 & -33.8 & 7.51 & 4.00E+10 & 0.076879 & 2.07E+11 & 1.23375E+13 & 0.2485 & 9.94E+09  \\ \hline
        2020-10-22 17:39:32 & GLM-16 & 17.6 & 22.1 & -134.1 & 8.14 & 2.38E+10 & 0.070092 & 6.70E+10 & 4.43891E+12 & 0.987 & 2.35E+10  \\ \hline
        2020-12-28 17:27:50 & GLM-16 & 15.2 & 37.2 & -55.1 & 6.46 & 3.08E+10 & 0.075667 & 6.39E+10 & 2.54725E+13 & 0.7381 & 2.27E+10  \\ \hline
        2021-01-29 16:12:46 & GLM-16 & 14.3 & 38.9 & -50.9 & 6.81 & 4.16E+10 & 0.076663 & 1.67E+11 & 2.14027E+13 & 0.5957 & 2.48E+10  \\ \hline
        2021-04-13 2:16:47 & GLM-16 & 14.1 & 26.9 & -79.1 & 5.57 & 1.14E+10 & 0.066025 & 1.62E+10 & 3.73836E+13 & 0.8493 & 9.64E+09  \\ \hline
        2021-09-06 17:55:41 & GLM-16 & 13.6 & -2.2 & -111.9 & 7.90 & 1.68E+10 & 0.067340 & 2.51E+10 & 3.12627E+13 & 0.9334 & 1.56E+10 \\ \hline
        2022-01-30 2:06:18 & GLM-16 & 20 & 50.4 & -37.4 & 8.05 & 2.92E+10 & 0.071685 & 1.45E+11 & 5.50724E+12 & 0.1274 & 3.72E+09  \\ \hline
        2022-04-04 0:30:38 & GLM-16 & 19.7 & -3.1 & -64.3 & 3.21 & 1.46E+10 & 0.067912 & 7.38E+09 & 4.22727E+13 & 1.1147 & 1.63E+10  \\ \hline
        2022-07-20 10:56:51 & GLM-16 & 16 & -43 & -59.3 & 6.86 & 4.65E+10 & 0.076110 & 2.85E+10 & 2.02965E+13 & 0.5957 & 2.77E+10  \\ \hline
        2022-07-22 0:16:18 & GLM-16 & 17.4 & -23.2 & -20.1 & 7.95 & 3.24E+10 & 0.075457 & 8.02E+10 & 6.51148E+12 & 0.2485 & 8.06E+09  \\ \hline
        2022-07-28 1:36:07 & GLM-16 & 29.9 & -5.9 & -87 & 2.30 & 1.36E+11 & 0.088200 & 4.81E+10 & 1.58335E+13 & 1.0892 & 1.48E+11  \\ \hline
        2022-11-20 13:53:52 & GLM-16 & 21.7 & 14.9 & -110 & 1.53 & 1.84E+10 & 0.067702 & 2.95E+10 & 3.9298E+13 & 0.9313 & 1.71E+10  \\ \hline
        2018-11-17 21:48:23 & GLM-17 & 19.1 & 47.3 & -172.9 & 8.61 & 1.19E+11 & 0.086178 & 5.70E+11 & 8.66336E+12 & 1.015 & 1.21E+11  \\ \hline
        2019-02-01 18:17:09 & GLM-17 & 16.3 & 22.6 & -83.5 & 5.33 & 3.13E+11 & 0.098822 & 2.55E+12 & 3.95466E+13 & 1.0589 & 3.31E+11  \\ \hline
        2019-06-30 16:52:47 & GLM-17 & 42.3 & 21.5 & -130 & 2.63 & 1.73E+10 & 0.069512 & 7.31E+09 & 4.31887E+13 & 0.8493 & 1.47E+10  \\ \hline
        2019-07-23 20:42:56 & GLM-17 & 16.1 & 44.9 & -147.6 & 7.87 & 1.38E+11 & 0.088307 & 2.04E+11 & 2.14594E+13 & 0.978 & 1.35E+11  \\ \hline
        2019-09-28 10:40:18 & GLM-17 & 20.4 & -12.6 & -107.5 & 0.85 & 1.24E+10 & 0.067022 & 1.72E+10 & 3.80755E+13 & 0.04228 & 5.26E+08  \\ \hline
        2019-11-05 11:24:49 & GLM-17 & 27.4 & -10.7 & -143.3 & 5.91 & 5.95E+10 & 0.079650 & 5.65E+10 & 4.54308E+13 & 0.987 & 5.87E+10  \\ \hline
        2020-09-18 8:05:25 & GLM-17 & 11.7 & 2.5 & -169.9 & 8.19 & 2.22E+10 & 0.069978 & 4.78E+10 & 3.78744E+13 & 0.88934 & 1.97E+10  \\ \hline
        2020-10-18 10:52:42 & GLM-17 & 16.2 & -11.5 & -135.9 & 5.11 & 1.95E+10 & 0.071685 & 2.64E+10 & 4.542E+13 & 0.9334 & 1.82E+10  \\ \hline
        2020-10-22 17:39:32 & GLM-17 & 17.6 & 22.1 & -134.1 & 5.51 & 2.38E+10 & 0.070092 & 1.09E+10 & 4.33315E+13 & 0.1274 & 3.03E+09  \\ \hline
        2021-07-05 3:46:22 & GLM-17 & 15.7 & 44.6 & -164.5 & 8.41 & 4.00E+11 & 0.098234 & 1.37E+12 & 1.5214E+13 & 0.88934 & 3.56E+11  \\ \hline
        2021-09-06 17:55:41 & GLM-17 & 13.6 & -2.2 & -111.9 & 5.06 & 1.68E+10 & 0.067340 & 2.80E+10 & 4.30651E+13 & 0.2485 & 4.16E+09  \\ \hline
        2021-09-29 10:50:58 & GLM-17 & 21.2 & 54.6 & -148.3 & 8.27 & 7.41E+10 & 0.081840 & 1.53E+11 & 9.76666E+12 & 0.88934 & 6.59E+10  \\ \hline
        2022-07-28 1:36:07 & GLM-17 & 29.9 & -5.9 & -87 & 3.68 & 1.36E+11 & 0.088200 & 1.30E+11 & 4.22681E+13 & 1.186 & 1.61E+11  \\ \hline
        2022-11-20 13:53:52 & GLM-17 & 21.7 & 14.9 & -110 & 5.02 & 1.84E+10 & 0.067702 & 1.71E+10 & 3.06051E+13 & 1.2368 & 2.27E+10 \\ \hline
    \end{tabular}
    \caption{Details of each simultaneously detected USG and GLM fireball. Velocity (Vel), location information and total optical energy from USG are from the CNEOS webpage. LA is the GLM Look Angle. C is the Correction Factor Normalization using the continuum calibration look-up table (LUT) provided by Lockheed Martin (Tillier, private communication). Here the USG $\tau$ is as given in \cite{Brown2002} and the GLM Total E0 is the total energy found from summation of the raw GLM L2 intensity and applying the correction to estimate total intensity as described in section \ref{sec:cali}.C is the correction applied to the raw L2 GLM data and C GLM Total E (J) is the corrected total energy from GLM using this correction value.}
\end{sidewaystable}

\newpage

\begin{figure}
  \includegraphics[width=\linewidth]{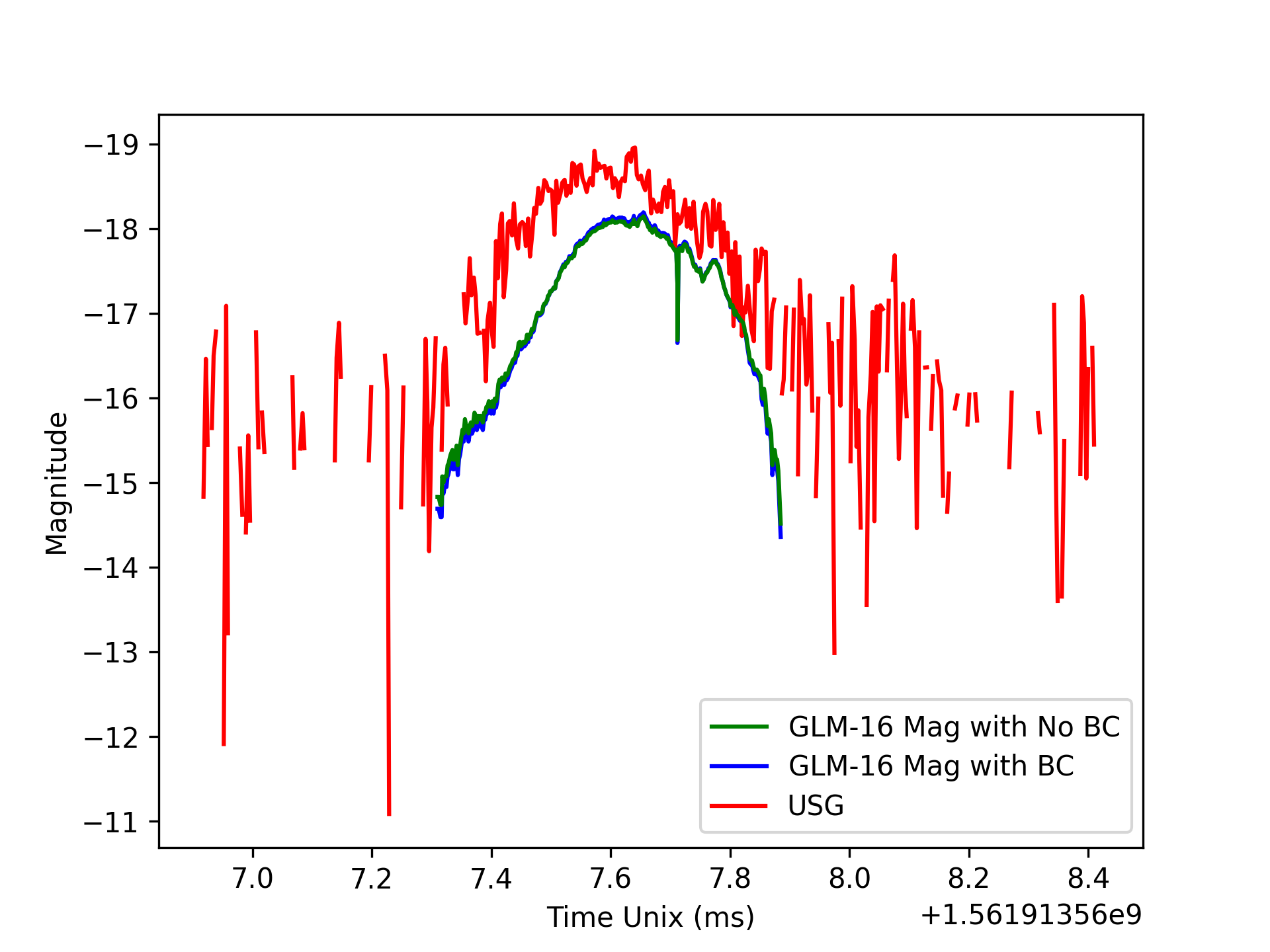}
  \caption{A comparison between the USG and GLM-16 magnitude curve for GLM-16 intensity, brightness correction magnitude for GLM-16 and the USG magnitude. Note that the uncorrected and brightness corrected light curve in this instance are basically the same. This event occurred on 2019-06-30 16:52:47 UTC at 2.5S 168.7E}
  \label{fig:usg_glm_magapp1}
\end{figure}

\begin{figure}
  \includegraphics[width=\linewidth]{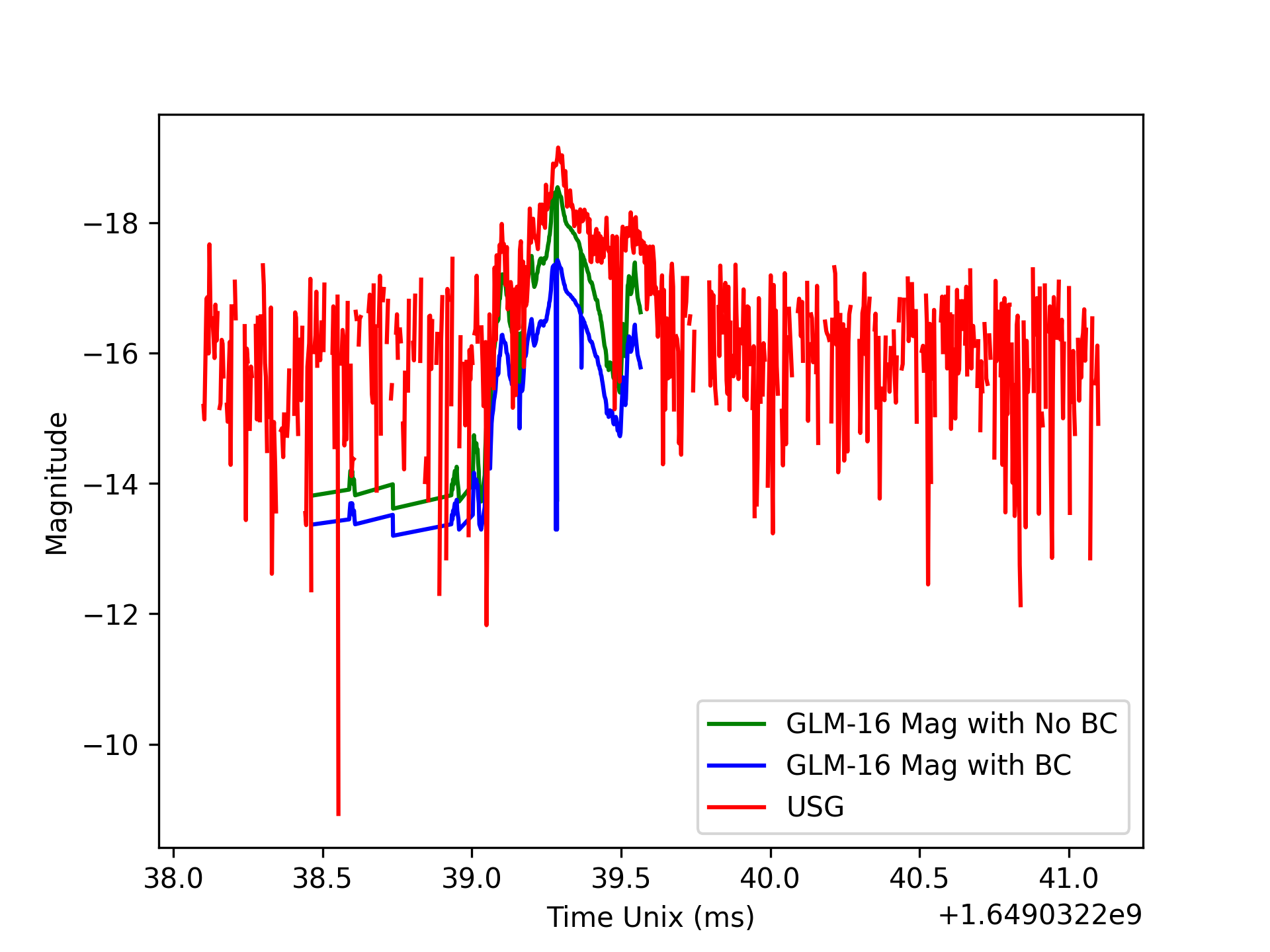}
  \caption{A comparison between the USG and GLM-16 magnitude curve for the visual magnitude for GLM-16, brightness correction magnitude for GLM-16 and the USG magnitude.  This event occurred on 2022-04-04 00:30:38 UTC at 3.2S 64.3W.}
  \label{fig:usg_glm_magapp2}
\end{figure}

\begin{figure}
  \includegraphics[width=\linewidth]{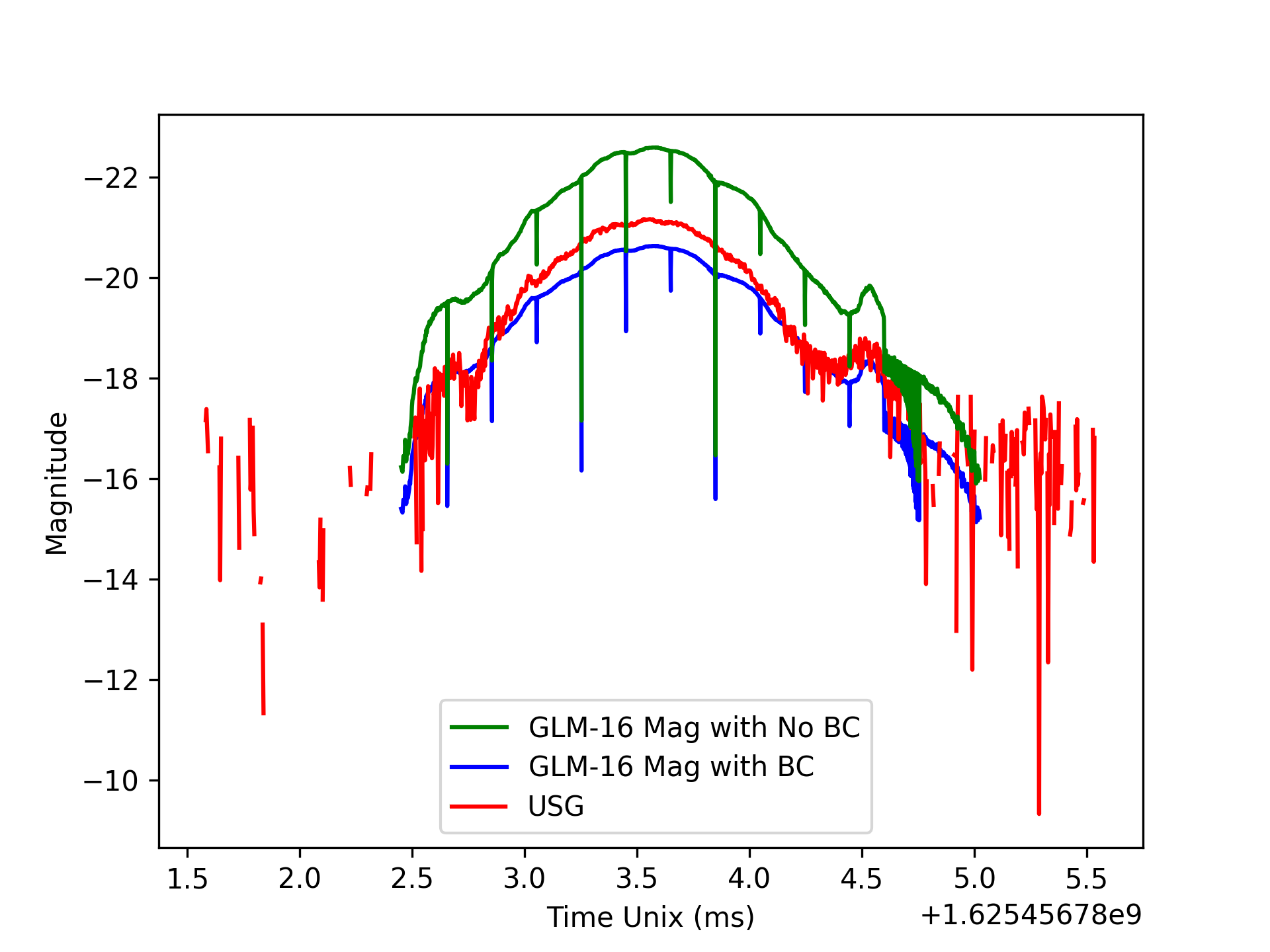}
  \caption{A comparison between the USG and GLM-17 magnitude curve for the visual magnitude for GLM-17, brightness correction magnitude for GLM-17 and the USG magnitude this observation occurred on 2021-07-05 03:45:22 at 44.3N 164.2W.}
  \label{fig:usg_glm_magapp3}
\end{figure}
\clearpage
\newpage
\section{Analysis of Taurids and Orionids with GLM} 
\label{app:tab_tau_ori}

It should be noted that over the years of the fireball detection, the  pipeline has been improved that there has been an increase in detection rate as time goes on. As our data are an average of 2019-2022, some showers may be artificially suppressed due to lower efficiency in early years of GLM operation.

As discussed in the main text, from among the showers we chose to examine (summarized in Table \ref{tab:whipple}) only three showed a strong signal above the normal GLM background rate (LEO, PER and ETA). These formed the core of our analysis described in this paper. However, two others had marginal signals (the Taurids and Orionids). All other showers showed no significant signal. 

All further filtering followed the same process as used for the three main showers. 

For the Orionids we examine the cumulative mass distribution from GLM-16 and 17 separately. We can see the steep fit is not far off the slope of the shallow fit due to the lack of larger events. The biggest individual GLM fireball event associated with the Orionids summarized in Table E.11 was approximately 2.44 kg. It had good velocity and bearing angle agreement and the look angle suggests the energy is accurate. For the Poisson masses for the Orionids we get 17.7 kg and 4.1 kg for GLM-16 and GLM-17 respectively. We note that these values are uncertain due to the small number of potential Orionids (only 30 total for each sensor) making up the steep section of the power-law (less than 10). 
Comparing these GLM estimates to the largest literature-reported Orionids (Figure \ref{fig:lit_mass_with_GLM}), these are more than an order of magnitude larger. However, they are comparable to the largest ETA and are roughly consistent with the Whipple gas-drag limit for 1P/Halley of 10 kg.

 \begin{figure}[ht!]
  \includegraphics[width=\linewidth]{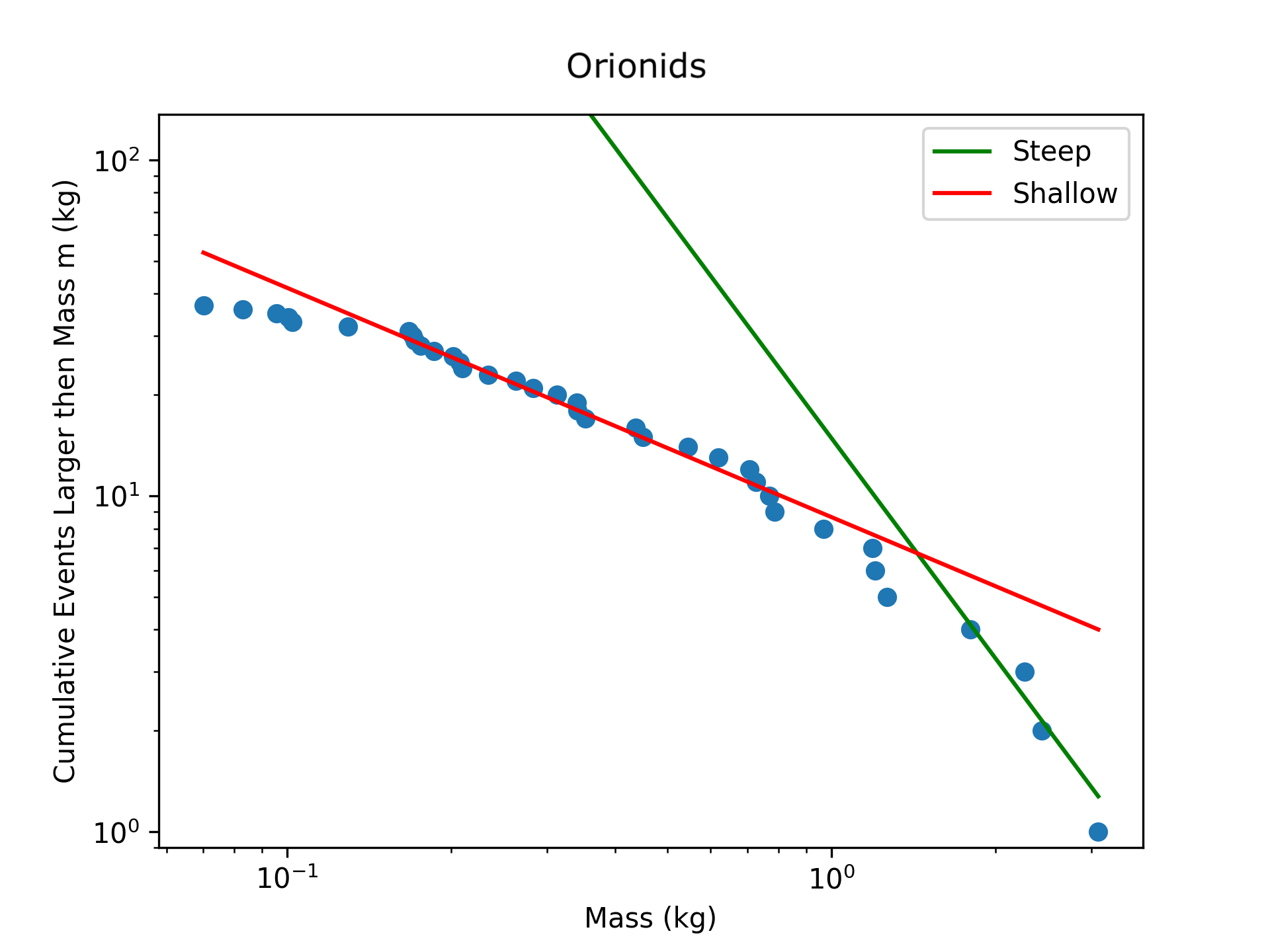}
  \caption{Cumulative Mass Distribution for Orionids  for the GLM-16 Sensor.}
  \label{fig:cm_ori}
\end{figure}

 \begin{figure}[ht!]
  \includegraphics[width=\linewidth]{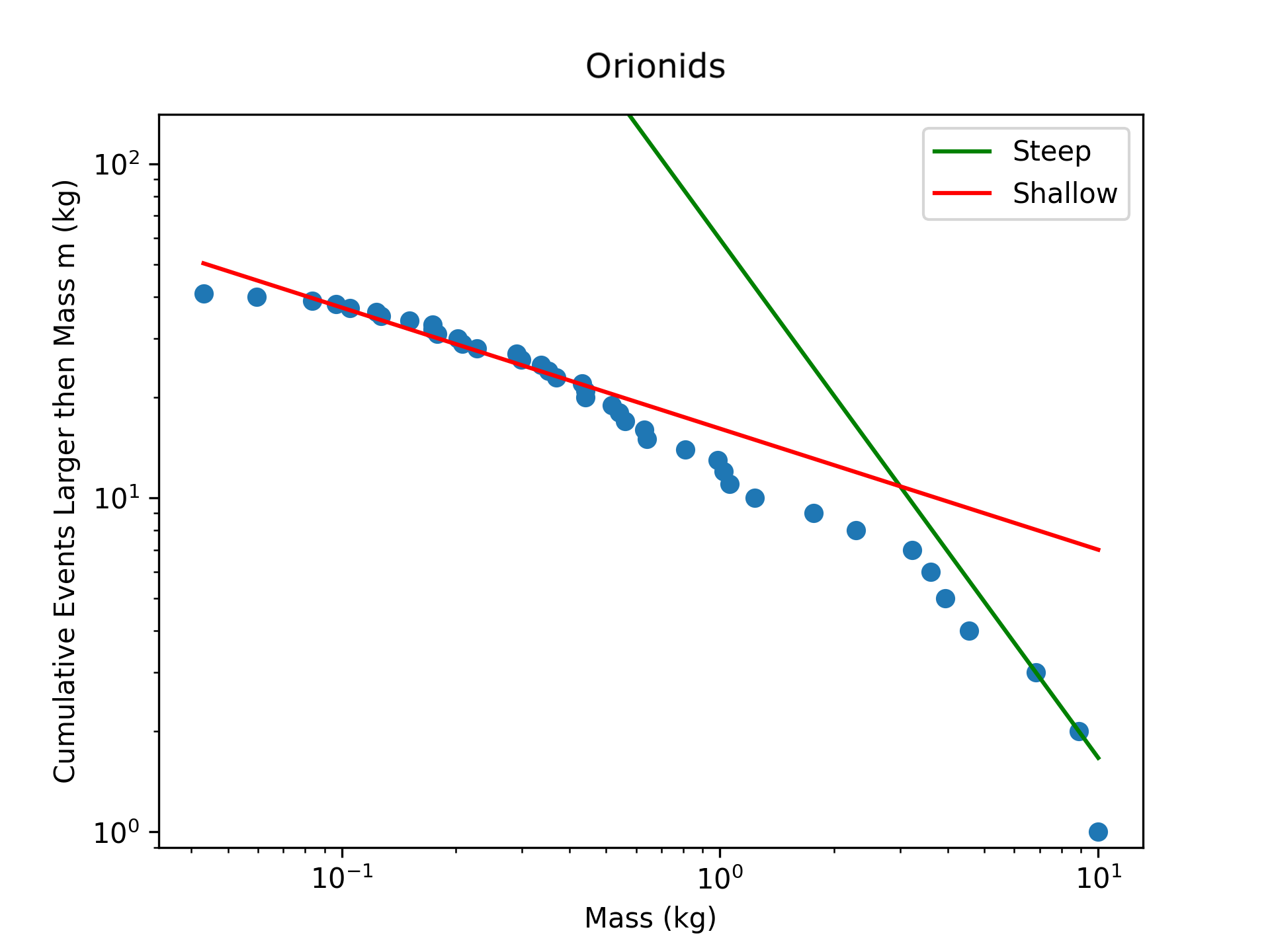}
  \caption{Cumulative Mass Distribution for Orionids for the GLM-17 Sensor.}
  \label{fig:cm_ori}
\end{figure}

Finally, we examine the signature of the Taurid shower in GLM data.  The overlap between the Northern and Southern Taurids made it difficult to isolate each separately, and thus they were combined together using the assumed velocity of 27 km/s. The Taurids have one of the longest activity periods spanning from September until December. However, they did not show a significant signal above the background except for a few days around the peak. The biggest event reported with confidence described in Table E.11 has a mass of 146.6 kg. This event has a lower GLM velocity speed then most showers observed, and although there is some difference between the observed and expected transverse velocity, it is within the range adopted in our study, which recognizes the severe limitations in speed measurements by GLM given its very coarse pixel scale.  This is the only case from among our showers where there are ground-based optical observations having a larger maximum meteoroid mass for a shower than detected by GLM.  We also have the two cumulative distributions for the Taurids, but with the few events possibly linked to the shower, the statistics are poor. For the Poisson Mass we find 304.3 kg and 467 kg for GLM-16 and GLM-17 respectively.

 \begin{figure}[ht!]
  \includegraphics[width=\linewidth]{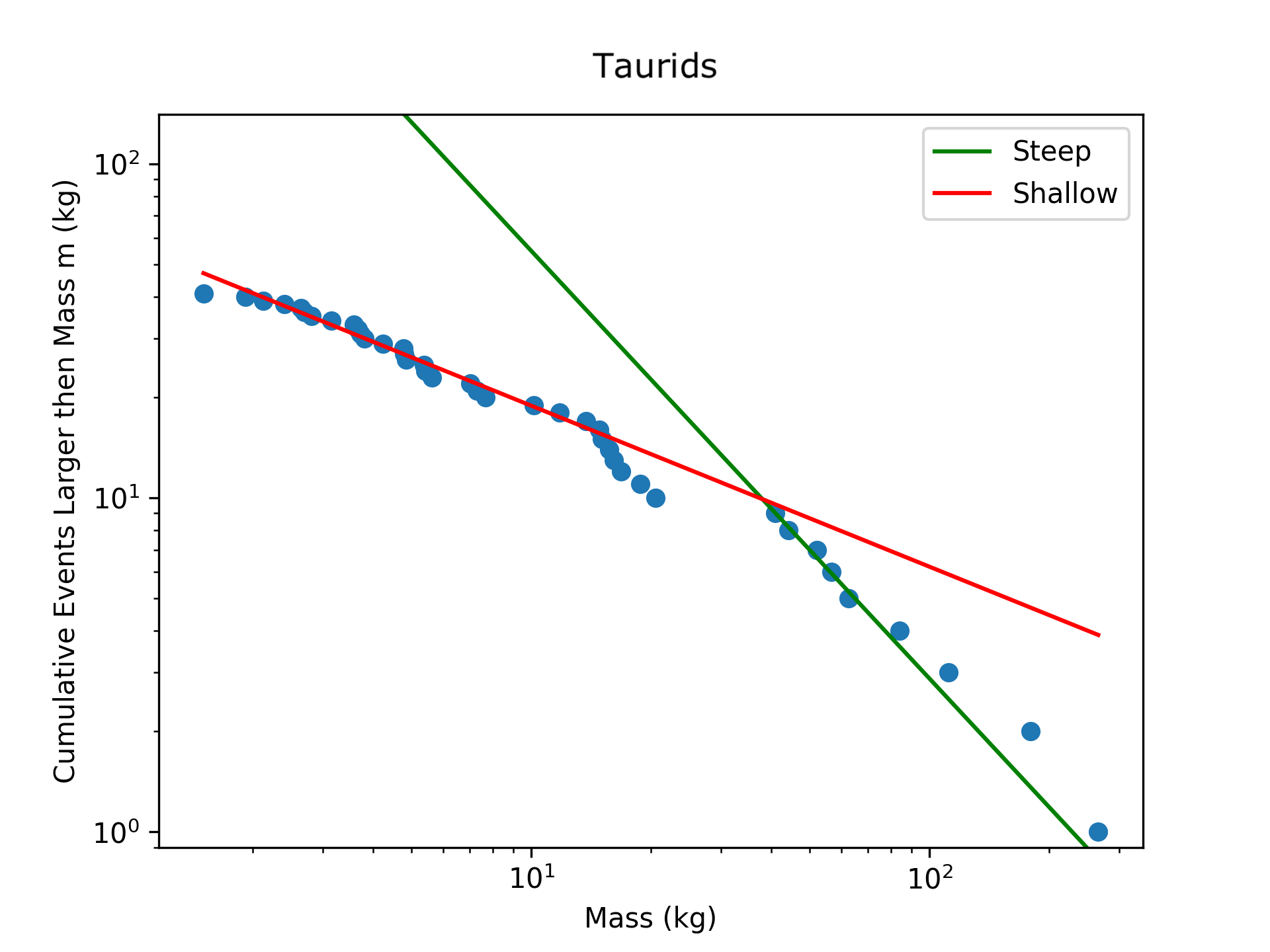}
  \caption{Cumulative Mass Distribution for the Taurids for the GLM-16 Sensor.}
  \label{fig:cm_tau}
\end{figure}

 \begin{figure}[ht!]
  \includegraphics[width=\linewidth]{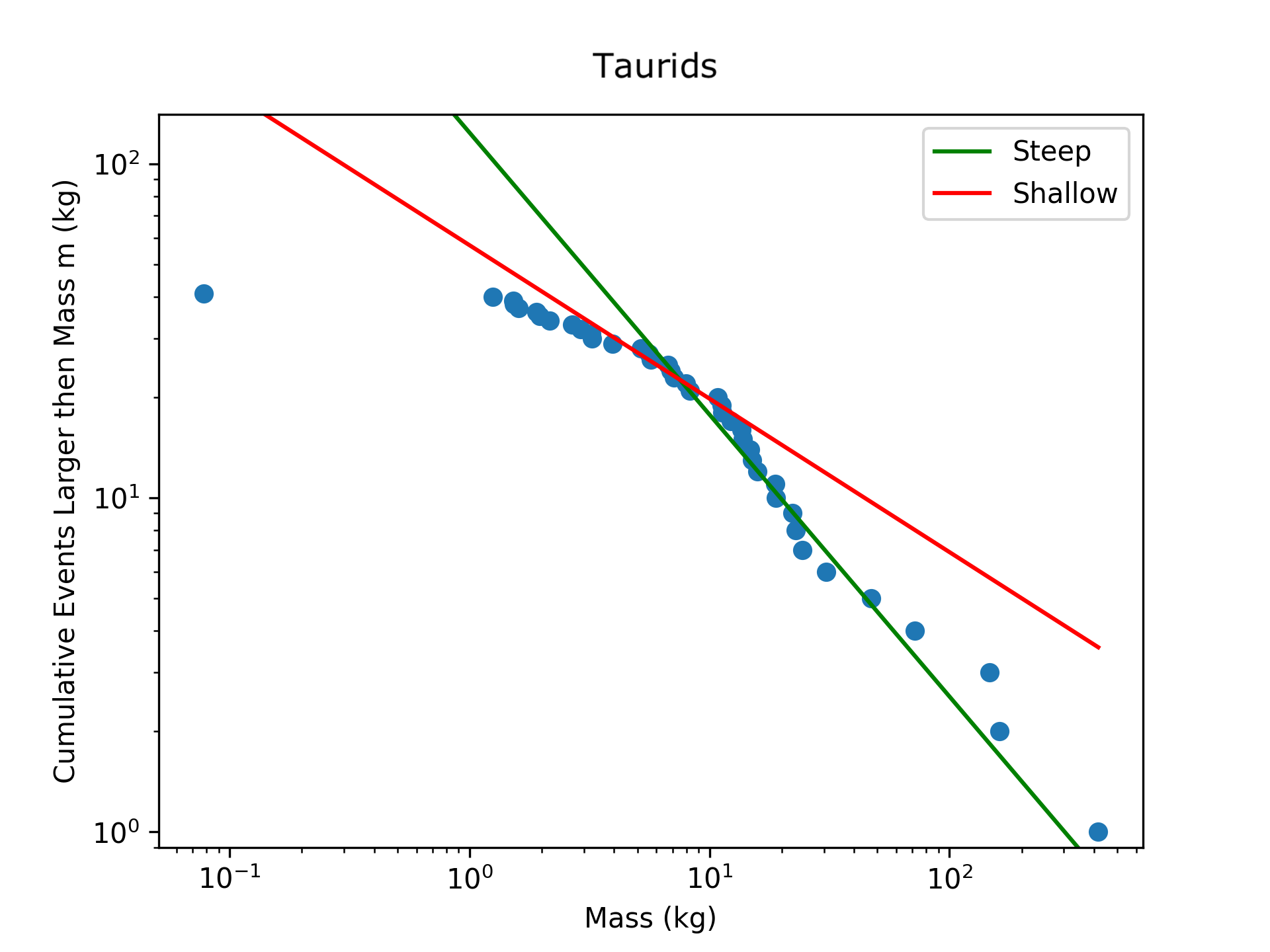}
  \caption{Cumulative Mass Distribution for the Taurids for the GLM-17 Sensor.}
  \label{fig:cm_tau}
\end{figure}

\begin{table}[!ht]
    \centering
    \begin{tabular}{|l|l|l|}
    \hline
        \textbf{Shower} & \textbf{Orionids} & \textbf{Taurids} \\ \hline
        \textbf{Date} & 2022-10-20 09:35 & 2022-10-31 07:34 \\ \hline
        \textbf{Solar Longitude (°)} & 206.66  & 217.54 \\ \hline
        \textbf{Latitude (°)} & 21.9 & 1.7 \\ \hline
        \textbf{Longitude (°)} & -107 & -97 \\ \hline
        \textbf{Sensor } & Stereo - GLM-16 & Stereo - GLM-17 \\ \hline
        \textbf{Look Angle (°)} & 6.0 & 1.9 \\ \hline
        \textbf{Height (km)} & 92.0 & 85.0 \\ \hline
        \textbf{Radiant Altitude (°)} & 61.6 & 69.7 \\ \hline
        \textbf{Total Radiated Energy (j)} & 1.28E+08 & 2.67E+09 \\ \hline
        \textbf{Photometric Mass (kg)} & 2.44 & 146.6 \\ \hline
        \textbf{Magnitude} & -13.2 & -16.9 \\ \hline
        \textbf{Shower Velocity (km/s)} & 66.3 & 27 \\ \hline
        \textbf{Transverse Velocity (km/s)} & 31.5 & 9.3 \\ \hline
        \textbf{GLM Velocity (km/s)} & 61.1 & 23.3 \\ \hline
        \textbf{Radiant Azimuth (°)} & 98.77 & 328.9 \\ \hline
        \textbf{GLM Bearing Angle (°)} & 127.9 & 269.7\\ \hline

    \end{tabular}
    \label{tab:big_app}
    \caption{The most energetic GLM fireballs that are potentially associated to major meteor showers found in our survey using the methodology described in the text. Here the shower radiant altitude is given at the time of the fireball and the total radiated energy is found from the GLM light curve following the procedure described in section \ref{sec:methods}. Here height refers to the height above the Earth surface of the detection. Note the GLM Velocity is the calculated velocity using the Haversine formula, which is also used to compute the GLM bearing angle.}
\end{table}

 \begin{figure}[ht!]
  \includegraphics[width=\linewidth]{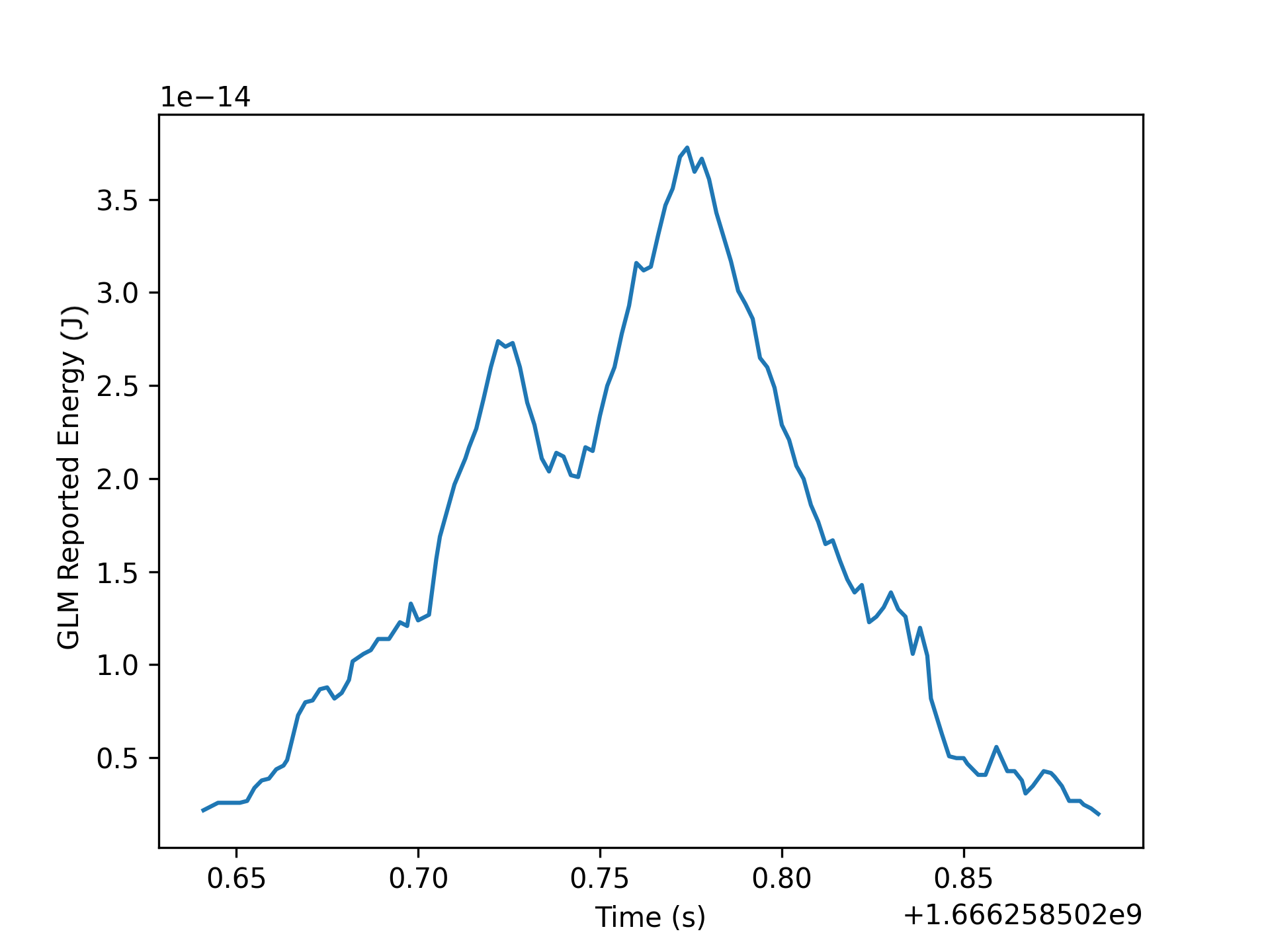}
  \caption{The light curve for the most energetic Orionid fireball as measured by GLM-16 detected on 2022-10-20 09:35:00 with the timescale starting at T0=09:35:00 UT. See Table E.11 for more detail.}
  \label{fig:ori_light}
\end{figure}

 \begin{figure}[ht!]
  \includegraphics[width=\linewidth]{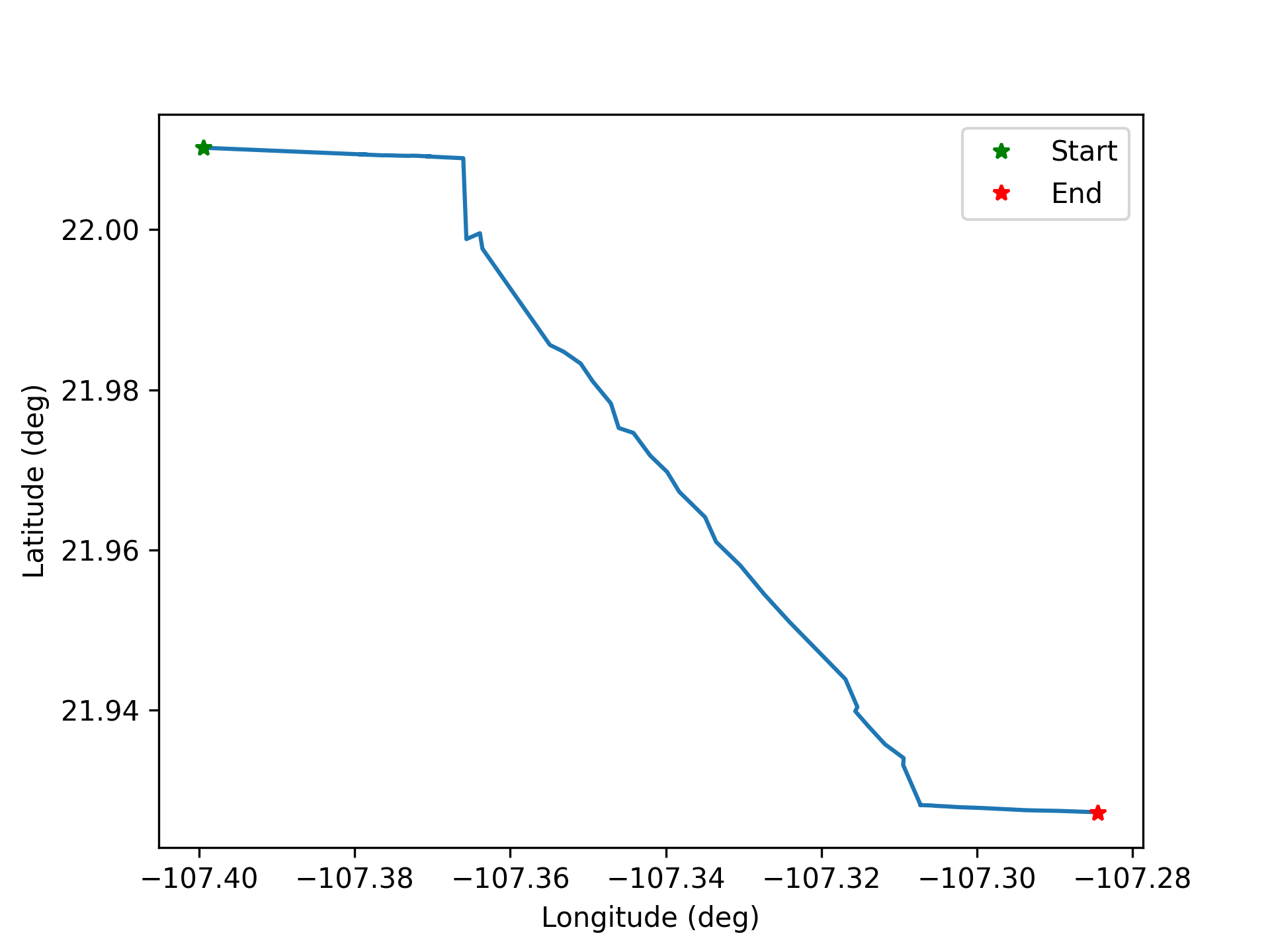}
  \caption{The ground track most energetic Orionid fireball as measured by GLM-16 detected on 2022-10-20 09:35:00. See Table E.11 for more detail.}
  \label{fig:ori_pos}
\end{figure}

 \begin{figure}[ht!]
  \includegraphics[width=\linewidth]{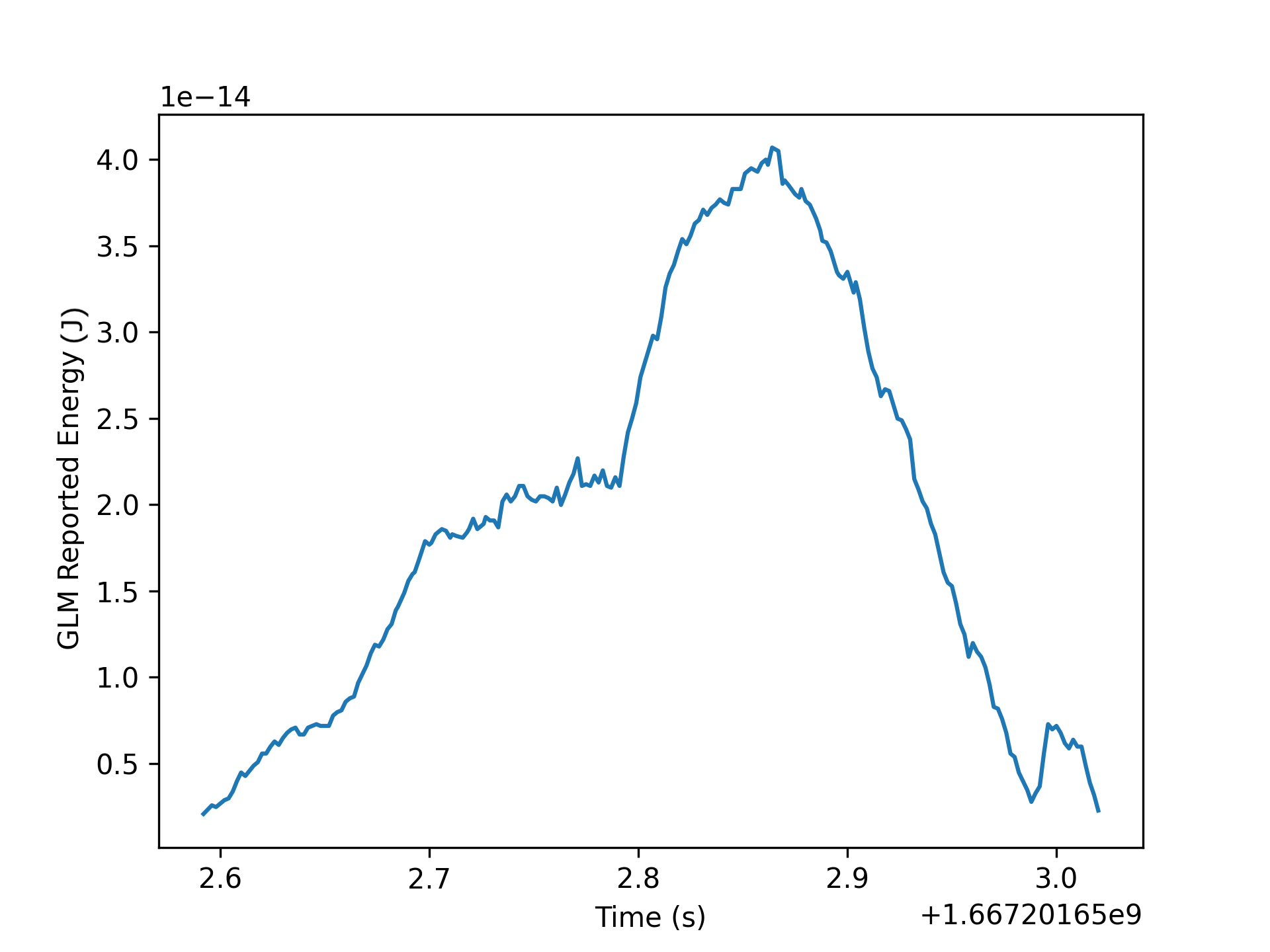}
  \caption{The light curve for the most energetic Taurid fireball as measured by GLM-17 detected on 2022-10-31 07:34:00 with the timescale starting at T0=07:34:00 UT. See Table E.11 for more detail.}
  \label{fig:tau_light}
\end{figure}

 \begin{figure}[ht!]
  \includegraphics[width=\linewidth]{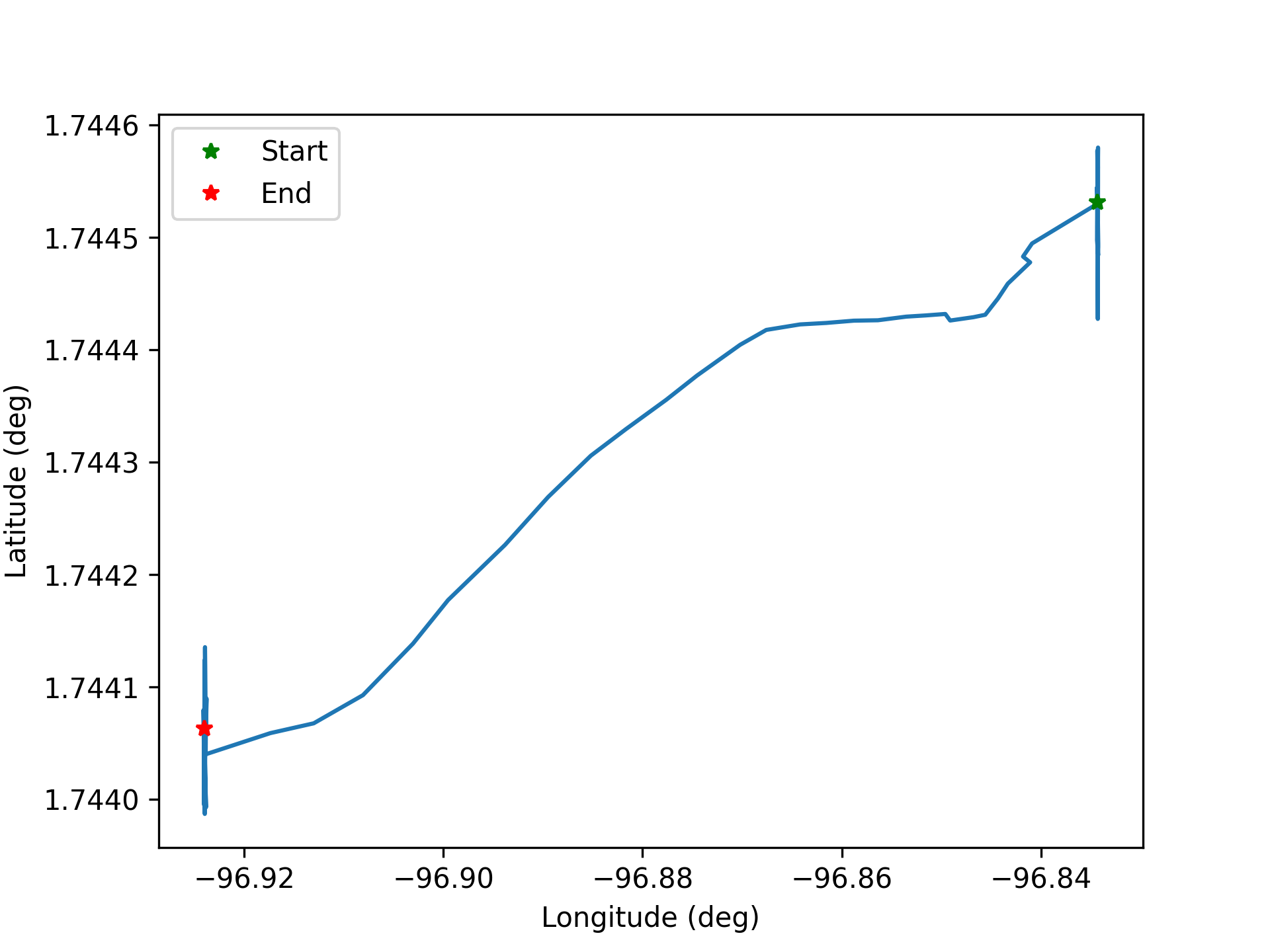}
  \caption{The ground track most energetic Taurid fireball as measured by GLM-16 detected on 2022-10-31 07:34:00. See Table E.11 for more detail.}
  \label{fig:tau_pos}
\end{figure}
\end{document}